\documentclass[%prl,
aps,
 ,twocolumn
]{revtex4}
\usepackage{tabularx}
\usepackage{graphicx}
\usepackage{amsmath}
\usepackage{amssymb}
\usepackage{amsthm}
\usepackage{xcolor}
\usepackage{hyperref}
\usepackage{slashed}
\hypersetup{colorlinks,linkcolor={blue},citecolor={red},urlcolor={blue}}

\usepackage[linesnumbered,ruled,vlined]{algorithm2e}

\newcommand{\ket}[1]{| #1 \rangle}

\DeclareMathOperator{\sech}{sech}

\newcommand{\vt}[1]{{\boldsymbol{#1}}}

\makeatletter
\newsavebox\myboxA
\newsavebox\myboxB
\newlength\mylenA

\newcommand*\xoverline[2][0.75]{%
    \sbox{\myboxA}{$\m@th#2$}%
    \setbox\myboxB\null% Phantom box
    \ht\myboxB=\ht\myboxA%
    \dp\myboxB=\dp\myboxA%
    \wd\myboxB=#1\wd\myboxA% Scale phantom
    \sbox\myboxB{$\m@th\overline{\copy\myboxB}$}%  Overlined phantom
    \setlength\mylenA{\the\wd\myboxA}%   calc width diff
    \addtolength\mylenA{-\the\wd\myboxB}%
    \ifdim\wd\myboxB<\wd\myboxA%
       \rlap{\hskip 0.5\mylenA\usebox\myboxB}{\usebox\myboxA}%
    \else
        \hskip -0.5\mylenA\rlap{\usebox\myboxA}{\hskip 0.5\mylenA\usebox\myboxB}%
    \fi}
\makeatother

\theoremstyle{plain}
\newtheorem{thm}{Theorem}
\theoremstyle{plain}

\newtheorem{conjecture}[thm]{Conjecture}

\begin{document}

\title{Conversion of Gaussian states to non-Gaussian states using photon number-resolving detectors}

\author{Daiqin Su}
\email{daiqin@xanadu.ai}
\affiliation{Xanadu, Toronto, Ontario, M5G 2C8, Canada}
\author{Casey R.~Myers}
\affiliation{Xanadu, Toronto, Ontario, M5G 2C8, Canada}
\author{Krishna Kumar Sabapathy}
\affiliation{Xanadu, Toronto, Ontario, M5G 2C8, Canada}

%\pacs{03.70.+k, 03.65.Ud, 04.62.+v}

\date{\today}

\begin{abstract}
{Generation of high fidelity photonic non-Gaussian states is a crucial ingredient for universal quantum computation using continous-variable platforms, 
yet it remains a challenge to do so efficiently.  We present a general framework for a probabilistic production of multimode non-Gaussian states by 
measuring few modes of multimode Gaussian states via photon-number-resolving detectors. We use Gaussian elements consisting of squeezed displaced 
vacuum states and interferometers, the only non-Gaussian elements consisting of photon-number-resolving detectors. We derive analytic expressions for the 
output Wigner function, and the probability of generating the states in terms of the mean and the covariance matrix of the Gaussian state and the photon 
detection pattern. We find that the output states can be written as a Fock basis superposition state followed by a Gaussian gate, and we derive explicit expressions 
for these parameters. These analytic expressions show exactly what non-Gaussian states can be generated by this probabilistic scheme. Further, 
it provides a method to search for the Gaussian circuit and measurement pattern that produces a target non-Gaussian state with optimal fidelity and success 
probability. We present specific examples such as the generation of cat states, ON states, Gottesman-Kitaev-Preskill states, NOON states and bosonic code states. 
The proposed framework has potential far-reaching implications for the generation of bosonic error-correction codes that require non-Gaussian states, 
resource states for the implementation of non-Gaussian gates needed for universal quantum computation, among other applications requiring non-Gaussianity. 
The tools developed here could also prove useful for the quantum resource theory of non-Gaussianity.}
\end{abstract}

\maketitle

\tableofcontents

\section{Introduction}

Quantum information processing based on continuous-variable systems \cite{RevModPhys.84.621, RevModPhys.77.513} can be broadly divided into the Gaussian and the non-Gaussian domains, consisting of the corresponding states and gates.  
 The distribution of quadratures in phase space of a Gaussian state follows Gaussian statistics. A Gaussian unitary, or more generally a Gaussian operation, transforms a Gaussian state into another Gaussian state. In quantum information architectures based on photonic platforms, the Gaussian states and Gaussian unitaries can be generated and implemented deterministically and thus are easily achievable experimentally. However, generating non-Gaussian states and implementing non-Gaussian gates deterministically are extremely challenging due to the weak nature of interaction Hamiltonians that are polynomials of quadrature operators with order $>2$, e.g., the optical  Kerr nonlinearity is far smaller than what would be required to implement a non-Gaussian gate. Since non-Gaussian states and gates are essential or advantageous to many applications, such as quantum 
optical lithography \cite{boto2000quantum}, quantum metrology \cite{dowling2008quantum},  entanglement distribution~\cite{sabapathy2011robustness}, error correction~\cite{niset2009nogo}, phase estimation~\cite{adesso2009optimal}, bosonic codes~\cite{chuang1997bosonic, PhysRevA.94.012311, bosonicterhal, michael2016new, niu2018hardware,li2017cat,heeres2017implementing}, quantum communication and optical non-classicality~\cite{sabapathy2017nongaussian}, cloning \cite{ngcloning}, 
and in particular to universal quantum computation \cite{knill2001scheme, PhysRevLett.82.1784}, a systematic approach must be found to produce non-Gaussianity.

One potential scheme is to generate non-Gaussian states by performing photon-number detection on a subsystem and post selecting a particular photon-number pattern. 
The requirement of post selection makes this scheme probabilistic, and so increasing the success probability is crucial. It is well known that
a single photon state can be generated by detecting a two-mode squeezed vacuum state via a photon-number-resolving (PNR)
detector with one photon registered \cite{PhysRevLett.56.58, PhysRevLett.87.050402}. More complicated non-Gaussian states like a superposition of several Fock states can be generated
by using the quantum scissor device \cite{pegg1998optical, dakna1999generation, lee2010quantum, resch2002quantum, ozdemir2001quantum, miranowicz2005optical,paris2019}, 
which also uses PNR detectors. However, the quantum scissor device requires non-Gaussian resource states as inputs, e.g.,
single photon states, making it experimentally more challenging. In principle, generation of a single-mode state in the form of a superposition of Fock states up to an arbitrary photon number is 
possible \cite{fiuravsek2005conditional, koniorczyk2000general, villas2001recurrence}. 

An alternative, which is known as photon subtraction \cite{PhysRevA.55.3184}, is a commonly used method for the production of non-Gaussian states.  The generation of Schr\"odinger's cat state, a superposition of two coherent states with opposite phases, by 
measuring a Gaussian state with PNR detectors has been proposed theoretically \cite{PhysRevA.55.3184} and implemented experimentally 
\cite{PhysRevA.55.3184, ourjoumtsev2006generating, PhysRevLett.97.083604, PhysRevLett.101.233605, PhysRevA.82.031802, morin2014remote, jeong2014generation}.
The generation of other non-Gaussian states, such as NOON states \cite{sanders1989quantum, boto2000quantum} %(\textcolor{red}{add more}) 
and small superpositions of Fock states, by photon subtraction
have also been investigated \cite{PhysRevA.75.063803}. The photon subtraction can also be used to tailor more complicated Gaussian states such as the continuous-variable cluster
states \cite{PhysRevLett.121.220501,ra2019}. 

Earlier methods lacked a systematic approach to  know whether a certain protocol is optimal to generate a given target non-Gaussian state. 
By  ``optimal" we mean to generate a target state with the highest fidelity and success probability. Recently~\cite{sabapathy2018near}, a machine learning scheme (also using Gaussian states and PNR detectors) was proposed to 
search the best input states and interferometers that could generate a given target non-Gaussian state, in particular, a superposition of Fock states up to three photons. A very high fidelity target state can be obtained with a substantially enhanced success probability  over previous methods \cite{sabapathy2018near}. Another machine learning method using a genetic algorithm and allowing for certain non-Gaussian input states was also recently investigated \cite{genetic}. 
In this paper, we present a thorough study of the conditional generation of non-Gaussian states by measuring multimode Gaussian states via PNR detectors. The main motivation for this is to study 
the ultimate limit of generating non-Gaussian states by measuring Gaussian states using PNR detectors and to maximize the success probability. This work is also motivated by recent experimental success in the generation of multiphoton states with PNR detectors~\cite{magana2019multiphoton, tiedau2019scalability}. 

\begin{figure}
\includegraphics[width=\columnwidth]{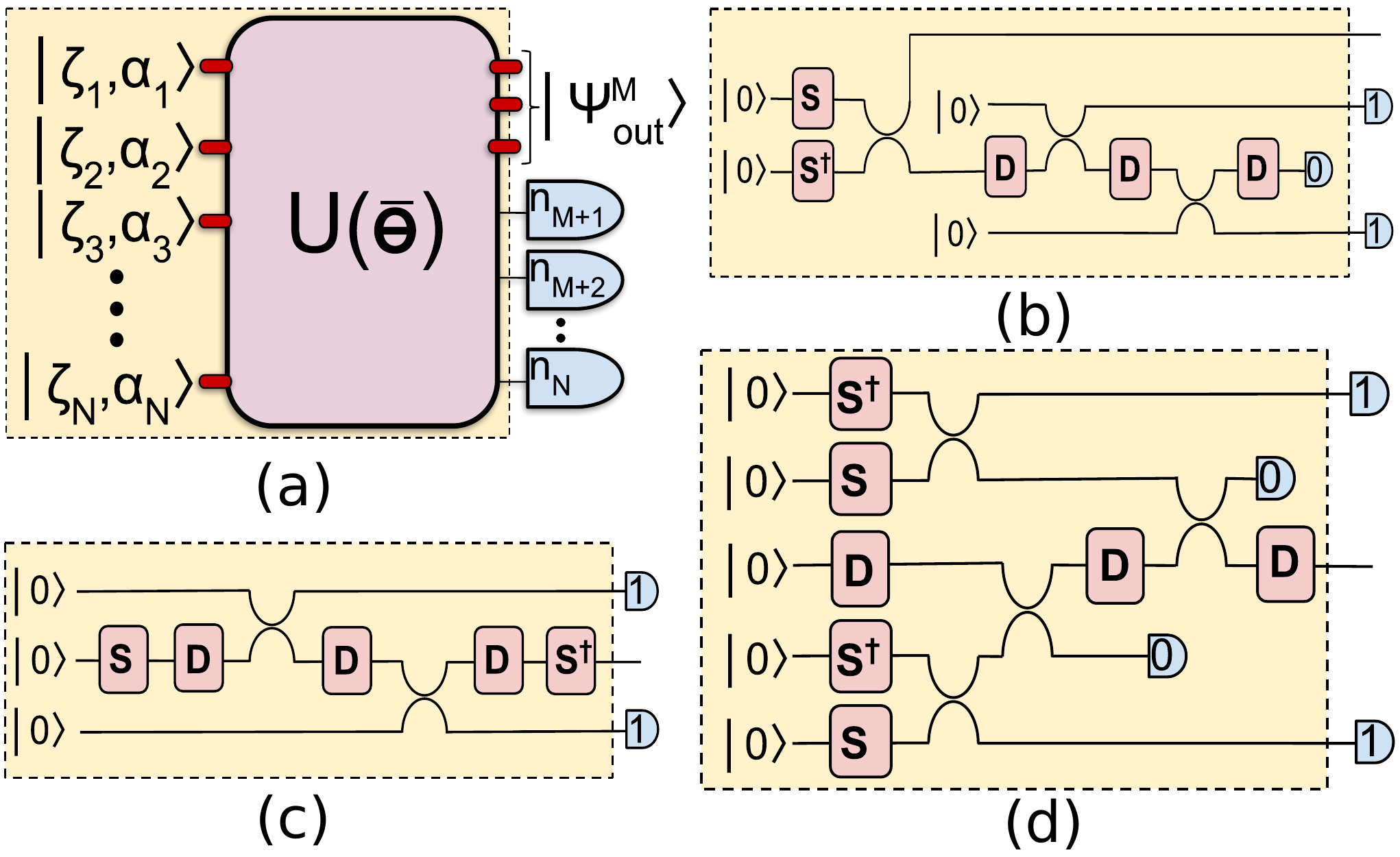}
\caption{Optical schemes for the generation of non-Gaussian states. (a) Our method to measure few modes of a multimode pure Gaussian state.  $\ket{\zeta_i,\alpha_i}$ is a squeezed displaced vacuum state in the $i^{th}$  mode, $U(\bar{\theta})$ is an interferometer, $n_j$ are photon-number-resolving-detector (PNRD) outcomes. (b) Application of repeated displacements and photon subtractions to one arm of a two-mode squeezed vacuum state \cite{yukawa2013}. (c) Utilization of repeated photon subtractions and displacements on a squeezed vacuum state \cite{fiuravsek2005conditional}. (d) Application of repeated displacements and photon additions \cite{dakna1999generation}. The dashed regions in methods (b)-(d) can be mapped to a particular instance of the dashed region in (a). Thus our scheme is the most general heralding scheme using input pure Gaussian states and photon-number-resolving (PNR) measurements.  }
\label{comparisons}
\end{figure}

The general setup we consider is schematically shown in Fig.~\ref{fig:ng-GBS-onemode-mixed} (single-mode output) and Fig.~\ref{fig:ng-GBS-multimode-mixedM} (multimode output). 
We assume that a general multimode Gaussian state (pure or mixed) has been prepared.
Some of the modes of the multimode Gaussian states are measured by PNR detectors, resulting in various photon number patterns. If one post selects a particular
photon number pattern, the heralded state in the remaining modes is generally a non-Gaussian state. There have been many previous universal schemes that use repeated photon-subtraction/photon-addition, along with displacements, for non-Gaussian state generation \cite{dakna1999generation,fiuravsek2005conditional,yukawa2013}. However, our scheme generalizes all of these methods as shown in Fig.~\ref{comparisons}, and therefore provides a concrete way to improve fidelity and success probability.

In this paper, we derive analytic expressions for the Wigner function and 
the probability of generating the heralded non-Gaussian state in terms of the mean and covariance matrix of the multimode Gaussian state, and the measurement outcomes. The resulting heralded state is a superposition of a finite number of Fock states, followed by a Gaussian operation. We provide a procedure to determine the Gaussian 
operation and the coefficients of the superposition of Fock states from the mean and covariance matrix of the multimode Gaussian states. This then answers the question of the type of 
 non-Gaussian states that can be generated. More importantly, we also try to address the inverse problem, namely, to find a Gaussian circuit and a photon detection pattern to generate a given target state with the highest fidelity and 
success probability. We partially solve the inverse problem by optimizing the success probability for specific multimode Gaussian states and measurement patterns 
under certain constraints. These constraints are directly related to the given target states. We demonstrate the proposed formalism by considering example states that are of interest to the wider quantum information community.

The rest of the paper is organized as follows. In Sec. \ref{sec:background}, we briefly introduce some of the required tools, such as the covariance matrix and Wigner function, that are important for the 
rest of the paper. In Sec. \ref{sec:SM-GeneralFormalism}, we derive general analytic expressions for the Wigner function and the success probability of generating 
single-mode non-Gaussian states. We then focus on discussing heralded single-mode non-Gaussian states by detecting multimode 
pure Gaussian states in Sec. \ref{sec:SM-PureFormalism}. Illustrative and relevant examples of  single-mode non-Gaussian states are discussed in Sec. \ref{sec:SM-Example}. 
In Sec. \ref{sec:MM-GeneralFormalism}, we generalize all single-mode results to 
the multimode case. We then focus on discussing heralded multimode 
non-Gaussian states by detecting multimode pure Gaussian states in Sec. \ref{sec:MM-PureFormalism}. We provide some examples of generating multimode non-Gaussian states, such as the W state and NOON states in Sec. \ref{sec:MM-Example}. Finally, we conclude in Sec. \ref{sec:conclusion}.

\section{Phase space methods}\label{sec:background}

We briefly review some background material on continuous-variable (CV) quantum systems that will be used in this paper. An $N$-mode optical field can be described 
by either the creation and annihilation operators, or the position and  momentum  quadratures. We define an operator vector 
$\hat{\vt{\xi}}^{(c)} = (\hat{\vt{a}}^{\dag}, \hat{\vt{a}} )^{\top} = (\hat a_1^{\dag}, \cdots, \hat a_N^{\dag}, \hat a_1, \cdots, \hat a_N)^{\top}$, where  $\hat a_k^{\dag} (\hat a_k)$ 
are the creation (annihilation) operators of the $k$-th optical mode, that satisfy the boson commutation relation $[\hat a_j, \hat a_k^{\dag}] = \delta_{j k}$. 
We also define another operator vector $\hat{\vt{\xi}}^{(r)} = (\hat{\vt{p}}, \hat{\vt{q}} )^{\top} = (\hat p_1, \cdots, \hat p_N, \hat q_1, \cdots, \hat q_N)^{\top}$, where
$\hat q_k$ and $\hat p_k$ are the position and momentum quadratures of the $k$-th optical mode, respectively. In this paper, we set $\hbar = 1$, so the 
position and momentum quadratures satisfy the commutation relation $[\hat q_j, \hat p_k] = i \delta_{j k}$, and they are related to the creation and annihilation 
operators via
\begin{eqnarray}\label{eq:BasisTransformation}
\hat p_k = \frac{i}{\sqrt{2}} (\hat a_k^{\dag} - \hat a_k), ~~~~~~ \hat q_k = \frac{1}{\sqrt{2}} (\hat a_k + \hat a_k^{\dag}). 
\end{eqnarray}
Let us  define a $2N \times 2N$ unitary matrix ${\bf \Omega}$ as
\begin{eqnarray}
\vt{\Omega} = \frac{1}{\sqrt{2}}
	\begin{pmatrix}
	i {\bf I}_N & -i {\bf I}_N \\
	{\bf I}_N &  ~~  {\bf I}_N
	\end{pmatrix},
\end{eqnarray}
where ${\bf I}_N$ is an $N \times N$ identity matrix, and we have 
\begin{eqnarray}\label{eq:BasisTransform}
	\begin{pmatrix}
	\hat{\vt{p}} \\
	\hat{\vt{q}}
	\end{pmatrix}
= {\bf \Omega}
	\begin{pmatrix}
	\hat{\vt{a}}^{\dag} \\
	\hat{\vt{a}}
	\end{pmatrix} ~~ \Leftrightarrow  ~~~ \vt{\hat{\xi}}^{(r)} = {\bf \Omega} \vt{\hat{\xi}}^{(c)}.
\end{eqnarray}

Gaussian states are fully characterized by the first and second moments of the mode operators \cite{RevModPhys.84.621}. 
In the basis $\hat{\vt{\xi}}^{(c)}$, the first moments are the displacements $\vt{Q}^{(c)} = \big\langle \hat{\vt{\xi}}^{(c)} \big\rangle$ and 
the second moments are represented by a covariance matrix ${\bf V}^{(c)}$, defined as
\begin{eqnarray}\label{eq:CMaadag}
V^{(c)}_{jk} = \frac{1}{2} \big\langle \big\{ \hat{\xi}^{(c)}_j, \, \hat{\xi}^{(c) \dag}_k \big\} \big\rangle - \big \langle \hat{\xi}^{(c)}_j \big\rangle \big \langle \hat{\xi}^{(c) \dag}_k \big\rangle,
\end{eqnarray}
where $\{ \cdot, \cdot\}$ represents the anticommutator. 
To be a valid physical covariance matrix, it must satisfy the uncertainty relation\,\cite{PhysRevA.49.1567} 
\begin{align}\label{eq:Cuncertainty}
    {\bf V}^{(c)} + \frac{{{\bf \Sigma}_3}}{2} \geq 0, 
    ~~~~~~
    {\bf \Sigma}_3 =
    \begin{pmatrix}
        {\bf I}_N & {\bf 0} \\
        {\bf 0} & -{\bf I}_N
    \end{pmatrix}.
\end{align}
In terms of $\big\{\hat{\vt{\xi}}^{(r)} \big\}$, the first moments are the displacements $\vt{Q}^{(r)} = \big\langle \hat{\vt{\xi}}^{(r)} \big\rangle$ and 
the second moments are represented by a covariance matrix ${\bf V}^{(r)}$ defined as
\begin{eqnarray}\label{eq:CMpq}
V^{(r)}_{jk} = \frac{1}{2} \big\langle \big\{ \hat{\xi}^{(r)}_j, \, \hat{\xi}^{(r)}_k \big\} \big\rangle - \big \langle \hat{\xi}^{(r)}_j \big\rangle \big \langle \hat{\xi}^{(r)}_k \big\rangle.
\end{eqnarray}
By using Eq.~\eqref{eq:BasisTransform}, we have 
\begin{eqnarray}\label{eq:CMDtransformation}
{\bf V}^{(r)} = {\bf \Omega} {\bf V}^{(c)} {\bf \Omega}^{\dag}, ~~~~~~ \vt{Q}^{(r)} = {\bf \Omega} \vt{Q}^{(c)}. 
\end{eqnarray}
Using Eq.~\eqref{eq:CMDtransformation}, we find that the uncertainty relation in Eq.~\eqref{eq:Cuncertainty} transforms  to 
\begin{align}\label{eq:Runcertainty}
    {\bf V}^{(r)} + \frac{i{\bf \Theta}}{2} \geq 0, ~~~~~~ {\bf \Theta} = \begin{pmatrix}
        {\bf 0} & -{\bf I}_N\\
       {\bf I}_N & {\bf 0}
    \end{pmatrix}.
\end{align}

The picture is different for non-Gaussian states where the first and second moments alone are not enough to describe the non-Gaussian state. The Wigner function is thus a useful representation  to completely characterize all CV quantum states. 
In the coherent state basis, the Wigner function for an $N$-mode state is defined as
\begin{eqnarray}\label{eq:WignerCoherentBasis}
W(\vt \alpha;\rho) = \frac{1}{\pi^{2N}} \int \mathrm d^2 \vt{\beta} ~ e^{- \vt{\beta}^{\top} \vt{\alpha}^* + \vt{\alpha}^{\top} \vt{\beta}^* } \chi(\vt{\beta};\rho),
\end{eqnarray}
where $\vt{\alpha} = (\alpha_1, \cdots, \alpha_N)^{\top}$, $\vt{\beta} = (\beta_1, \cdots, \beta_N)^{\top}$,
$\mathrm d^2 \beta_k = \mathrm d \beta_k^R \mathrm d \beta_k^I $, with $\beta_k^R$ and $\beta_k^I$ the real and imaginary parts of $\beta_k$, 
and $\chi(\vt{\beta};\rho)$ is the characteristic function,
\begin{eqnarray}\label{eq:CharacteristicF-single}
\chi(\vt{\beta};\rho)
= \text{Tr}\big[ \hat D(\vt \beta) \rho \big]
\end{eqnarray}
with $\rho$ the density matrix and $\hat D(\vt \beta) = e^{\vt{\beta}^{\top} {\hat{\vt a}}^{\dag} - \vt{\beta}^{\dag} \hat{\vt a}}$ the Weyl-Heisenberg displacement operators. The Wigner function $W(\vt{\alpha};\rho)$ is a real function on the phase space and is normalized to one:
\begin{eqnarray}\label{eq:WignerAnormalization}
\int \mathrm d^2 \vt{\alpha} ~ W(\vt{\alpha};\rho) = \text{Tr} (\rho) = 1. 
\end{eqnarray}

There are two conventions to obtain the Wigner function $W(\vt{\alpha};\rho)$ in terms of $\vt{p}$ and $\vt{q}$, 
where $\vt p = (p_1, \cdots, p_N)^{\top}$ and $\vt q = (q_1, \cdots, q_N)^{\top}$. First, analogous to Eq.~\eqref{eq:BasisTransformation}, we define the relation between the pairs $\vt{\xi}^{(r)}:= (\vt{p}, \vt{q})^{\top}$ and $\vt{\xi}^{(c)}:= (\vt{\alpha}^*, \vt{\alpha})^{\top}$ as $p_k = i (\alpha_k^* - \alpha_k)/\sqrt{2}$, $q_k = (\alpha_k^* + \alpha_k)/\sqrt{2}$. Using these relations one can write down $W(\vt{\alpha};\rho)$ in terms of 
$\vt p$ and $\vt q$ as $W(\vt p, \vt q;\rho)$. The second convention is to work in the $q$-$p$ basis where the  Wigner function for an $N$-mode state is defined as
\begin{eqnarray}\label{eq:WignerXPgeneral}
\xoverline{W}(\vt p, \vt q;\rho) = \frac{1}{\pi^N} \int \mathrm d \vt y \, e^{- 2i \vt{p}^{\top} \vt{y}} \langle \vt q - \vt y | \rho | \vt q + \vt y \rangle,
\end{eqnarray}
where $\vt y = (y_1, \cdots, y_N)^{\top}$ is a real vector. %and $\vt q = (q_1, \cdots, q_N)^{\top}$. 
The Wigner function $\xoverline{W}(\vt p, \vt q;\rho)$ is normalized to one in the following way,
\begin{eqnarray}\label{eq:WignerXPnormalization}
\int \mathrm d \vt{p} \, \mathrm d \vt{q} ~ \xoverline{W}(\vt p, \vt q;\rho) = \text{Tr} (\rho) = 1. 
\end{eqnarray}
 However, due to the convention we use, we find by comparing Eqs.~\eqref{eq:WignerAnormalization} and \eqref{eq:WignerXPnormalization} that 
\begin{align}\label{eq:convention}
W(\vt p, \vt q;\rho) = 2^N \xoverline{W}(\vt p, \vt q;\rho). 
\end{align}

For Gaussian states  the Wigner function is Gaussian and is fully determined by the displacements and the covariance matrix. In the coherent 
state basis with $\Delta \vt{\xi}_1 = \big[ {\vt{\xi}}^{(c)} - \vt{Q}^{(c)} \big]$,
\begin{eqnarray}\label{eq:WignerCoherent}
 W(\vt \alpha;\rho)  
=  \frac{2^N}{\pi^N} \exp\bigg\{ - \frac{1}{2} \big( \Delta \vt{\xi}_1 \big)^{\dag} \big[ {\bf V}^{(c)} \big]^{ -1} \big( \Delta \vt{\xi}_1 \big)  \bigg\}; %\nonumber\\
\end{eqnarray}
in the $q$-$p$ basis with $\Delta \vt{\xi}_2 = \big[ {\vt{\xi}}^{(r)} - \vt{Q}^{(r)} \big]$,
\begin{eqnarray}\label{eq:wbarxp}
\xoverline{W}(\vt p, \vt q;\rho)  
=  \frac{1}{\pi^N} \exp\bigg\{ - \frac{1}{2} \big( \Delta \vt{\xi}_2 \big)^{\top} \big[ {\bf V}^{(r)} \big]^{ -1} \big( \Delta \vt{\xi}_2 \big)  \bigg\}. %\nonumber\\
\end{eqnarray}

Any Gaussian unitary can be described in the complex basis through the associated symplectic transformation ${\bf S}^{(c)}$ and a displacement $ \vt{d}^{(c)}$. Under the action of this Gaussian unitary operator, the covariance matrix and the Wigner function %, and the first moments of a state $\rho$ 
transform as 
\begin{align}\label{eq:WignerTF}
&\hat{\vt{\xi}}^{(c)} \rightarrow  {\bf S}^{(c)} \hat{\vt{\xi}}^{(c)} + \vt{d}^{(c)},\nonumber\\
&{\bf V}^{(c)} \rightarrow {\bf S}^{(c)} {\bf V}^{(c)} {\bf S}^{(c) \dag},\nonumber\\
&W\big({\vt{\xi}}^{(c)};\rho \big) \rightarrow  W\bigg( \big[{\bf S}^{(c)}\big]^{-1} \big( {\vt{\xi}}^{(c)} - \vt{d}^{(c)} \big);\rho \bigg). 
\end{align}
When the Gaussian transformation is described in real form through ${\bf S}^{(r)}$ and $ \vt{d}^{(r)}$, the analogous transformations of the phase space properties can be written as % which transform $\hat{\vt{\xi}}^{(r)}$ as
\begin{align}
&\hat{\vt{\xi}}^{(r)} \rightarrow  {\bf S}^{(r)} \hat{\vt{\xi}}^{(r)} + \vt{d}^{(r)}, \nonumber\\
&{\bf V}^{(r)} \rightarrow {\bf S}^{(r)} {\bf V}^{(r)} {\bf S}^{(r) \top}, \nonumber\\
&\xoverline{W} \big({\vt{\xi}}^{(r)} ;\rho\big) \rightarrow  \xoverline{W} \bigg( \big[{\bf S}^{(r)}\big]^{-1} \big( {\vt{\xi}}^{(r)} - \vt{d}^{(r)} \big);\rho \bigg). 
\end{align}
With this background material, we next move on to the preparation of single-mode non-Gaussian states using multimode Gaussian states.

\section{General formalism for single-mode output states}\label{sec:SM-GeneralFormalism}

\begin{figure}
    \centering
    \scalebox{0.7}{\includegraphics{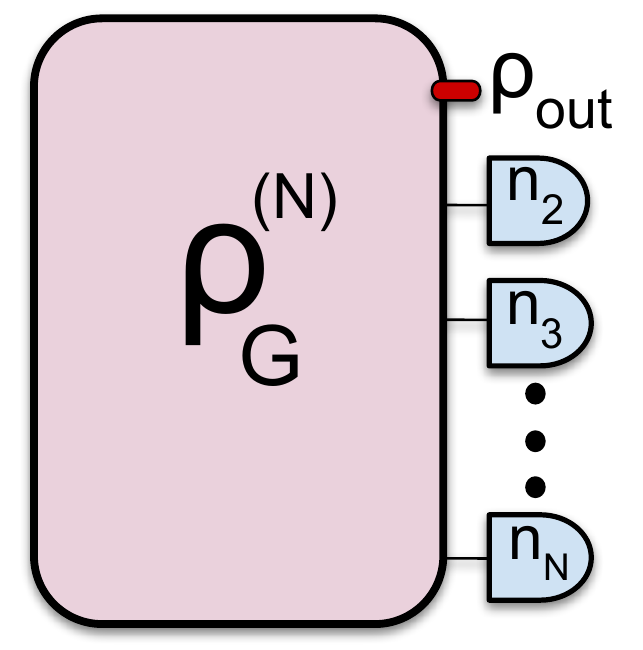}}
    \caption{Probabilistic generation of single-mode non-Gaussian states. Here, we consider a general multimode Gaussian state $\rho^{(N)}_G$ of $N$-modes. All but one of the modes are measured using PNRDs giving values $n_k$ ($k=2, 3, \cdots, N$), resulting in a conditional output state $\rho_{\rm out}$ in the remaining mode. }
    \label{fig:ng-GBS-onemode-mixed}
\end{figure}

We now discuss the generation of single-mode non-Gaussian states when all but one of the modes of a multimode Gaussian state are measured using photon-number-resolving detectors (PNRDs) as schematically depicted in Fig.~\ref{fig:ng-GBS-onemode-mixed}.  This is the simplest case to begin with and we consider multimode output states later in  Sec.~\ref{sec:MM-GeneralFormalism}.  If all the PNRDs register no photons then the output corresponds to a Gaussian state, otherwise it is non-Gaussian. This single-mode case includes some very important non-Gaussian
states such as Schr\"odinger's cat state, ON state, the cubic phase state and the Gottesmann-Kitaev-Preskill (GKP) state.  

We are now going to derive the Wigner function of the single-mode non-Gaussian state in the coherent state basis. The derivation is summarized as follows. First, we
expand the density matrix $\rho^{(N)}_G \equiv \rho$ of the $N$-mode Gaussian state in the coherent state basis. Second, we project the density matrix $\rho$ onto the Fock state
$| \bar{\vt{n}} \rangle = | n_2, n_3, \cdots, n_N \rangle$ and obtain the unnormalized density matrix of the first mode: $\tilde \rho_1 = \langle \bar{\vt{n}} | \rho \ket{\bar{\vt{n}}}$.
Without loss of generality, we assume that the last $(N-1)$ modes were detected and $n_k$ is the number of photons registered at the $k$-th photon-number-resolving detector (PNRD). Third, by using the transformation 
between the coherent state basis and the Fock state basis, and the relation between the density matrix and Wigner function, we find the unnormalized 
Wigner function $ W(\alpha;\tilde{\rho}_1)$. Finally, the measurement probability $P(\bar{\vt{n}})$ is obtained from the trace of the unnormalized density matrix, i.e., $P(\bar{\vt{n}}) = \text{Tr}(\tilde \rho_1) $. 

\subsection{Single-mode output Wigner function}

Coherent states form an over-complete basis. We can expand the density matrix  $\rho$ of an $N$-mode Gaussian state in the coherent state basis as
\begin{eqnarray}\label{eq:DMcoherent}
\rho = \frac{1}{\pi^{2N}} \int \mathrm d^2 \vt{\alpha} \int \mathrm d^2 \vt{\beta} ~ | \vt{\beta} \rangle \langle \vt{\beta} | \rho | \vt{\alpha} \rangle \langle \vt{\alpha} |,
\end{eqnarray}
where $| \vt{\alpha} \rangle = |\alpha_1, \alpha_2, \cdots, \alpha_N \rangle $ and $| \vt{\beta} \rangle = |\beta_1, \beta_2, \cdots, \beta_N \rangle$.
It can be shown that $\langle \vt{\beta} | \rho | \vt{\alpha} \rangle$ can be expressed in terms of the Wigner function as \cite{dodonov1994photon},
\begin{eqnarray}\label{eq:DOcoherent}
\langle \vt{\beta} | \rho | \vt{\alpha} \rangle = \frac{1}{(2 \pi)^{N}} \int \mathrm d \vt{p} \, \mathrm d \vt{q} ~
\xoverline{W} (\vt{p}, \vt{q};\rho) W_{\vt{\alpha \vt{\beta}}}(\vt{p}, \vt{q};\rho),
\end{eqnarray}
where 
$W_{\vt{\alpha \vt{\beta}}}(\vt{p}, \vt{q};\rho)$
is the Wigner-Weyl transformation of the operator $| \vt{\alpha} \rangle \langle \vt{\beta} |$ given by \cite{dodonov1994multidimensional} 
\begin{eqnarray}
&&W_{\vt{\alpha \vt{\beta}}}(\vt{p}, \vt{q};\rho) = 2^N \exp \bigg\{ -\frac{|\vt{\alpha}|^2+|\vt{\beta}|^2}{2} - \vt{\alpha}^{\top} \vt{\beta}^*  
\nonumber\\
&& 
 - \vt{p}^{\top} \vt{p} - \vt{q}^{\top} \vt{q} + \sqrt{2} \, \vt{\alpha}^{\top} (\vt{q} - i \vt{p}) + \sqrt{2} \, \vt{\beta}^{\dag}(\vt{q} + i \vt{p}) 
\bigg\}.  \nonumber
\end{eqnarray} 
Using the expression in Eq.~\eqref{eq:wbarxp} and performing a Gaussian integration in Eq.~\eqref{eq:DOcoherent}, one obtains \cite{dodonov1994multidimensional}
\begin{eqnarray}\label{eq:DOcoherent-Matrix}
\langle \vt{\beta} | {\rho} | \vt{\alpha} \rangle = \mathcal{P}_0 \exp\bigg(-\frac{| \tilde{\vt{\gamma}} |^2}{2} + \frac{1}{2} \tilde{\vt{\gamma}}^{\top} \tilde{\bf R} \tilde{\vt{\gamma}}
+ \tilde{\vt{\gamma}}^{\top} \tilde{\vt{y}}
\bigg),
\end{eqnarray} 
where $\tilde{\vt{\gamma}} = (\vt{\beta}^*, \vt{\alpha})^{\top}$ and 
\begin{eqnarray}\label{eq:RYpq}
\tilde{\bf R} &=& {\bf \Omega}^{\top} \big[ 2{\bf V}^{(r)} - {\bf I}_{2N}][2{\bf V}^{(r)} + {\bf I}_{2N} \big]^{-1} {\bf \Omega}, \nonumber\\
\tilde{\vt{y}} &=& 2 \, {\bf \Omega}^{\top} \big[ 2 {\bf V}^{(r)} + {\bf I}_{2N} \big]^{-1} \vt{Q}^{(r)}, \nonumber\\
\mathcal{P}_0 &=&  \frac{2^{N} \, \exp\big( - \frac{1}{2}  \vt{Q}^{(r) \top} {\bf \Omega} ^* \tilde{\vt{y}} \big)}{\sqrt{\text{det}\big( 2 {\bf V}^{(r)} + {\bf I}_{2N} \big) }}. 
\end{eqnarray}
Here, $\tilde{\bf R}$ is a $2N \times 2N$ symmetric complex matrix and $\tilde{\vt{y}}$ is a vector with $2N$ components. 
By using the relation $\vt{\Omega}^{\top} \vt{\Omega} = {\bf X}_{2N}$ and Eq.~\eqref{eq:CMDtransformation}, we can rewrite the
quantities $\tilde{\bf R}$, $\tilde{\vt{y}}$ and $\mathcal{P}_0$ in terms of ${\bf V}^{(c)}$ and $\vt{Q}^{(c)}$ as
\begin{eqnarray}\label{eq:RYcoherent}
\tilde{\bf R} &=& {\bf X}_{2N} \big[ 2 {\bf V}^{(c)} -  {\bf I}_{2N} \big] \big[ 2{\bf V}^{(c)} + {\bf I}_{2N} \big]^{-1}, \nonumber\\
\tilde{\vt{y}} &=& 2 \, {\bf X}_{2N} \big[  2{\bf V}^{(c)} + {\bf I}_{2N} \big]^{-1} \vt{Q}^{(c)}, \nonumber\\
\mathcal{P}_0 &=&  \frac{2^{N} \, \exp\big( - \frac{1}{2}  \vt{Q}^{(c) \top} \tilde{\vt{y}} \big)}{\sqrt{\text{det}\big( 2 {\bf V}^{(c)} + {\bf I}_{2N} \big) }}.
\end{eqnarray}

Let us measure the last $(N-1)$ modes using PNRDs and obtain a photon number pattern $\bar{\vt{n}} = (n_2, n_3, \cdots, n_N)$, namely, the projected state in the 
detected modes is $| \bar{\vt{n}} \rangle = | n_2, n_3, \cdots, n_N \rangle$. By using Eqs.~\eqref{eq:DMcoherent} and \eqref{eq:DOcoherent-Matrix}
we find that the unnormalized density matrix $\tilde \rho_1$ of the heralded mode is 
\begin{align}\label{eq:DM-1}
&\tilde \rho_1 = \langle \bar{\vt{n}} | \rho | \bar{\vt{n}} \rangle 
\nonumber\\ 
&= \frac{1}{\pi^{2N}} \int \mathrm d^2 \vt{\alpha} \int \mathrm d^2 \vt{\beta} ~
\langle \bar{\vt{n}} | \vt{\beta} \rangle \langle \vt{\alpha} | \bar{\vt{n}} \rangle \, \langle \vt{\beta} | \rho | \vt{\alpha} \rangle \nonumber\\
&=
%\frac{1}{\pi^{2N}} \int \vt{\mathrm{d} x}\, 
\frac{1}{\pi^{2N}} \int \, \mathrm d^2 {\alpha}_1 \,\mathrm d^2 \beta_1  \,\mathrm d^2 \bar{\vt{\alpha}}  \,\mathrm d^2 \bar{\vt{\beta}} ~
| \beta_1 \rangle \langle \alpha_1 | \, \langle \bar{\vt{n}} | \bar{\vt{\beta}} \rangle \langle \bar{\vt{\alpha}} | \bar{\vt{n}} \rangle \, \nonumber\\
&~~~~~~~~~\times \mathcal{P}_0 \exp\bigg(-\frac{| \tilde{\vt{\gamma}} |^2}{2} + \frac{1}{2} \tilde{\vt{\gamma}}^{\top} \tilde{\bf R} \tilde{\vt{\gamma}} + \tilde{\vt{\gamma}}^{\top} \tilde{\vt{y}} \bigg), %\nonumber\\ 
%&\text{with } \vt{\mathrm{d} x} =  \mathrm d^2 {\alpha}_1 \,\mathrm d^2 \beta_1  \,\mathrm d^2 \bar{\vt{\alpha}}  \,\mathrm d^2 \bar{\vt{\beta}}, ~
\end{align}
where we have defined $| \bar{\vt{\alpha}} \rangle = | \alpha_2, \alpha_3, \cdots, \alpha_N \rangle$ and $| \bar{\vt{\beta}} \rangle = | \beta_2, \beta_3, \cdots, \beta_N \rangle$. 
The inner product $\langle n_k \ket{\alpha_k}$ represents the transformation between the Fock state basis and the coherent state basis, and can be calculated using the 
Fock state expansion of the coherent state. A coherent state $\ket{\alpha_k}$ is given by 
\begin{eqnarray}\label{eq:CoherentState}
\ket{\alpha_k} = e^{-|\alpha_k|^2/2} \sum_{n_k=0}^{\infty} \frac{\alpha_k^{n_k}}{\sqrt{n_k!}} \ket{n_k},
\end{eqnarray}
so we have 
\begin{eqnarray}\label{eq:CoherentFock}
\langle \bar{\vt{n}} | \bar{\vt{\beta}} \rangle \langle \bar{\vt{\alpha}} | \bar{\vt{n}} \rangle 
= \frac{1}{\bar{\vt{n}}!} \, e^{-(|\bar{\vt{\alpha}}|^2 + |\bar{\vt{\beta}}|^2 )/2} \prod_{k=2}^{N} \big(\alpha_k^* \beta_k \big)^{n_k},
\end{eqnarray}
where $\bar{\vt{n}}! \equiv n_2! n_3! \cdots n_N!$.

In Eq.~\eqref{eq:DM-1}, the integration variables have been divided into two sets, one of which corresponds to the heralded mode $\alpha_1, \beta_1$ and the other corresponds to the detected modes $\vt{\bar{\alpha}}, \vt{\bar{\beta}}$. To perform the integration, we also need to decompose the exponential term in Eq.~\eqref{eq:DM-1} into parts corresponding to the
heralded mode, detected modes and their overlap. To do that we 
define a new vector $\vt{\gamma} = (\beta_1^*, \alpha_1, \beta_2^*, \beta_3^*, \cdots, \beta_N^*, \alpha_2, \alpha_3, \cdots, \alpha_N)^{\top} 
= (\vt{\gamma}_h, \vt{\gamma}_d)^{\top}$, where $\vt{\gamma}_h$ and $\vt{\gamma}_d$ are vectors corresponding to the heralded mode and detected modes, respectively.
The vectors $\vt{\gamma}$ and $\tilde{\vt{\gamma}}$ are  related by a permutation matrix ${\bf P}$, namely, $\vt{\gamma} = {\bf P} \tilde{\vt{\gamma}}$. 
The action of ${\bf P}$ is to permute the ($N+1$)-th component of $\tilde{\vt{\gamma}}$ to the second component.
Correspondingly, we define a new symmetric matrix ${\bf R}$ and a new vector $\vt{y}$ as 
\begin{align} \label{eq:Ry}
    {\bf R} = {\bf P} \tilde{\bf R} {\bf P}^{\top}, ~~~~~~ {\vt{y}} = {\bf P}\tilde{\vt{y}}.
\end{align}
The matrix ${\bf R}$ can be partitioned into
\begin{eqnarray}\label{eq:R}
{\bf R} =
	\begin{pmatrix}
	{\bf R}_{hh} & {\bf R}_{hd} \\
	{\bf R}_{dh} & {\bf R}_{dd}
	\end{pmatrix},
\end{eqnarray}
where ${\bf R}_{hh}$ is a $2 \times 2$ symmetric matrix corresponding to the heralded mode, ${\bf R}_{dd}$ is a $(2N-2) \times (2N-2)$ symmetric matrix
corresponding to the detected modes and ${\bf R}_{hd}$ is a $2 \times (2N-2)$ matrix that represents the connections between the detected modes and 
heralded mode. Since ${\bf R}$ is symmetric, ${\bf R}_{dh} = {\bf R}_{hd}^{\top}$. Similarly, the vector $\vt{y}$ is partitioned into $(\vt{y}_h, \vt{y}_d)^{\top}$, 
where $\vt{y}_h$ corresponds to the heralded mode and $\vt{y}_d$ corresponds to the detected modes. 

The three terms in the exponential in Eq.~\eqref{eq:DM-1} become
\begin{eqnarray}\label{eq:ModeSaperation}
| \tilde{\vt{\gamma}} |^2 &=& %| \vt{\gamma} |^2 = 
| \vt{\gamma}_h |^2 + | \vt{\gamma}_d |^2, 
\nonumber\\
\tilde{\vt{\gamma}}^{\top} \tilde{\vt{y}} &=& %{\vt{\gamma}}^{\top} {\vt{y}} = 
\vt{\gamma}_h^{\top} \vt{y}_h + \vt{\gamma}_d^{\top} \vt{y}_d, 
%\nonumber\\
\\
\tilde{\vt{\gamma}}^{\top} \tilde{\bf R} \tilde{\vt{\gamma}} &=& %\vt{\gamma}^{\top} {\bf R} \vt{\gamma} =
	\vt{\gamma}_h^{\top} {\bf R}_{hh} \vt{\gamma}_h + \vt{\gamma}_d^{\top} {\bf R}_{dd} \vt{\gamma}_d 
	+ 2 \, \vt{\gamma}_h^{\top} {\bf R}_{hd} \vt{\gamma}_d. \nonumber
\end{eqnarray}
Substituting Eqs.~\eqref{eq:CoherentFock} and \eqref{eq:ModeSaperation} into Eq.~\eqref{eq:DM-1}, we find that the unnormalized density matrix $\tilde \rho_1$ can be written as
\begin{eqnarray}\label{eq:DM-3}
\tilde \rho_1 = \frac{1}{\pi^{2}} \int \mathrm d^2 {\alpha}_1 \int \mathrm d^2 \beta_1 ~ | \beta_1 \rangle \langle \alpha_1 | F(\alpha_1, \beta_1),
\end{eqnarray}
where
\begin{align}\label{eq:Ffunction}
&F(\alpha_1, \beta_1) \nonumber \\ 
&=
\frac{\mathcal{P}_0 \,\exp( L_2)}{\pi^{2N-2}\bar{\vt{n}}!}
\,  \int \mathrm d^2 \bar{\vt{\alpha}} \,\mathrm d^2 \bar{\vt{\beta}} ~
\prod_{k=2}^{N} \big(\alpha_k^* \beta_k \big)^{n_k}
\exp (L_3)
\nonumber\\
&=
\frac{\mathcal{P}_0}{\bar{\vt{n}}!}
\, \exp (L_2)
\prod_{k=2}^{N} \bigg(\frac{\partial^2}{\partial \alpha_k \partial \beta_k^*} \bigg)^{n_k}
\exp (L_3)
\bigg|_{\vt{\gamma}_d = {\bf 0}},\nonumber\\
&L_2 = -\frac{1}{2} | \vt{\gamma}_h |^2 + \frac{1}{2} \vt{\gamma}_h^{\top} {\bf R}_{hh} \vt{\gamma}_h + \vt{\gamma}_h^{\top} \vt{y}_h , \nonumber\\
& L_3 = - | \vt{\gamma}_d |^2 + \frac{1}{2} \vt{\gamma}_d^{\top} {\bf R}_{dd} \vt{\gamma}_d + \vt{\gamma}_d^{\top} \vt{y}_d + \vt{\gamma}_d^{\top} {\bf R}_{dh} \vt{\gamma}_h.
\end{align}
In the second equality of Eq.~\eqref{eq:Ffunction}, we have performed integration by parts over $\bar{\vt{\alpha}}$ and $\bar{\vt{\beta}}$, the details of which 
are given in Eq.~\eqref{eq:IntegralDerivative} of Appendix~\ref{appedix:IntegralDerivative}. 

From the unnormalized density matrix $\tilde \rho_1$ we can calculate the unnormalized characteristic function $ \chi(\beta;\tilde \rho_1)$
and the unnormalized Wigner function $ W(\alpha;\tilde \rho_1)$. By substituting  $\tilde \rho_1$ into Eq.~\eqref{eq:CharacteristicF-single} we have
\begin{align}
& \chi(\beta;\tilde \rho_1) 
=
e^{-|\beta|^2/2} \text{Tr}\big( e^{ -\beta^* \hat a} \tilde \rho_1 e^{\beta {\hat a}^{\dag}}  \big) \nonumber\\
&=\frac{1}{\pi^{2}}  e^{-|\beta|^2/2} \int \mathrm d^2 {\alpha}_1 \, \mathrm d^2 \beta_1 ~ 
 e^{ \beta \alpha_1^* - \beta^* \beta_1 } \langle \alpha_1  | \beta_1 \rangle F(\alpha_1, \beta_1), \nonumber
\end{align}
where we have used the fact that the coherent state is the eigenstate of the annihilation operator, $\hat a \ket{\alpha} = \alpha \ket{\alpha}$.
Substituting  $ \chi(\beta;\tilde \rho_1)$ into Eq.~\eqref{eq:WignerCoherentBasis} we find the unnormalized Wigner function as
\begin{align}
& W(\alpha; \tilde{\rho}_1) =\frac{1}{\pi^{4}}   \int \mathrm d^2 {\alpha}_1 \int \mathrm d^2 \beta_1 ~ \langle \alpha_1  | \beta_1 \rangle F(\alpha_1, \beta_1) \nonumber\\
&~~~~~\times 
	\int \mathrm d^2 \beta ~ e^{-|\beta|^2/2} e^{ - \beta^* (\beta_1-\alpha) +  \beta (\alpha_1^*-\alpha^*)}  \nonumber\\
&=
\frac{2}{\pi^{3}}  e^{-2|\alpha|^2} \int \mathrm d^2 {\alpha}_1 \int \mathrm d^2 \beta_1 ~ F(\alpha_1, \beta_1) \, \nonumber\\
&\times 
	\exp\bigg[-\frac{|\alpha_1|^2}{2} - \frac{|\beta_1|^2}{2} - \alpha_1^* \beta_1 + 2 \, (\alpha \alpha_1^*+\alpha^* \beta_1) \bigg],
\label{eq:Wigner-2}
\end{align}
where in the last equality we have performed the integration over $\beta$ and used the relation 
$\langle \alpha_1  | \beta_1 \rangle = e^{-|\alpha_1|^2/2 - |\beta_1|^2/2 + \alpha_1^* \beta_1}$. 
By substituting the function $F(\alpha_1, \beta_1)$ of Eq.~\eqref{eq:Ffunction} into Eq.~\eqref{eq:Wigner-2}, interchanging the order of partial derivatives
and integration, and then performing the integration over $\alpha_1$ and $\beta_1$ (which is  a Gaussian integration), 
we arrive at the final expression for the unnormalized Wigner function (see Appendix~\ref{app:Wigner-Function-Mmode} for more details) as
\begin{eqnarray}\label{eq:Wigner-single-general}
&& W(\alpha;\tilde \rho_1)
=
\frac{2 \, \mathcal{ P}_0}{\pi \, \bar{\vt{n}}!}
\frac{\exp \big( \frac{1}{2} \vt{y}_h^{\top} {\bf L}_4 {\bf X}_2 \vt{y}_h \big)}{\sqrt{\text{det} ({\bf I}_2 + {\bf X}_2 {\bf R}_{hh} )}}
\exp \big( - \vt{v}^{\dag} {\bf L}_5 \vt{v} \big) \nonumber\\
&& \times 
\prod_{k=2}^{N} \bigg(\frac{\partial^2}{\partial \alpha_k \partial \beta_k^*} \bigg)^{n_k}
\exp\bigg( \frac{1}{2} \vt{\gamma}_d^{\top} {\bf A} \vt{\gamma}_d + \vt{z}^{\top} \vt{\gamma}_d \bigg) \bigg|_{\vt{\gamma}_d = {\bf 0}},\nonumber\\
&&{\bf L}_4 = ({\bf I}_2 - {\bf X}_2 {\bf R}_{hh} )^{-1}, \nonumber\\
&& {\bf L}_5 =  ({\bf I}_2 + {\bf X}_2 {\bf R}_{hh})^{-1} ({\bf I}_2 - {\bf X}_2 {\bf R}_{hh}),
\end{eqnarray}

where we have defined
\begin{eqnarray}
\vt{v} &=& (\alpha^*, \alpha)^{\top}
- ({\bf I}_2 - {\bf X}_2 {\bf R}_{hh})^{-1} {\bf X}_2 \vt{y}_h, \nonumber\\
{\bf A} &=& {\bf R}_{dd} - {\bf R}_{dh} ({\bf I}_2 + {\bf X}_2 {\bf R}_{hh} )^{-1} {\bf X}_2 {\bf R}_{hd}, \nonumber\\
\vt{z} &=& \vt{Y} + 2\, {\bf R}_{dh} ({\bf I}_2 + {\bf X}_2 {\bf R}_{hh})^{-1} \vt{v}, \nonumber\\
 \vt{Y} &=&\vt{y}_d + {\bf R}_{dh} ({\bf I}_2 - {\bf X}_2 {\bf R}_{hh})^{-1} {\bf X}_2 \vt{y}_h. 
\end{eqnarray}
In the following, we define the vector $\vt{Y}$ as $\vt{Y} = (Y_2^*, Y_3^*, \cdots, Y_N^*, Y_2, Y_3, \cdots, Y_N)^{\top}$ for convenience. 

The unnormalized Wigner function in Eq.~\eqref{eq:Wigner-single-general} is factorized into two parts: the first part is  a 
Gaussian function of $\vt{v}$; the second part is the partial derivatives of a Gaussian function evaluated at $\vt{\gamma}_d=0$, which results in a polynomial of $\vt{v}$. 
The maximum order of the 
polynomial depends on the detected photon number pattern $\vt{\bar{n}}$. If $n_k = 0$ for all $k$, i.e., all PNRDs register no photons, the polynomial is trivially equal to one. 
The unnormalized Wigner function is then a Gaussian distribution, which implies that the heralded state in the first mode is a Gaussian state. 
By comparing Eq.~\eqref{eq:Wigner-single-general} with Eq.~\eqref{eq:WignerCoherent}, 
we find that the displacement of the heralded state is 
\begin{eqnarray}\label{eq:DisNoPhoton-Single}
\vt{d} = ({\bf I}_2 - {\bf X}_2 {\bf R}_{hh})^{-1} {\bf X}_2 \vt{y}_h
\end{eqnarray}
and the covariance matrix is
\begin{eqnarray}\label{eq:CMnoPhoton-Single}
{\bf V}^{(c)} (\bar{\vt{n}} = {\bf 0}) =  \frac{1}{2} ({\bf I}_2 + {\bf X}_2 {\bf R}_{hh}) ({\bf I}_2 - {\bf X}_2 {\bf R}_{hh})^{-1}. 
\end{eqnarray}
To generate a non-Gaussian state, the polynomial that results from the action of the partial derivatives in Eq.~\eqref{eq:Wigner-single-general} should be nontrivial. For this, two conditions need to be satisfied: (1) PNRDs should register photons; (2) the matrix ${\bf R}_{hd} \ne {\bf 0}$, which means that the heralded mode must have 
some connections with the detected modes as viewed through the ${\bf R}$ matrix.

\subsection{Measurement probability}

\begin{figure*}
    \centering
    \scalebox{0.6}{\includegraphics{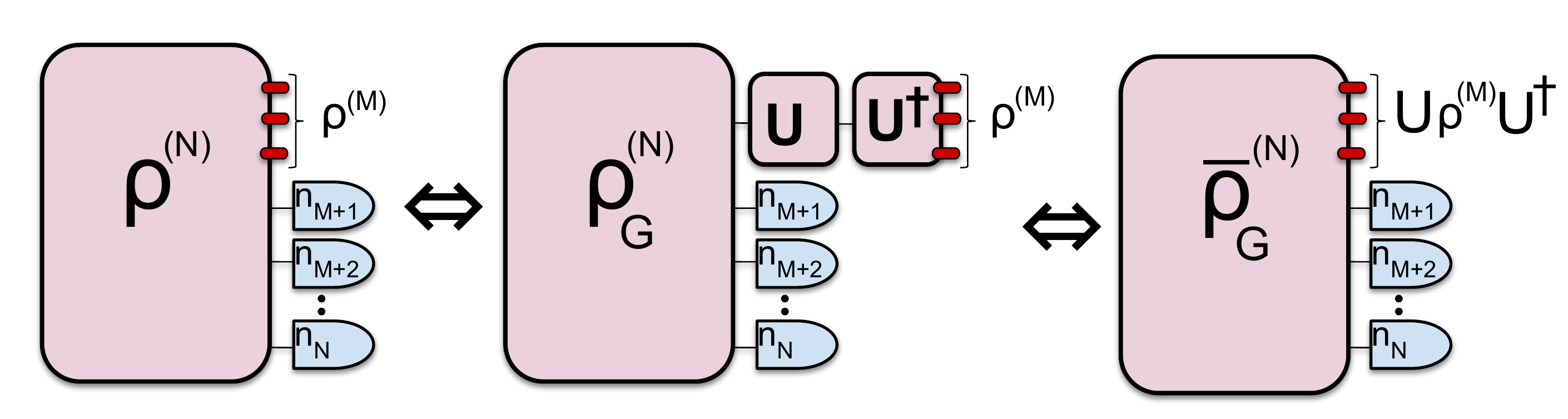}} \caption{Scheme to obtain a Gaussian gate applied to a particular state given the circuit parameters to generate said state. The diagram on the left depicts an $N$-mode Gaussian state to which the last $(N-M)$ modes are detected to obtain an $M$-mode output state. Suppose we want to obtain the same output state but now followed by an $M$-mode  Gaussian gate $U$, all we need to do is update the initial Gaussian gate by the $U^{\dag}$ on the first $M$ modes while retaining the same measurement pattern as before. This results in  an output state with the unitary gate applied to it with the same success probability as compared to the case without the gate.} 
    \label{fig:symmetry}
\end{figure*}

We have derived the expression for the unnormalized Wigner function, but have yet to determine the success probability of producing the output state.
Obtaining the photon number distribution of
a multimode Gaussian state was studied by Refs.~\cite{dodonov1994photon, dodonov1994multidimensional} and recently became an important topic known as \textsf{Gaussian BosonSampling} \cite{PhysRevLett.119.170501}.
Here, the measurement probability $P(\bar{\vt{n}})$ can be obtained by performing a trace of the unnormalized density operator $\tilde \rho_1$, 
which corresponds to integrating the unnormalized Wigner function $ W(\alpha;\tilde \rho_1)$ over the arguments $\alpha$, giving
\begin{eqnarray}
P(\bar{\vt{n}}) = \text{Tr}(\tilde \rho_1) = \int \mathrm d^2 \alpha \,  W(\alpha;\tilde \rho_1). 
\end{eqnarray}
It is evident from Eq.~\eqref{eq:Wigner-single-general} that the integration over $\alpha$ is a straightforward Gaussian integration. Using the equality
\begin{align*}
\int \mathrm d^2 \alpha \, \exp \big[- \vt{v}^{\dag} {\bf L}_5  \vt{v} \big] 
= \frac{\pi}{2} \, \left[\sqrt{\text{det}\big[ {\bf L}_5  \big]}\right]^{-1}, 
\end{align*}
we obtain the measurement probability
\begin{align}\label{eq:ProbabilitySinlgemode}
&P(\bar{\vt{n}}) = %\frac{\mathcal{ P}_0}{\bar{\vt{n}}!}
\frac{\mathcal{ P}_0}{\bar{\vt{n}}! \, \sqrt{\text{det}({\bf I}_2 - {\bf X}_2 {\bf R}_{hh})}}\nonumber\\
&\times \exp \bigg\{ \frac{1}{2} \vt{y}_h^{\top} ({\bf I}_2 - {\bf X}_2 {\bf R}_{hh} )^{-1} {\bf X}_2 \vt{y}_h \bigg\} \nonumber\\
& \times \prod_{k=2}^{N} \bigg(\frac{\partial^2}{\partial \alpha_k \partial \beta_k^*} \bigg)^{n_k}
\exp\bigg( \frac{1}{2} \vt{\gamma}_d^{\top} {\bf A}_p \vt{\gamma}_d + \vt{z}_p^{\top} \vt{\gamma}_d \bigg) \bigg|_{\vt{\gamma}_d = 0}, \nonumber\\
\end{align}
where
\begin{eqnarray}\label{eq:probApZp}
{\bf A}_p &=& {\bf R}_{dd} + {\bf R}_{dh} ({\bf I}_2 - {\bf X}_2 {\bf R}_{hh} )^{-1} {\bf X}_2 {\bf R}_{hd}, \nonumber\\
\vt{z}_p &=& \vt{y}_d + {\bf R}_{dh} ({\bf I}_2 - {\bf X}_2 {\bf R}_{hh})^{-1} {\bf X}_2 \vt{y}_h.
\end{eqnarray}

The general scheme has a particular symmetry that we could exploit for our purposes. Let us begin with a particular initial $N$-mode Gaussian state $\rho^{(N)}$ and we measure $(N-M)$ modes to obtain a measurement pattern $\vt{\bar{n}}$ and an $M$-mode output state $\rho^{(M)}$. This same setup could be used to obtain an output state $U \rho^{(M)} U^{\dag}$, where $U$ is an $M$-mode Gaussian unitary as depicted in Fig.~\ref{fig:symmetry}. All we need to do is to to update the initial Gaussian state to $\bar{\rho}^{(N)} = [U \otimes 1\!\!1_{N-M}] \rho^{(N)} [U \otimes 1\!\!1_{N-M}]^{\dag}$ and retain the same measurement pattern as before. This will then herald a state  $U \rho^{(M)} U^{\dag}$ with the same success probability as before. We see that obtaining an output state with additional Gaussian gates applied to it has a  straightforward method. In the next section, we investigate the particular case when the measured $N$-mode Gaussian state is pure.

\section{Single-mode output states by measuring pure Gaussian states}\label{sec:SM-PureFormalism}
 
\begin{figure}
    \centering
    \includegraphics[width=\columnwidth]{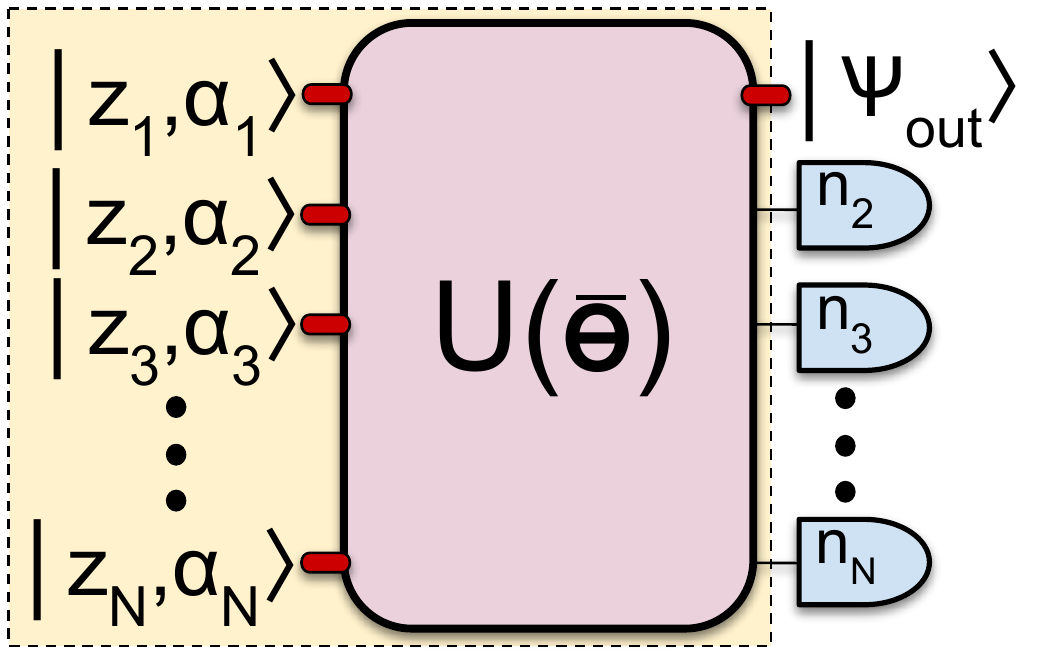}    \caption{Probabilisitic generation of single-mode non-Gaussian states. A general pure multimode Gaussian state  can be decomposed into displaced squeezed states,  $\ket{z_i,\alpha_i} =\hat D(\mathsf{\alpha}_i) \hat S(z_i)  \ket{0} $, on the $i^{\rm th}$ mode with $i=1~ {\rm to} ~ N$, followed by an interferometer $\textsf{U}(\bar{\theta})$. The last $(N-1)$ modes are measured using PNRDs giving values $\{n_k\}_{2}^{N}$, resulting in a conditional output state $\ket{\psi_{\rm out}}$ in the first mode.} 
    \label{fig:ng-GBS-one-mode-pure}
\end{figure}

Any pure Gaussian state can be prepared by sending displaced squeezed vacuum states into a multiport interferometer \cite{PhysRevA.71.055801}. In this section we consider the case when all but one mode of a pure Gaussian state are measured using PNRDs, as depicted in Fig.~\ref{fig:ng-GBS-one-mode-pure}. Note that when measuring a pure Gaussian state, the heralded non-Gaussian state is also pure. This section will 
clarify the significance of each part in the unnormalized Wigner
function in Eq.~\eqref{eq:Wigner-single-general}. The heralded non-Gaussian state is a finite superposition of Fock states, acted on by a single mode Gaussian unitary (such as a phase shift, squeezing operator, displacement, or any combinations of these). The relationship between the parameters of the output state and the parameters of the  measured Gaussian state will be derived. We also study in detail the relationship between the number of independent coefficients in the output Fock state superposition and the number of modes of the Gaussian state, which provides insight into what non-Gaussian states can be generated using multimode Gaussian states. 

\subsection{Output Wigner function}

As mentioned above, an arbitrary $N$-mode pure Gaussian state can be generated by injecting $N$ single-mode displaced squeezed vacuum states into a linear interferometer. 
The covariance matrix of $N$ independent single-mode displaced squeezed states is 
\begin{eqnarray}
{\bf V}^{(c)}_{\text{sq}} = \frac{1}{2}
	\begin{pmatrix}
	{\bf D}_c & {\bf D}_s \\
	{\bf D}_s & {\bf D}_c
	\end{pmatrix},
\end{eqnarray}
where we have defined two diagonal matrices ${\bf D}_c = \bigoplus_{j=1}^{N} \cosh(2 r_j)$ and ${\bf D}_s = \bigoplus_{j=1}^{N} \sinh(2 r_j)$
with $r_j$ the squeezing parameter of the $j$-th input mode. The symplectic matrix representing the transformation of a linear interferometer can be written as a block diagonal form,
\begin{eqnarray}
{\bf S}^{(c)} = 
	\begin{pmatrix}
	{\bf U}^* & {\bf 0} \\
	{\bf 0} & {\bf U}
	\end{pmatrix},
\end{eqnarray}
where the unitary matrix ${\bf U}$ satisfies
\begin{eqnarray}
\hat a_j \rightarrow \sum_{j=1}^N U_{jk} \hat a_k. 
\end{eqnarray}
The covariance matrix of a pure Gaussian state can be written as \cite{PhysRevLett.119.170501}
\begin{eqnarray}\label{eq:CMpureSingle}
{\bf V}^{(c)} = {\bf S}^{(c)}  {\bf V}^{(c)}_{\text{sq}} {{\bf S}^{(c)} }^{\dag}  
=\frac{1}{2}
	\begin{pmatrix}
	{\bf U}^* {\bf D}_c \, {\bf U}^{\top} & {\bf U}^{*} {\bf D}_s \, {\bf U}^{\dag}\\
	{\bf U} \, {\bf D}_s \, {\bf U}^{\top} & {\bf U} \, {\bf D}_c \, {\bf U}^{\dag}
	\end{pmatrix}. 
\end{eqnarray}
By substituting Eq.~\eqref{eq:CMpureSingle} into Eq.~\eqref{eq:RYcoherent} and using the blockwise inversion formula, we find that the matrix
$\tilde {\bf R}$ is in a block diagonal form, i.e., $\tilde {\bf R} = {\bf B} \oplus {\bf B}^*$, where ${\bf B}$ (with entries $b_{ij}$) is an $N \times N$ symmetric matrix. ${\bf B}$ 
is completely determined by the input squeezing and the linear interferometer (not the input displacements) as \cite{PhysRevLett.119.170501}
\begin{eqnarray}\label{eq:Bmatrix-Signle}
{\bf B} = {\bf U} \displaystyle \bigoplus_{j=1}^{N} \tanh(r_j) \, {\bf U}^{\top}. 
\end{eqnarray}
By applying the permutation ${\bf P}$ we can obtain the matrix ${\bf R}$ of Eq.~\eqref{eq:R}. It is easy to see that ${\bf R}_{hh}$ is diagonal and only depends on $b_{11}$,
\begin{eqnarray}\label{eq:Rhh-Single}
{\bf R}_{hh} = 
	\begin{pmatrix}
	b_{11} & 0 \\
	0 & b_{11}^*
	\end{pmatrix}.  
\end{eqnarray}
Similarly, we have
\begin{align}
    &{\bf R}_{hd} =
	\begin{pmatrix}
	b_{12} & b_{13} & \cdots & b_{1N} & 0 & 0 & \cdots & 0 \\
	0 & 0 & \cdots & 0 & b_{12}^* & b_{13}^* & \cdots & b_{1N}^* 
	\end{pmatrix} =   {\bf R}_{hd}^{\top}, \nonumber\\
	&{\bf R}_{dd} = {\bf B_1} \oplus {\bf B_1^*},    
\end{align}
where ${\bf B_1}$ is the ${\bf B}$ matrix with the first row and column deleted. \\

\noindent
{\bf Zero photon detection} ($\vt{\bar{n}}={\bf 0}$) \,: We first consider the Gaussian factor in the unnormalized Wigner function in Eq.~\eqref{eq:Wigner-single-general}, 
which fully characterizes the heralded Gaussian state when all PNRDs register no photons. The covariance matrix can be obtained by substituting Eq.~\eqref{eq:Rhh-Single}
into Eq.~\eqref{eq:CMnoPhoton-Single} as
\begin{eqnarray}\label{eq:CM0photon}
{\bf V}^{(c)}_1(\bar{\bf n}= {\bf 0}) = \frac{1}{2(1 - |b_{11}|^2)}
	\begin{pmatrix}
	1+|b_{11}|^2 & 2\, b_{11}^* \\
	2\, b_{11} & 1+ |b_{11}|^2
	\end{pmatrix}.
\end{eqnarray}
It is easy to check that the determinant of ${\bf V}^{(c)}_1(\bar{\bf n}={\bf 0})$ is $1/4$, indicating that the heralded state is pure. The squeezing parameter of a pure single-mode Gaussian state  can be obtained from the eigenvalues of the covariance matrix. The eigenvalues of 
${\bf V}^{(c)}_1(\bar{\bf n}={\bf 0})$ are $\frac{\lambda_1}{2}$ and $\frac{1}{2\lambda_1}$, where $\lambda_1 = \frac{1+|b_{11}|}{1 - |b_{11}|}$. This implies that the squeezing parameter is 
\begin{eqnarray}\label{eq:b11Squeezing}
r_1 = \frac{1}{2} \ln \bigg( \frac{1+|b_{11}|}{1 - |b_{11}|} \bigg). 
\end{eqnarray}
Other than the squeezing, there is also a rotation (phase shift) included in the covariance matrix of Eq.~\eqref{eq:CM0photon}. If we define $b_{11} = |b_{11}| e^{i\phi_1}$,  then for the rotation angle  we have 
\begin{align} \label{eq:angle-singlemode}
    \varphi_1 = - \phi_1/2.
\end{align}  
This means that the heralded squeezed state has a squeezing amplitude $\zeta_1 = r_1 e^{i \varphi_1} =r_1 e^{-i \phi_1/2}$. 
To determine the displacement we define $\vt{y}_h = (y_{1}^*, y_{1})^{\top}$ and $\vt{y}_d = (y_{2}^*, y_{3}^*, \cdots, y_{N}^*, y_{2}, y_{3}, \cdots, y_{N})^{\top}$. Substituting Eq.~\eqref{eq:Rhh-Single} into Eq.~\eqref{eq:DisNoPhoton-Single} we obtain the displacement vector as
\begin{eqnarray}%\label{eq:WignerVariable}
\label{eq:disp-singlemode}
\vt{d} = 
\frac{1}{1 - |b_{11}|^2}
	\begin{pmatrix}
	1 & b_{11}^* \\
	b_{11} & 1
	\end{pmatrix}
	\begin{pmatrix}
	y_1 \\
	y_1^*
	\end{pmatrix}
&\equiv&
    \begin{pmatrix}
	d_1^* \\
	d_1
	\end{pmatrix}.
\end{eqnarray}
It is evident that $b_{11}$ and $\vt{y}_h$ uniquely determine the heralded Gaussian state when the PNRDs register no photons. \\

\noindent
{\bf Non-zero photon detection} ($\vt{\bar{n}} \neq {\bf 0}$)\,: 
When the PNRDs register photons, the heralded state is generally a non-Gaussian state. The non-Gaussianity is dictated by the polynomial factor in the 
unnormalized Wigner function in Eq.~\eqref{eq:Wigner-single-general}. The Gaussian factor involving  the squeezing and the displacement has to be interpreted as  Gaussian operations acting on a finite superposition of Fock states. To tranparently demonstrate this point we define a new vector $\vt{w} = (\delta^*, \delta)^{\top}$ as
\begin{eqnarray}
\vt{w} = \sqrt{1 - |b_{11}|^2} \, ({\bf I}_2 + {\bf X}_2 {\bf R}_{hh})^{-1} \vt{v}.
\end{eqnarray}
Then we find
\begin{eqnarray}
\vt{v}^{\dag} ({\bf I}_2 + {\bf X}_2 {\bf R}_{hh})^{-1} ({\bf I}_2 - {\bf X}_2 {\bf R}_{hh}) \vt{v} 
= \vt{w}^{\dag} \vt{w}.
\end{eqnarray}
The output Wigner function now can be written as 
\begin{align}\label{eq:WignerSignlePure}
&W(\alpha; \rho_1) \propto e^{- \vt{w}^{\dag}  \vt{w}}
\prod_{k=2}^{N} \bigg(\frac{\partial^2}{\partial \alpha_k \partial \beta_k^*} \bigg)^{n_k} \nonumber\\
& ~~~~~~\times \exp\bigg( \frac{1}{2} \vt{\gamma}_d^{\top} {\bf A} \vt{\gamma}_d + \vt{z}^{\top} \vt{\gamma}_d \bigg) \bigg|_{\vt{\gamma}_d = 0}, 
\end{align}
where 
\begin{eqnarray}\label{eq:Zpure}
\vt{z} 
= {\vt Y} + \frac{2}{\sqrt{1-|b_{11}|^2}} \, {\bf R}_{dh} \vt{w}. 
\end{eqnarray}
It is clear from Eq.~\eqref{eq:WignerSignlePure} that the heralded non-Gaussian state is a superposition of a finite number of Fock states, followed by a squeezing 
operation and a displacement. In other words, the output state is of the form 
\begin{align}\label{eq:StateSingle}
    \ket{\psi_{\text{out}}} = \hat{D}(d_1)  \hat{S}(\zeta_1)  \sum_{n=0}^{n_{\rm max}} c_{n} \ket{n}. 
\end{align}
This can also be understood in the following way: according to the transformation rule Eq.~\eqref{eq:WignerTF}, we first apply a displacement and then a squeezing operation to the state in Eq.~\eqref{eq:WignerSignlePure}, which transforms the Wigner function back to the one corresponding to only a finite superposition of Fock states. The explicit expressions for the coefficients $\{ c_n\}$ are dealt with in the following subsection.

\subsection{Coefficients $\{ c_n\}$ in the Fock basis superposition}

The coefficients $\{ c_n\}$  of the superposition of Fock states remain to be determined. 
Suppose the position space wave function of a quantum state $\ket{\psi}$ is $\psi(q)$, it can be expanded in the Fock basis as 
\begin{eqnarray}
\psi(q) = \sum_{n=0}^{\infty} c_n \psi_n (q).
\end{eqnarray}
Here, $c_n$ is the coefficient, and $\psi_n (q)$ is the wave function of the Fock state $\ket{n}$ given by
\begin{eqnarray}\label{eq:FockWF}
\psi_n (q) = \frac{1}{\pi^{1/4}\sqrt{ 2^n \, n!}} e^{-q^2/2} H_n (q),
\end{eqnarray}
with $H_n (q)$ the Hermite polynomials. From Eq.~\eqref{eq:WignerXPgeneral}, the single-mode Wigner function is 
\begin{eqnarray}\label{eq:WignerFockSuper}
\xoverline{W}(p, q) &=& \frac{1}{\pi} \int \mathrm d y \, e^{- 2i py} \langle q - y \ket{\psi} \langle \psi \ket{q + y} 
\nonumber\\
&=&
\frac{1}{\pi} \sum_{m,n=0}^{\infty} c_m c_n^* \, W_{m n}(p, q),
\end{eqnarray}
where $W_{mn}(p, q)$ is defined as 
\begin{eqnarray}\label{eq:Wigner-Ito}
W_{mn}(p, q) &=& \int \mathrm d y \, e^{- 2i py} \langle q - y \ket{\psi_m} \langle \psi_n \ket{q + y}
\nonumber\\
&=&
\frac{1}{\sqrt{n! \, m!}} e^{-q^2 - p^2} H_{mn}(2 \alpha, 2 \alpha^*). 
\end{eqnarray}
Here, $H_{mn}(2 \alpha, 2 \alpha^*)$ is known as Ito's 2D-Hermite polynomial \cite{ismail2017review} (see Appendix \ref{app:Wigner-Ito-Hermite} for details). 

By using the orthogonality relation of Ito's 2D-Hermite polynomials we can find a systematic way to evaluate the coefficients of the heralded states. 
Ito's 2D-Hermite polynomials satisfy the following orthogonality relation \cite{intissar2006spectral, ismail2017review}:
\begin{align}\label{eq:HermiteOrthogonality-alpha}
 &\int \mathrm d^2 \alpha \, H_{m_1 n_1}(2\alpha, 2\alpha^*) H^*_{m_2 n_2}(2\alpha, 2\alpha^*) e^{- 4 |\alpha|^2}  
\nonumber\\
 &~~~~~~=\frac{\pi}{2} m_1! \, n_1! \, \delta_{m_1, m_2} \delta_{n_1, n_2}. 
\end{align}
The Wigner function of a quantum state can be expressed in terms of Ito's 2D-Hermite polynomials, as can be seen from 
Eqs.~\eqref{eq:WignerFockSuper} and \eqref{eq:Wigner-Ito} for a pure state. Therefore, the Fock-state coefficients of a quantum state can be written as an overlap 
integral between the Wigner function and Ito's 2D-Hermite polynomials,
\begin{eqnarray}\label{eq:CoeffWignerHermite}
c_m c_n^* = \frac{1}{\sqrt{m! n!}} \int \mathrm d^2 \alpha \, W(\alpha) H^*_{m n}(2\alpha, 2\alpha^*) e^{- 2 |\alpha|^2},
\end{eqnarray}
where we have taken into account the convention that $W(\alpha) = 2 \, \xoverline{W} (p, q)$ [Eq.~\eqref{eq:convention}].

If the quantum state $\ket{\psi}$ is squeezed and displaced, according to the transformation rule of the Wigner function and from Eq.~\eqref{eq:WignerFockSuper} we see that
the coefficients $c_n$ are unchanged while the arguments of the Wigner function are changed. This change can be taken into account by replacing $\alpha$ by $\delta$,
where $\delta$ contains the squeezing and displacement information. Now by substituting the Wigner function \eqref{eq:WignerSignlePure} into Eq.~\eqref{eq:CoeffWignerHermite} and performing the integration over $\delta$, we find (see Appendix \ref{app:Fock-Coefficients} for more details)
\begin{align}\label{eq:CmCn-1-single}
&c_m c_n^* = \frac{1}{\sqrt{m! n!}} \int \mathrm d^2 \delta \, W(\alpha) H^*_{m n}(2\delta, 2\delta^*) e^{- 2 |\delta|^2}
\nonumber\\
&=
\frac{\pi \, \mathcal{N}_1}{4 \sqrt{m! n!}} 
\prod_{k=2}^{N} \bigg(\frac{\partial^2}{\partial \alpha_k \partial \beta_k^*} \bigg)^{n_k}
\exp\bigg( \frac{1}{2} \vt{\gamma}_d^{\top} {\bf C} \vt{\gamma}_d + \vt{Y}^{\top} \vt{\gamma}_d \bigg) \nonumber\\
&~~~~~~ \times 
\bigg( \sum_{j=2}^N \kappa_j^* \alpha_j \bigg)^m \bigg(  \sum_{i=2}^N \kappa_i \beta_i^* \bigg)^n
\bigg|_{\vt{\gamma}_d = 0},
\end{align}
where $\mathcal{N}_1$ is the normalization factor of the Wigner function in Eq.~\eqref{eq:WignerSignlePure}, whose exact value is irrelevant to the coefficients $c_n$. 
Here, we have defined
\begin{eqnarray}\label{eq:DefkappaC}
\kappa_j &=& \frac{b_{1j}}{\sqrt{1-|b_{11}|^2}}, ~~~~~~ j = 2, 3, \cdots, N,
\nonumber\\
{\bf C} &=& {\bf A} + \frac{1}{1 - |b_{11}|^2}{\bf R}_{dh} {\bf X}_2 {\bf R}_{hd} 
\nonumber\\
&=&
 {\bf R}_{dd} + \frac{1}{1 - |b_{11}|^2} {\bf R}_{dh} 
	\begin{pmatrix}
	b_{11}^* & 0\\
	0 & b_{11}
	\end{pmatrix}
{\bf R}_{hd}. 
\end{eqnarray}
Equation~\eqref{eq:CmCn-1-single} can also be written in an equivalent form, which only involves partial derivatives of a Gaussian function. 
To do that we first introduce a two-component vector $\vt{t} = (t_1, s_1)^{\top}$ and a $2N \times 2N$ matrix ${\bf M}$ given by
\begin{eqnarray}
{\bf M} = 
	\begin{pmatrix}
	{\bf 0} & \frac{1}{\sqrt{1 - |b_{11}|^2}}{\bf X}_2 {\bf R}_{hd}  \\
	%\\
	\frac{1}{\sqrt{1 - |b_{11}|^2}}{\bf R}_{dh} {\bf X}_2  & {\bf C}
	\end{pmatrix}. 
\end{eqnarray}
The product of the two coefficients $c_m c_n^*$  of Eq.~\eqref{eq:CmCn-1-single} can be rewritten as
\begin{align}\label{eq:CoeffGeneral}
&c_m c_n^* =
\frac{\pi \, \mathcal{N}_1}{4 \sqrt{m! n!}} \, 
\frac{\partial^m}{\partial t_1^m}  \frac{\partial^n}{\partial s_1^n} \prod_{k=2}^{N} \bigg(\frac{\partial^2}{\partial \alpha_k \partial \beta_k^*} \bigg)^{n_k} \nonumber\\
& \times \exp\bigg\{ \frac{1}{2} (\vt{t}^{\top}, \vt{\gamma}_d^{\top}) \, {\bf M} \begin{pmatrix} \vt{t} \\ \vt{\gamma}_d \end{pmatrix} + \vt{Y}^{\top} \vt{\gamma}_d \bigg\}
\bigg|_{\vt{\gamma}_d = 0, \, t_1=s_1 = 0}. 
\end{align}

Equations~\eqref{eq:CmCn-1-single} and \eqref{eq:CoeffGeneral} provide all the information one needs to evaluate the coefficients $\{c_n\}$ \cite{niconote}. Although the product of two coefficients is given and the normalization factor $\mathcal{N}_1$ remains unknown, one can still determine $\{c_n\}$ as
follows. The first step is to determine the maximal $n$, denoted by $n_{\text{max}}$, whose corresponding coefficient is nonzero. 
From Eq.~\eqref{eq:CmCn-1-single} it can be shown that $n_{\text{max}} \le n_T$, where 
\begin{align} \label{eq:nt}
    n_T = n_2 + n_3 + \cdots + n_N
\end{align} is the total number of 
detected photons. The equality occurs when $\kappa_j \ne 0$ in Eq.~\eqref{eq:DefkappaC}  for all $j$ from 2 to $N$, which indicates that the heralded mode has full connections with 
all detected modes. When $\kappa_j$ is zero, which means the $j$-th mode has no connection to the heralded mode, the detection of photons in the $j$-th mode 
does not help to increase the order of the polynomial, implying $n_{\text{max}} < n_T$. 

There is no  upper bound for the total photon number $n_T$ because the detected state is an $N$-mode Gaussian state, which implies that there is also no  upper bound for
$n_{\text{max}}$. The value of $n_T$ is in fact fixed when we post-select a particular measurement outcome. However, on the other hand, the number of independent coefficients should be finite because these coefficients are determined by an $N$-mode Gaussian state which
is fully characterized by the finite number of parameters in the covariance matrix and mean vector. We are going to derive the relation between the maximal number of 
independent coefficients and the size of the detected Gaussian state. The first step is to assume $\kappa_j \ne 0$ for all $j$ to guarantee $n_{\text{max}} = n_T$. 
By setting $m = n = n_T$ in Eq.~\eqref{eq:CmCn-1-single}, we find that 
\begin{eqnarray}\label{eq:CoeffNT}
|c_{n_T}|^2 
&=&
\frac{1}{4} \pi \, \mathcal{N}_1 n_T! \, |\kappa_2|^2 |\kappa_3|^2 \cdots |\kappa_N|^2 \ne 0,
\end{eqnarray}
which is consistent with the assumption $\kappa_j \ne 0$. To determine a state, it is sufficient to fix the ratios between other coefficients and $c_{n_T}$ because 
taking into account the normalization condition will uniquely determine the state. The ratio $c_n/c_{n_T}$  can be obtained by calculating $c_n c_{n_T}^*/|c_{n_T}|^2$,
where the numerator is from Eq.~\eqref{eq:CmCn-1-single} and the denominator is from Eq.~\eqref{eq:CoeffNT}. By defining new variables 
$\omega_i = \kappa_i^* \alpha_i$, $\sigma_i = \kappa_i \beta_i^*$, we find
\begin{align}\label{eq:CoeffRatio-single-1}
&\frac{c_n}{c_{n_T}} 
=\prod_{k=2}^{N} \bigg(\frac{\partial^2}{\partial \omega_k \partial \sigma_k^*} \bigg)^{n_k}
\frac{\exp (\mathcal U_1 )
\mathcal V_1 \mathcal W_1}{\sqrt{n! \, (n_T!)^3}}  \bigg|_{\vt{\omega} = \vt{\sigma} = {\bf 0}}, 
\nonumber\\
&\mathcal U_1 = \frac{1}{2} (\vt{\sigma}^{* \top}, \vt{\omega}^{\top}) \, {\bf C}_{\text{rn}} \begin{pmatrix} \vt{\sigma}^{*} \\ \vt{\omega} \end{pmatrix} 
	+ (\vt{\mu}^{*\top}, \vt{\mu}^{\top}) \begin{pmatrix} \vt{\sigma}^{*} \\ \vt{\omega} \end{pmatrix}, \nonumber\\
& \mathcal V_1 = \bigg( \sum_{j=2}^N \omega_j \bigg)^n, ~~ \mathcal W_1 =  \bigg(  \sum_{i=2}^N \sigma_i^* \bigg)^{n_T},
\end{align}
where $ \mu_i = Y_i/\kappa_i^*$, ${\bf C}_{\text{rn}} = {\bf F} \oplus {\bf F}^*$ and ${\bf F}$ is an $(N-1) \times (N-1)$ symmetric matrix with entries $f_{ij}$ defined as 
\begin{eqnarray}\label{eq:matrixF}
f_{ij} = b_{11}^* + \frac{b_{ij}}{\kappa_i \kappa_j}, ~~~~~~ i, j = 2, 3, \cdots, N. 
\end{eqnarray}
As in the earlier case, $c_n/c_{n_T}$ can be written in an equivalent form where there are only partial derivatives acting on a Gaussian function, and we have
\begin{align}\label{eq:CoeffRatio-single-2}
&\frac{c_n}{c_{n_T}} 
=
\frac{\partial^n}{\partial t_1^n}  \frac{\partial^{n_T}}{\partial s_1^{n_T}} \prod_{k=2}^{N} \bigg(\frac{\partial^2}{\partial \omega_k \partial \sigma_k^*} \bigg)^{n_k} \mkern-13mu \frac{\exp ( \mathcal U_2 )}{\sqrt{n! \, (n_T!)^3}}
\bigg|_{\vt{\omega}, \vt{\sigma},t_1, s_1 = 0}, 
\nonumber\\
& \mathcal U_2 = \frac{1}{2} (\vt{t}^{\top}, \vt{\sigma}^{* \top}, \vt{\omega}^{\top}) \, {\bf M}_{\text{rn}} \begin{pmatrix} \vt{t} \\ \vt{\sigma}^{*} \\ \vt{\omega} \end{pmatrix} 
	+ (\vt{\mu}^{*\top}, \vt{\mu}^{\top}) \begin{pmatrix} \vt{\sigma}^{*} \\ \vt{\omega} \end{pmatrix},\nonumber\\ 
&{\bf M}_{\text{rn}} = 
	\begin{pmatrix}
	{\bf 0} & {\bf R}^{\text{(rn)}} _{hd} \\
	%\\
	{\bf R}^{\text{(rn)}} _{dh}  & {\bf C}_{\text{rn}}
	\end{pmatrix}, \nonumber\\ 
&{\bf R}^{\text{(rn)}} _{hd} = 
	\begin{pmatrix}
	0 & 0 & \cdots & 0 & 1 & 1 & \cdots & 1 \\
	1 & 1 & \cdots & 1 & 0 & 0 & \cdots & 0
	\end{pmatrix}. 
\end{align}
Equations~\eqref{eq:CoeffRatio-single-1} and \eqref{eq:CoeffRatio-single-2} provide a systematic way to evaluate the coefficients of the heralded superposition of Fock states. 
By explicitly evaluating the partial derivatives in Eqs.~\eqref{eq:CoeffRatio-single-1} and \eqref{eq:CoeffRatio-single-2}, we find that the ratios $c_n/c_{n_T}$ are polynomials 
of $\mu_i$ and $f_{ij}$.  
Noting that {\bf F} is symmetric, so the total number of independent parameters is equal to $ \mathfrak{D} = (N+2)(N-1)/2$, comprised of the components of $\vt{\mu}$ and the entries of {\bf F}.

The problem of determining the number of independent $\{c_n\}$'s can be formulated as follows. Let us assume that $\mu_j$ and $f_{ij}$ are unknown
and have to be solved from $n_T$ nonlinear polynomial equations, which come from Eq.~\eqref{eq:CoeffRatio-single-1} or \eqref{eq:CoeffRatio-single-2} by taking $n = 0, 1, \cdots, n_T-1$. If $n_T < \mathfrak{D}$, 
the nonlinear equations are underdetermined, which means that for a given set of $\{c_n\}$ there is an infinite number of solutions. This implies that there are many initial Gaussian states
that can generate the same non-Gaussian state. If $n_T > \mathfrak{D}$, the nonlinear equations are overdetermined and there is no guarantee for the existence of a 
solution for an arbitrary given set of $\{c_n\}$, which means that they are not independent. The situation is subtle for the case of $n_T = \mathfrak{D}$. If there exists solutions,
the number of solutions is finite. It is also possible that there exists no solutions. We checked cases when $N$ is 2,3 and found that when $n_T = \mathfrak{D}$ there
always exists a finite number of solutions. We thus propose the following 
\begin{conjecture}
\label{conj}
Measuring $(N-1)$ modes of an $N$-mode pure Gaussian state using PNRDs outputs a superposition of Fock states %with maximum Fock support given by $n_T = \sum n_i$ 
with  at most $(N+2)(N-1)/2$ independent coefficients. 
\end{conjecture}
Conjecture \ref{conj} demonstrates the extent and power of generating non-Gaussian states using the method of measuring multimode Gaussian states with PNRDs. We now summarize the methods in this subsection for obtaining the output state given an input pure Gaussian state and a measurement pattern, in the form of  Algorithm \ref{algo1}. 

\begin{algorithm}
\label{algo1}
\KwIn{${\bf V}^{(c)},{\bf Q}^{(c)} $ of a pure multimode Gaussian state and a photon detection pattern $\vt{\bar{n}}$ }
~Compute $\tilde {\bf R},  \vt{\tilde y}$ using  Eq.~\eqref{eq:RYcoherent}.\\
~Apply permutation ${\bf P} : (\tilde {\bf R},  \vt{\tilde y}) \to ( {\bf R},  \vt{ y})$ in Eq.~\eqref{eq:Ry}.\\
~Obtain final squeezing $\zeta_1$ using Eqs.~\eqref{eq:b11Squeezing} and \eqref{eq:angle-singlemode}.\\
~Compute the final displacement $\vt{d}$ by Eq.~\eqref{eq:disp-singlemode}. \\
~Evaluate coefficients $\{ c_n\}$ using Eqs.~\eqref{eq:CmCn-1-single} and \eqref{eq:DefkappaC}. \\
~~Note: If required, the Wigner function $W (\vt{\alpha};\tilde \rho_1)$ and the success probability $P(\vt{\bar{n}})$ can be computed using Eqs.~\eqref{eq:Wigner-single-general} and \eqref{eq:ProbabilitySinlgemode}, respectively, directly after Step 2.\\
\caption{Obtaining single-mode output states by measuring pure multimode Gaussian states}
\KwOut{Heralded state as represented in Eq.~\eqref{eq:StateSingle}}
\end{algorithm}

There is one more application of our general formalism. We can formulate the complementary problem of obtaining the input pure Gaussian state and measurement pattern such that one obtains the target single-mode output state with the highest fidelity and success probability. Note that in general, the mapping from Gaussian states and measurement patterns to the output state is in general many-to-one and also involves both continuous parameters for the Gaussian state and discrete parameters for the measurement patters. So this problem of obtaining the optimal Gaussian circuit and measurement pattern to generate a particular target state is more intricate and requires careful considerations. We summarize the steps necessary for the case when we assume that the input Gaussian state is pure as Algorithm \ref{algo2}. 

\begin{algorithm}
\label{algo2}
\KwIn{Target state $\sum_{k=0}^{n_0} \tilde c_{k} \ket{k} $ }
~Approximate the target state in the form of Eq.~\eqref{eq:StateSingle}. \\
~Use Conjecture \ref{conj} to estimate the number of input modes $N$ that are required from the relation $n_{\rm max} \leq \mathfrak{D}$.\\
~Working principle\,: choose measurement pattern $\vt{\bar{n}}= \{ n_j\}$ such that $\sum_j n_j = n_{\rm max}$.\\
~Assume a generic complex symmetric matrix ${\bf B}$ with ${\bf B}{\bf B}^{\dag} \leq 1\!\!1$, and a complex displacement vector ${\vt Y}$.\\
%with Eq.\,\eqref{eq:nt} with $n_T=n_0$.\\
~Obtain nonlinear constrained equations using  Eqs.~\eqref{eq:CmCn-1-single} and \eqref{eq:DefkappaC} to connect $({\bf B}, {\vt Y})$ and $\{c_n \}$. \\
~ Maximize the success probability in Eq.~\eqref{eq:ProbabilitySinlgemode} subject to constraints in Step 5.\\
~ Repeat Steps 3-6 over various discrete measurement patterns to obtain the best success probability and the optimal pair $({\bf B}, {\vt Y})$.\\
~ Compute $({\bf V}^{(c)},{\bf Q}^{(c)})$ from the optimal $({\bf B}, {\vt Y})$.\\
~ If required, the input squeezed states and the interferometer corresponding to the pure Gaussian state can be obtained from ${\bf B}$ using the Autonne-Takagi normal form.\\ 
~ Further, the interferometer in Step 9 can be broken down into beam splitters and phase shifters using, for example, the triangle \cite{reck} or square  \cite{clements} schemes.\\ 
\caption{Obtaining the optimal pure Gaussian state and measurement pattern that generates a given target state}
\KwOut{Optimal pure Gaussian state $({\bf V}^{(c)},{\bf Q}^{(c)})$ and measurement pattern $\vt{\bar{n}}$}
\end{algorithm}

We next present examples for the generation of useful  single-mode non-Gaussian states using our general formalism. 

\section{Examples of generating single-mode non-Gaussian states}\label{sec:SM-Example}
We begin with pure Gaussian states in two and three modes. We then detect all but one of the modes to generate single-mode non-Gaussian states at the output. 
A few examples are considered in each case. 

\subsection{Detecting two-mode pure Gaussian states}

In this subsection, we are going to use our formalism to study the generation of single-mode non-Gaussian states via
detecting one mode of a pure two-mode Gaussian state. This is the simplest nontrivial case which already includes some practically interesting 
examples, e.g., Schr\"odinger cat states. We will investigate two kinds of problems: (i) to derive the output non-Gaussian state given the interferometer, the input states and the choice of measurement patterns; and (ii) to identify optimal Gaussian states (in terms of the interferometers and input states) which give the highest success probability and fidelity,
for a particular target non-Gaussian state.

In the two-mode case, $\kappa_2 = 0$ corresponds to a trivial case where the two modes are uncorrelated and detecting one of them cannot generate a non-Gaussian state. 
So we always consider the case where $\kappa_2 \ne 0$ in this subsection. We list explicitly the coefficients of the superposition of Fock states which are calculated 
by using either Eq.~\eqref{eq:CoeffRatio-single-1} or Eq.~\eqref{eq:CoeffRatio-single-2}. Note that depending on the number of photons detected in the PNRD, say $n$, the heralded state has a Fock state superposition up to $\ket{n}$, apart from the possible follow-up with a Gaussian gate. 

We now list the relations between the output Fock coefficients $\{c_n\}$ and the parameters of the Gaussian state. For a single photon detection we have 
\begin{eqnarray}\label{eq:CE1photon-2mode}
\frac{c_0}{c_1} = \mu_2;
\end{eqnarray}
for two photon detection we obtain the relations 
\begin{eqnarray}\label{eq:CE2photon-2mode}
		\frac{c_1}{c_2} = \sqrt{2} \, \mu_2, ~~~~~~ \frac{c_0}{c_2} = \frac{1}{\sqrt{2}} \big( \mu_2^2 + f_{22}^* \big); 
		\end{eqnarray}
three photon detection leads to 
	\begin{eqnarray}\label{eq:CE3photon-2mode}
		\frac{c_2}{c_3} &=& \sqrt{3} \,\mu_2, ~~~~~~ \frac{c_1}{c_3} = \sqrt{\frac{3}{2}} \big( \mu_2^2 + f_{22}^* \big), \nonumber\\
		\frac{c_0}{c_3} &=& \frac{\mu_2}{\sqrt{6}} \big( \mu_2^2 + 3 f_{22}^* \big);
		\end{eqnarray} 
and finally, four photon case gives 
		\begin{eqnarray}\label{eq:CE4photon-2mode}
		\frac{c_3}{c_4} &=& 2 \, \mu_2, ~~~~~~ \frac{c_2}{c_4} = \sqrt{3} \big( \mu_2^2 + f_{22}^* \big), \nonumber\\
		\frac{c_1}{c_4} &=& \sqrt{\frac{2}{3}} \mu_2 \big( \mu_2^2 + 3 f_{22}^* \big), \nonumber\\
		\frac{c_0}{c_4} &=& \frac{1}{2\sqrt{6}} \big( \mu_2^4 + 6 \, \mu_2^2 \, f_{22}^* + 3 \, f_{22}^{*2}  \big). 
		\end{eqnarray}
Using these relations, we can solve for the explicit output state given the initial Gaussian state that is to be measured. We now look at a concrete and commonly used technique of photon-subtraction.

\subsubsection{Photon subtraction from a squeezed vacuum state}
\label{subsec:photon-subtraction}

%\subsection{Example: Photon subtracted squeezed state}

\begin{figure}
    \centering
    \scalebox{1}{\includegraphics{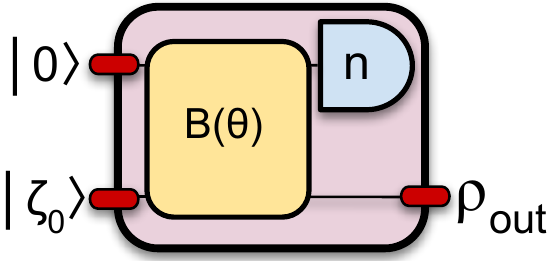}}
    \caption{Photon subtraction from a squeezed vacuum state.  A squeezed vacuum state $\ket{\zeta_0}$ is mixed with a vacuum via a beam splitter 
    $B(\theta)$. One of the output modes is detected by a PNRD, registering $n$ photons. The heralded state in the other mode is $\rho_{\text{out}}$. }
    \label{fig:photon-sub}
\end{figure}

Generating non-Gaussian states via photon subtraction from squeezed vacuum states  have been studied extensively. % in the literature. % (\textcolor{red}{add references}). 
Here, we consider photon subtraction for two purposes: the first is to show how to use our formalism to solve a specific problem, the second is to verify known
results via this new method. 
A setup to generate a photon subtracted state is shown in Fig.~\ref{fig:photon-sub}. A single-mode squeezed vacuum state $\ket{\zeta_0}$, with $\zeta_0 = r_0 e^{i \varphi_0}$, is combined with a
vacuum on a beam splitter, after which a PNRD measures one of the output modes and registers $n$ photons. 
Standard single photon subtraction uses a high transmission beam splitter and a single photon state is detected post measurement, however, here we do not restrict our beam splitter parameters and the photon-detection outcome. 

To simplify the problem, we assume that the phase of the squeezed vacuum state is $\varphi_0=0$, namely,
the covariance matrix is 
\begin{eqnarray}
{\bf V}_s^{(r)} = \frac{1}{2}
	\begin{pmatrix}
	e^{2r_0} & 0 \\
	0 & e^{-2r_0}
	\end{pmatrix},
\end{eqnarray}
where we use the basis $(\hat p_1, \hat q_1)$, which implies that the position quadrature is squeezed if $r_0>0$. The symplectic transformation of a beam splitter (and no additional phase) is chosen as
\begin{eqnarray}
{\bf S}_{\text{bs}}^{(r)} = 
	\begin{pmatrix}
	\cos \theta & - \sin \theta & 0 & 0 \\
	\sin \theta & \cos \theta & 0 & 0 \\
	0 & 0 & \cos \theta & - \sin \theta \\
	0 & 0 & \sin \theta & \cos \theta
	\end{pmatrix},
\end{eqnarray}
where we use the basis $(\hat p_1, \hat p_2, \hat q_1, \hat q_2)$ and $\cos^2 \theta$ is the transmission coefficient of the beam splitter. The output covariance matrix 
before detection is 
\begin{align}
&{\bf V}^{(r)} = {\bf S}_{\text{bs}}^{(r)} {\bf V}_s^{(r)} {\bf S}_{\text{bs}}^{(r)\top} = 
\frac{1}{2} \begin{pmatrix}
    V_{11} & 0 \\ 0 & V_{22}
\end{pmatrix}, \nonumber\\
& V_{11} = \begin{pmatrix}
    	e^{2r_0} c^2 + s^2 & (e^{2r_0} - 1) c s \\
    	(e^{2r_0} - 1) cs & c^2 + e^{2r_0} s^2
\end{pmatrix}, \nonumber\\
& V_{22} = \begin{pmatrix}
     e^{-2r_0}c^2 + s^2 & (e^{-2r_0} - 1) cs \\
     (e^{-2r_0} - 1) cs & c^2 + e^{-2r_0} s^2
\end{pmatrix},
\end{align}
where $c = \cos{\theta}$ and $s=\sin{\theta}$.
From Eq.~\eqref{eq:RYpq}, we obtain the matrix $\tilde{{\bf R}} = {\bf B} \oplus {\bf B}^*$ where ${\bf B}$ is given by 
\begin{eqnarray}\label{eq:BmatrixPhotonSubtract}
{\bf B} = \tanh r_0 
\begin{pmatrix}
	\cos^2 \theta & \cos \theta \sin \theta \\
	\cos \theta \sin \theta & \sin^2 \theta \\
	\end{pmatrix}.
\end{eqnarray}
%\begin{eqnarray}
%\tilde{\bf R} = \tanh r 
%\begin{pmatrix}
%	\cos^2 \theta & \cos \theta \sin \theta & 0 & 0 \\
%	\cos \theta \sin \theta & \sin^2 \theta & 0 & 0 \\
%	0 & 0 & \cos^2 \theta & \cos \theta \sin \theta \\
%	0 & 0 & \cos \theta \sin \theta & \sin^2 \theta
%	\end{pmatrix}
%\end{eqnarray}
%and 
%\begin{eqnarray}
%{\bf R} = \tanh r 
%\begin{pmatrix}
%	\cos^2 \theta & 0 & \cos \theta \sin \theta & 0 \\
%	0  & \cos^2 \theta & 0 & \cos \theta \sin \theta \\
%	\cos \theta \sin \theta & 0 & \sin^2 \theta & 0 \\
%	0 & \cos \theta \sin \theta & 0 & \sin^2 \theta
%	\end{pmatrix}. 
%\end{eqnarray}
By applying a permutation ${\bf P}$ on $\tilde{\bf R}$ we get the matrix ${\bf R}$ [Eq.~\eqref{eq:R}], whose block submatrices are
\begin{eqnarray*}
\begin{pmatrix}
	{\bf R}_{hh} & {\bf R}_{hd} \\
	{\bf R}_{dh} & {\bf R}_{dd}\\
	\end{pmatrix} = \tanh r_0 
\begin{pmatrix}
	\cos^2 \theta \,{\bf I}_2 & \cos \theta \sin \theta \,{\bf I}_2 \\
	\cos \theta \sin \theta \,{\bf I}_2 & \sin^2 \theta \,{\bf I}_2\\
	\end{pmatrix}.
\end{eqnarray*}

Now we have all the information to derive the heralded states. Since there is no displacement in the input, the heralded states do not contain any displacement, namely, $\vt{d} = {\bf 0}$.
Note that $b_{11} = \tanh r \cos^2 \theta \ne 0$ for nontrivial cases, which implies that the heralded states contain squeezing. The squeezing can be read out from the 
covariance matrix of the heralded state with zero photon detected ($n=0$), which is given by 
\begin{eqnarray}
{\bf V}_1^{(r)}(n=0) = \frac{1}{2}
	\begin{pmatrix}
	\lambda & 0 \\
	0 & 1/\lambda
	\end{pmatrix},
\end{eqnarray}
where $\lambda = \frac{1+\kappa}{1-\kappa} $ with $\kappa = \tanh r_0 \cos^2 \theta$. %$ =  \frac{1+\tanh r_0 \cos^2 \theta}{1 - \tanh r_0 \cos^2 \theta}$. 
This implies that the output state with zero photon detected is a single-mode
squeezed vacuum state. However, the amount of squeezing is smaller than the input squeezing.  

When the PNRD registers photons, the output state is a superposition of Fock states followed by a squeezing operation with squeezing factor $\lambda$. 
To determine the heralded state and the detection probability, we first have to calculate $\kappa_2, \mu_2, f_{22}, \vt{z}_p$ and ${\bf A}_p$.
Since there is no displacement in the input, $\mu_2 = 0$ and $\vt{z}_p = {\bf 0}$. From Eqs.~\eqref{eq:DefkappaC}, \eqref{eq:matrixF} and \eqref{eq:BmatrixPhotonSubtract},
\begin{eqnarray*}
\kappa_2 &=& \frac{\kappa \tan \theta}{\sqrt{1 - \kappa^2}}, ~~~~~~
f_{22} = \frac{1}{\kappa},
\end{eqnarray*}
and from Eq.~\eqref{eq:probApZp} we have 
\begin{eqnarray*}
{\bf A}_p = \frac{\kappa \tan^2 \theta}{1 - \kappa^2}
	\begin{pmatrix}
	1 & \kappa \\
	\kappa & 1
	\end{pmatrix}.
\end{eqnarray*}

When the PNRD registers one photon, the heralded state is of the form $\hat{S}(r_s) (c_0 \ket{0} + c_1 \ket{1})$, 
where $r_s = \frac{1}{2} \ln \lambda $. From Eq.~\eqref{eq:CE1photon-2mode}
we find $c_0/c_1 = \mu_2 = 0$. Therefore, the heralded state is a squeezed single-photon state,
\begin{eqnarray}
\ket{\psi(n=1)} = \hat{S}(r_s) \ket{1}.
\end{eqnarray}
From Eq.~\eqref{eq:ProbabilitySinlgemode}, the detection probability is found to be 
\begin{eqnarray}
P(1) = \frac{\kappa^2 \tan^2 \theta}{\cosh r_0 (1 - \kappa^2)^{3/2}}. 
\end{eqnarray}

\begin{figure*}
\scalebox{0.38}{\includegraphics{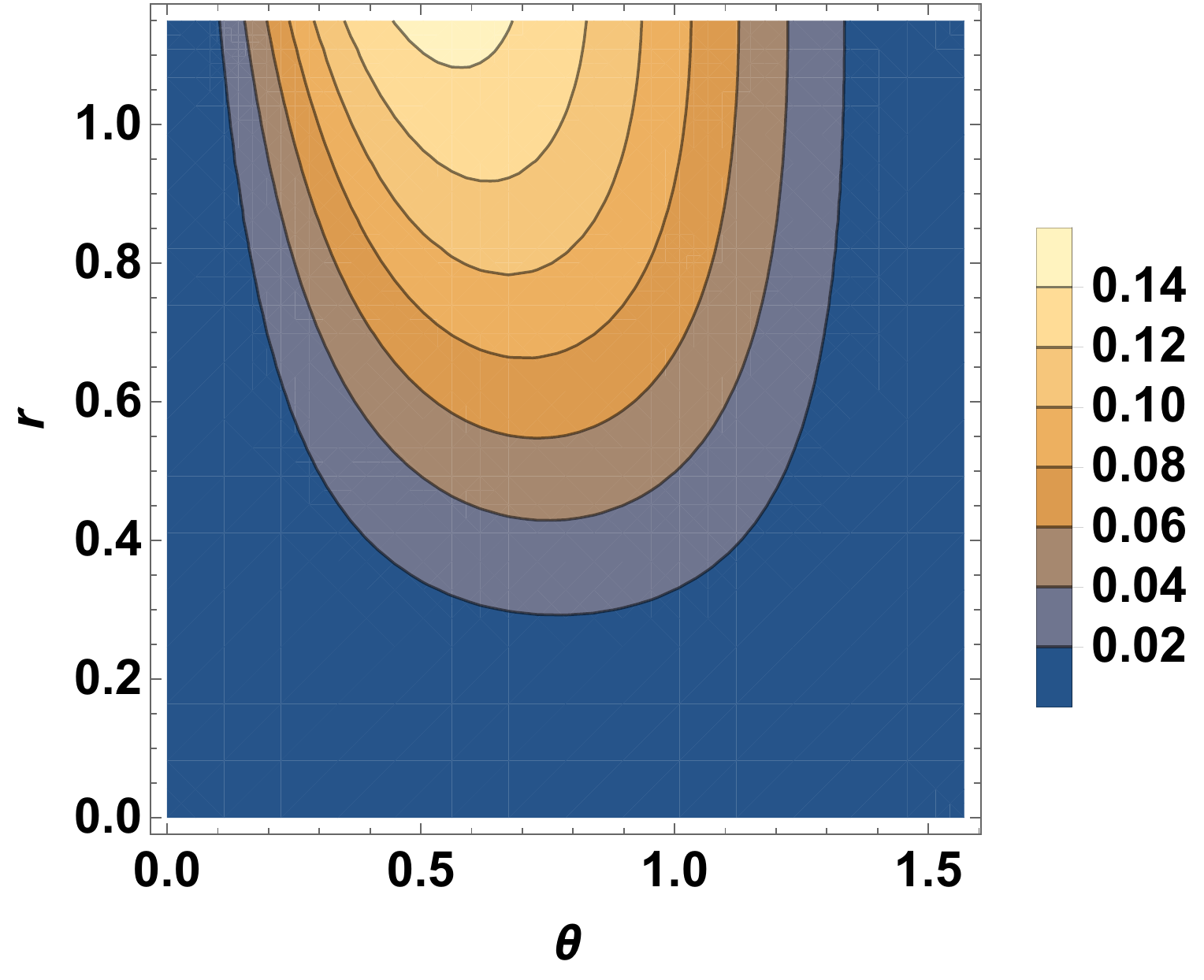}
\includegraphics{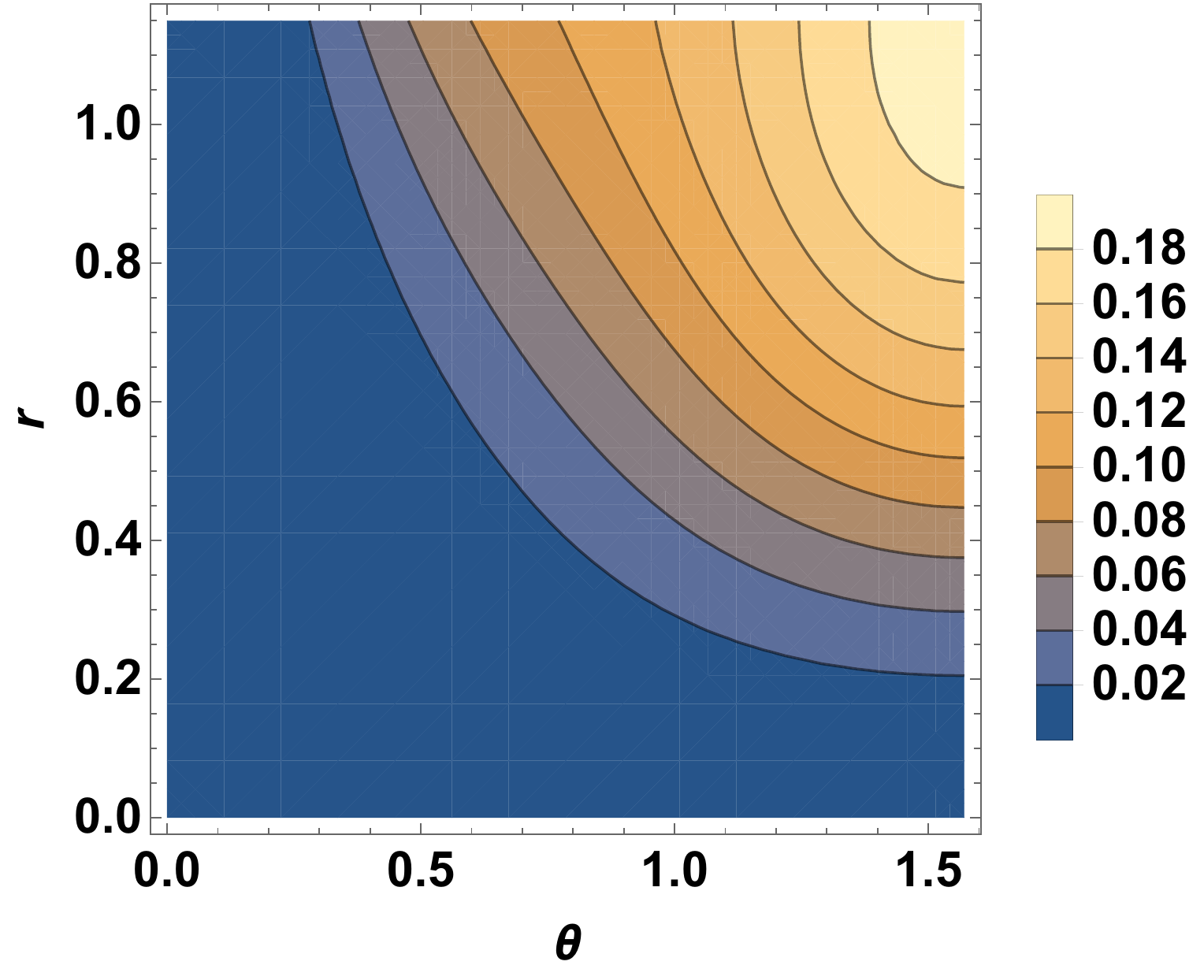}
\includegraphics{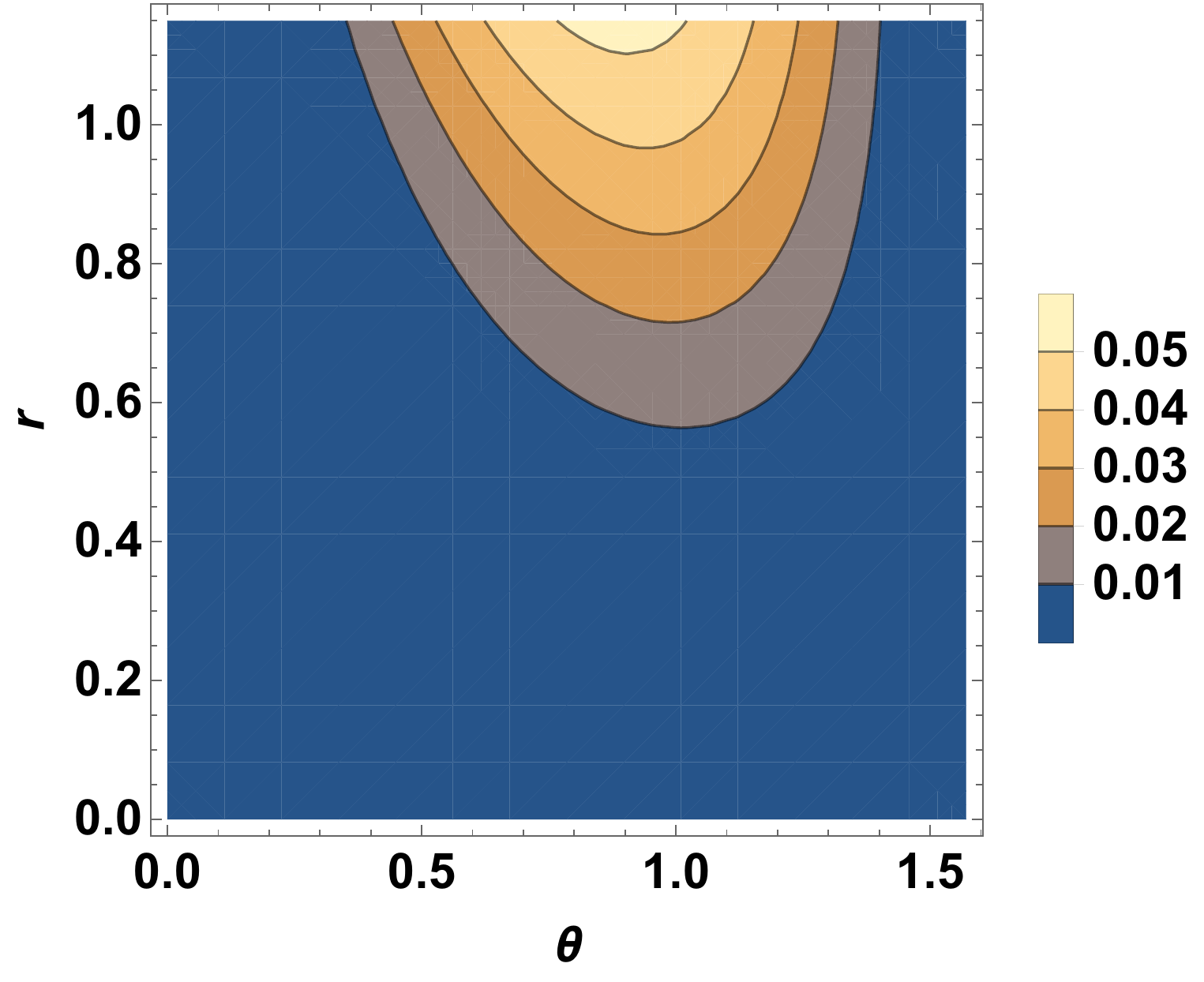}}
\caption{Contour plots of the success probabilities of detecting 1, 2 and 3 photons, respectively, in the optical scheme represented in Fig.~\ref{fig:photon-sub}, as a function of the input squeezing parameter $r_0 \in [0, 1.15]$, i.e., in the range $0-10$ dB and the beam splitter angle $\theta \in [0,\frac{\pi}{2}]$. The behaviour of the even photon detection $(n=2)$ is qualitatively different from that of the odd photon detection $(n=1,3)$. The three photon detection is an order of magnitude smaller than the 1 or 2 photon detection cases. The bottom (dark blue) regions of the contour plots correspond to near-zero success probability in the parameter space. }
\label{fig:fockprob}
\end{figure*}

When the PNRD detects two photons, the heralded state is of the form $\hat{S}(r_s) (c_0 \ket{0} + c_1 \ket{1} + c_2 \ket{2})$. From Eq.~\eqref{eq:CE2photon-2mode}
we find that $c_1/c_2 = 0$ and $c_0/c_2 = f_{22}^*/\sqrt{2}$. Taking into account the normalization condition $|c_0|^2+|c_2|^2=1$, we find
%\begin{eqnarray}
%c_0 = \frac{1}{\sqrt{1+2 \, \kappa^2}}, ~~~~~~ c_2 = \frac{\sqrt{2} \, \kappa}{\sqrt{1+2 \, \kappa^2}}
%\end{eqnarray}
the heralded state to be 
\begin{eqnarray}
\ket{\psi(n=2)} = \hat{S}(r_s) \bigg[ \frac{1}{\sqrt{1+2 \, \kappa^2}} \ket{0} + \frac{\sqrt{2} \, \kappa}{\sqrt{1+2 \, \kappa^2}} \ket{2} \bigg],
\end{eqnarray}
with measurement probability
\begin{eqnarray}
P(2) = \frac{\kappa^2 (1+2\kappa^2) \tan^4 \theta}{2 \, \cosh r_0 (1 - \kappa^2)^{5/2}}. 
\end{eqnarray}

When the PNRD registers three photons, the heralded state is of the form $\hat{S}(r_s) (c_0 \ket{0} + c_1 \ket{1} + c_2 \ket{2} + c_3 \ket{3})$. 
From Eq.~\eqref{eq:CE3photon-2mode}
we find that $c_0 = c_2 = 0$ and $c_1/c_3 = \sqrt{3} f_{22}^*/\sqrt{2}$. Taking into account the normalization condition $|c_1|^2+|c_3|^2=1$, we find that
%\begin{eqnarray}
%c_0 = \frac{1}{\sqrt{1+2 \, \kappa^2}}, ~~~~~~ c_2 = \frac{\sqrt{2} \, \kappa}{\sqrt{1+2 \, \kappa^2}}
%\end{eqnarray}
the heralded state and the success probability is 
\begin{eqnarray}
&&\ket{\psi(n=3)} = \hat{S}(r_s) \bigg[ \frac{\sqrt{3}}{\sqrt{3+2 \, \kappa^2}} \ket{1} + \frac{\sqrt{2} \, \kappa}{\sqrt{3+2 \, \kappa^2}} \ket{3} \bigg], \nonumber\\
&&P(3) = \frac{\kappa^4 (3+2\kappa^2) \tan^6 \theta}{2 \, \cosh r_0 (1 - \kappa^2)^{7/2}}. 
\end{eqnarray}
These results are consistent with those derived using a different method \cite{PhysRevA.55.3184} and we schmatically depict the dependence of the success probability as a function of the input squeezing parameter $r$ and the beam splitter angle $\theta$ in Fig.~\ref{fig:fockprob}. We next consider the case of generation of cat states. 

\subsubsection{Target Schr\"odinger cat state}

The goal of this section is complementary to that of Sec.~\ref{subsec:photon-subtraction}: we want to search for a multimode Gaussian state and a measurement scheme, to 
generate Schr\"odinger cat states with high fidelity and success probability. The same procedure 
can be generalized in a straightforward manner to target other non-Gaussian states, such as GKP states, which we consider in the next subsection. % and weak cubic phase states. 

A Schr\"odinger cat state is a superposition of two coherent states
with opposite phases: $\ket{\alpha}$ and $\ket{-\alpha}$. Two orthogonal cat states are of particular interest, the even cat state $\ket{\text{cat}_e}$ and the odd cat state
$\ket{\text{cat}_o}$, given by
\begin{eqnarray}\label{eq:CatStates}
\ket{\text{cat}_e} &=& \frac{1}{\sqrt{2(1+e^{-2|\alpha|^2})}} (\ket{\alpha} + \ket{-\alpha}),
\nonumber\\
\ket{\text{cat}_o} &=& \frac{1}{\sqrt{2(1-e^{-2|\alpha|^2})}} (\ket{\alpha} - \ket{-\alpha}). 
\end{eqnarray}
The even cat state is a superposition of only even Fock states, whilst the odd cat state is a superposition of only odd Fock states. 

When $\alpha$ is small, the even cat state can be well approximated by $c_0 \ket{0} + c_2 \ket{2}$, an example of an ON state \cite{ONstate}. If $\alpha$ is large then one needs to 
introduce a higher Fock state support to approximate the cat state. However, we find that by squeezing $c_0 \ket{0} + c_2 \ket{2}$ one can obtain a
very good approximation to an even cat state with a larger $\alpha$, namely, $\hat{S}(\zeta_1) (c_0 \ket{0} + c_2 \ket{2})$ could be a good approximation 
to $\ket{\text{cat}_e}$. This is due to the squeezing operator pulling apart the two peaks of the cat state. Table~\ref{tab:ApproxEvenCat} shows 
how well $\hat{S}(\zeta_1) (c_0 \ket{0} + c_2 \ket{2})$ approximates an even cat state. We see that the fidelity drops from perfect fidelity to $97\%$ as  $\alpha$ varies from $0$ to $2$.

\begin{table*}[tp]
\caption{ Target an even cat state by detecting a two-mode Gaussian state with a PNRD. The even cat state is approximated by $\hat{S}(\zeta_1) (c_0 \ket{0} + c_2 \ket{2})$. $\mathcal{F}_{\text{max}}$ is the highest fidelity between the cat state and the approximation, $P_{\text{max}}$ is the optimal success probability,
$\zeta_{01}$ and $\zeta_{02}$ are the squeezing parameters of input squeezed vacuum states of the two modes, and $\theta$ is the parameter of the beam splitter defined as 
$e^{\theta(\hat{a}_1 \hat{a}_2^{\dag} - \hat{a}_1^{\dag} \hat{a}_2)}$. We observe that the squeezing requirement on the first arm is substantially more than that of the second arm. The maximum success probability decreases with increasing cat state parameter $\alpha$.} 
\label{tab:ApproxEvenCat}
\centering
\begin{center}
\resizebox{0.9\textwidth}{!}{
    \begin{tabular}{| c | c | c | c | c | c | c | c |}
    \hline  \hline
    ~~~~~~$\alpha$~~~~~~ & ~~~~~~$\mathcal{F}_{\text{max}}$~~~~~~ & ~~~~~~$\zeta_1$~~~~~~& ~~~~~~$c_0/c_2$~~~~~~ &
   ~~~~~~$P_{\text{max}}$~~~~~~&~~~~~~$\zeta_{01}$~~~~~~&~~~~~~$\zeta_{02}$~~~~~~ &~~~~~~$\theta$~~~~~~ \\ \hline 
    0.25 & 1.0000 & 0.0115 & 27.717 & 18.12\% & 1.1587 & $-0.0136$ & $-1.3965$ \\ \hline 
    0.50 & 1.0000 & 0.0458 & 6.9428 & 15.49\% & 1.1936 & $-0.0499$ & $1.2351$ \\ \hline 
    0.75 & 0.9999 & 0.1025 & 3.1112 & 12.87\% & 1.2447 & $-0.0982$ & $-1.0927$ \\ \hline 
    1.00 & 0.9999 & 0.1796 & 1.7885 & 11.20\% & 1.3073 & $-0.1474$ & $-0.9686$ \\ \hline
    1.25 & 0.9991 & 0.2730 & 1.1932 & 10.55\% & 1.3780 & $-0.1898$ & $0.8606$ \\ \hline
    1.50 & 0.9958 & 0.3763 & 0.8841 & 10.51\% & 1.4546 & $-0.2228$ & $-0.7668$ \\ \hline
    1.75 & 0.9870 & 0.4832 & 0.7082 & 10.73\% & 1.5346 & $-0.2464$ & $-0.6859$ \\ \hline
    2.00 & 0.9709 & 0.5884 & 0.6011 & 11.01\% & 1.6150 & $-0.2626$ & $-0.6170$ \\ 
    \hline
    \end{tabular}
   }
\end{center}
\end{table*}

\begin{table*}[tp]
\caption{ Target an odd cat state by detecting a two-mode Gaussian state with a PNRD. The odd cat state is
approximated by $\hat{S}(\zeta_1) (c_1 \ket{1} + c_3 \ket{3})$. $\mathcal{F}_{\text{max}}$ is the highest fidelity between the cat state and the approximation, 
$P_{\text{max}}$ is the optimal success probability, $\zeta_{01}$ and $\zeta_{02}$ are the squeezing parameters of input squeezed vacuum states, 
and $\theta$ is the parameter of the beam splitter (as in the even cat case).  As for the even cat generation, we observe that the squeezing requirement on the first arm is substantially more than that of the second arm. However, the  maximum success probability has the opposite behaviour and increases with increasing cat state parameter $\alpha$.} 
\label{tab:ApproxOddCat}
\centering
\begin{center}
\resizebox{0.9\textwidth}{!}{
    \begin{tabular}{| c | c | c | c | c | c | c | c |}
    \hline  \hline
    ~~~~~~$\alpha$~~~~~~ & ~~~~~~$\mathcal{F}_{\text{max}}$~~~~~~ & ~~~~~~$\zeta_1$~~~~~~& ~~~~~~$c_1/c_3$~~~~~~ &
   ~~~~~~$P_{\text{max}}$~~~~~~&~~~~~~$\zeta_{01}$~~~~~~&~~~~~~$\zeta_{02}$~~~~~~ &~~~~~~$\theta$~~~~~~ \\ \hline 
    0.25 & 1.0000 & 0.0044 & 49.636 & 1.11\% & 1.3288 & $-0.0197$ & $1.4053$ \\ \hline 
    0.50 & 1.0000 & 0.0306 & 15.507 & 2.97\% & $1.3538$ & $-0.0444$ & $1.2813$ \\ \hline 
    0.75 & 1.0000 & 0.0687 & 6.9179 & 5.01\% & $1.3945$ & $-0.0903$ & $1.1554$ \\ \hline 
    1.00 & 0.9999 & 0.1213 & 3.9303 & 6.32\% & $1.4442$ & $-0.1414$ & $1.0445$ \\ \hline
    1.25 & 0.9999 & 0.1870 & 2.5664 & 6.95\% & $1.4998$ & $-0.1907$ & $0.9468$ \\ \hline
    1.50 & 0.9995 & 0.2633 & 1.8435 & 7.21\% & $1.5605$ & $-0.2339$ & $0.8603$ \\ \hline
    1.75 & 0.9979 & 0.3467 & 1.4229 & 7.35\% & $1.6242$ & $-0.2692$ & $0.7835$ \\ \hline
    2.00 & 0.9938 & 0.4336 & 1.1620 & 7.47\% & $1.6900$ & $-0.2967$ & $0.7153$ \\ 
    \hline
    \end{tabular}
   }
\end{center}
\end{table*}

The state $\hat{S}(\zeta_1) (c_0 \ket{0} + c_2 \ket{2})$ can be generated by detecting a two-mode Gaussian state with two photons registered in our general scheme. 
Let us target a state given by a particular $\zeta_1$ and $c_0/c_2$. For simplicity, we assume $\alpha$ is real, so $\zeta_1$, $c_0$ and
$c_2$ are also real. By using Eq.~\eqref{eq:b11Squeezing} we can derive $b_{11}$ from $\zeta_1$: $b_{11} = \tanh \zeta_1$. 
From Eq.~\eqref{eq:CE2photon-2mode} we find $\mu_2 = 0$ and $f_{22} = \sqrt{2} \, c_0/c_2$. Therefore, the matrix ${\bf B}$ can be written as
\begin{eqnarray}\label{eq:BmatrixEvenCat}
{\bf B} =  
\begin{pmatrix}
	\tanh \zeta_1 & \kappa_2 \sech \zeta_1 \\
	\kappa_2 \sech \zeta_1 & \kappa_2^2 \big(\sqrt{2} \, c_{02}- \tanh \zeta_1 \big) 
	\end{pmatrix},
\end{eqnarray}
where we have defined $c_{02} = c_0/c_2$ and $\kappa_2$ is an unknown parameter. The parameter $\kappa_2$ has to be chosen such that ${\bf B}$ corresponds 
to a physical two-mode Gaussian state, namely, the singular values of ${\bf B}$ should be smaller than one, a condition that is easily derived from Eq.~\eqref{eq:Bmatrix-Signle}.
Provided we have a physical state, the success probability of detecting two photons in the second mode is
\begin{eqnarray}\label{eq:ProbEvenCat02}
P(2) = \big(1+c_{02}^2 \big) \kappa_2^4 \sqrt{1-2 \, \kappa_2^2 + \big(1-2 \, c_{02}^2 \big)\kappa_2^4}. 
\end{eqnarray}
Note that the success probability is independent of $\zeta_1$. This can be understood as follows. Generating $\hat{S}(\zeta_1) (c_0 \ket{0} + c_2 \ket{2})$ can be performed in two steps: we first target $c_0 \ket{0} + c_2 \ket{2}$ with success probability given by Eq.~\eqref{eq:ProbEvenCat02}, and after the photon number detection we apply a 
squeezing gate $\hat{S}(\zeta_1)$. Since the order of performing photon number detection and applying a local unitary is irrelevant, we can absorb the local unitary
gate into the circuit without changing the detection probability. Recall that this fact was highlighted for a general case in Fig.~\ref{fig:symmetry}.

There is one free parameter, $\kappa_2$, in the success probability of Eq.~\eqref{eq:ProbEvenCat02}, that can be used to optimize. After the optimization, 
we substitute $\kappa_2$ back into Eq.~\eqref{eq:BmatrixEvenCat} to determine the optimal input squeezed states and the circuit. We target even 
cat states with representative values of $\alpha$, and calculate the maximal fidelity $\mathcal{F}_{\text{max}}$, maximal success probability $P_{\text{max}}$, 
input squeezing and circuit parameters, as summarized in Table~\ref{tab:ApproxEvenCat}. It shows that high fidelity ($> 97\%$) and high success probability ($>10\%$)
can be achieved for $\alpha \le 2$. This is the best one can achieve by detecting two-mode Gaussian states to generate an even cat state. The requirement for
input squeezing, $1.1587 < r_{01} < 1.6150$, is on the high side, which corresponds to squeezing in the range $\sim 10 - 14$ dB. However, this range of squeezing is within 
current technology since $15$ dB squeezing has been demonstrated experimentally~\cite{PhysRevLett.117.110801}. 
If the amount of input squeezing is limited to a certain value, one either obtains a lower fidelity and/or a lower success probability.
One useful application of squeezed cat states for suppressing decoherence was demonstrated in Ref.~\cite{PhysRevLett.120.073603}. Using our formalism, one can generate squeezed cat states in a transparent manner using only  offline squeezing, as alluded to earlier in Fig.~\ref{fig:symmetry}.

An odd cat state $\ket{\text{cat}_o}$ can be well approximated by a squeezed single-photon state: $\hat{S}(\zeta_1) \ket{1}$~\cite{PhysRevA.70.020101}. The fidelity is 
greater than $99\%$ for $\alpha<1.2$, but quickly drops to $87.8\%$ when $\alpha = 2.0$. 
From Eq.~\eqref{eq:CE1photon-2mode}
we find that $\mu_2 = 0$, indicating that there is no input displacement. The matrix ${\bf B}$ can be written as 
\begin{eqnarray}\label{eq:BmatrixOddCat1}
{\bf B} =  
\begin{pmatrix}
	\tanh \zeta_1 & \kappa_2 \sech \zeta_1 \\
	\kappa_2 \sech \zeta_1 & \kappa_2^2 \big(f_{22}- \tanh \zeta_1 \big) 
	\end{pmatrix}
\end{eqnarray}
and the success probability of detecting one photon in the second mode is 
\begin{eqnarray}\label{eq:ProbOddCat1}
P(1) = \kappa_2^2 \sqrt{1-2 \, \kappa_2^2 + \big(1- f_{22}^{\,2} \big)\kappa_2^4}. 
\end{eqnarray}
It is evident that the success probability $P(1)$ is optimized to be $25\%$ when $f_{22}=0$ and $\kappa_2 = 1/\sqrt{2}$. 

\begin{figure}
\includegraphics[width=\columnwidth]{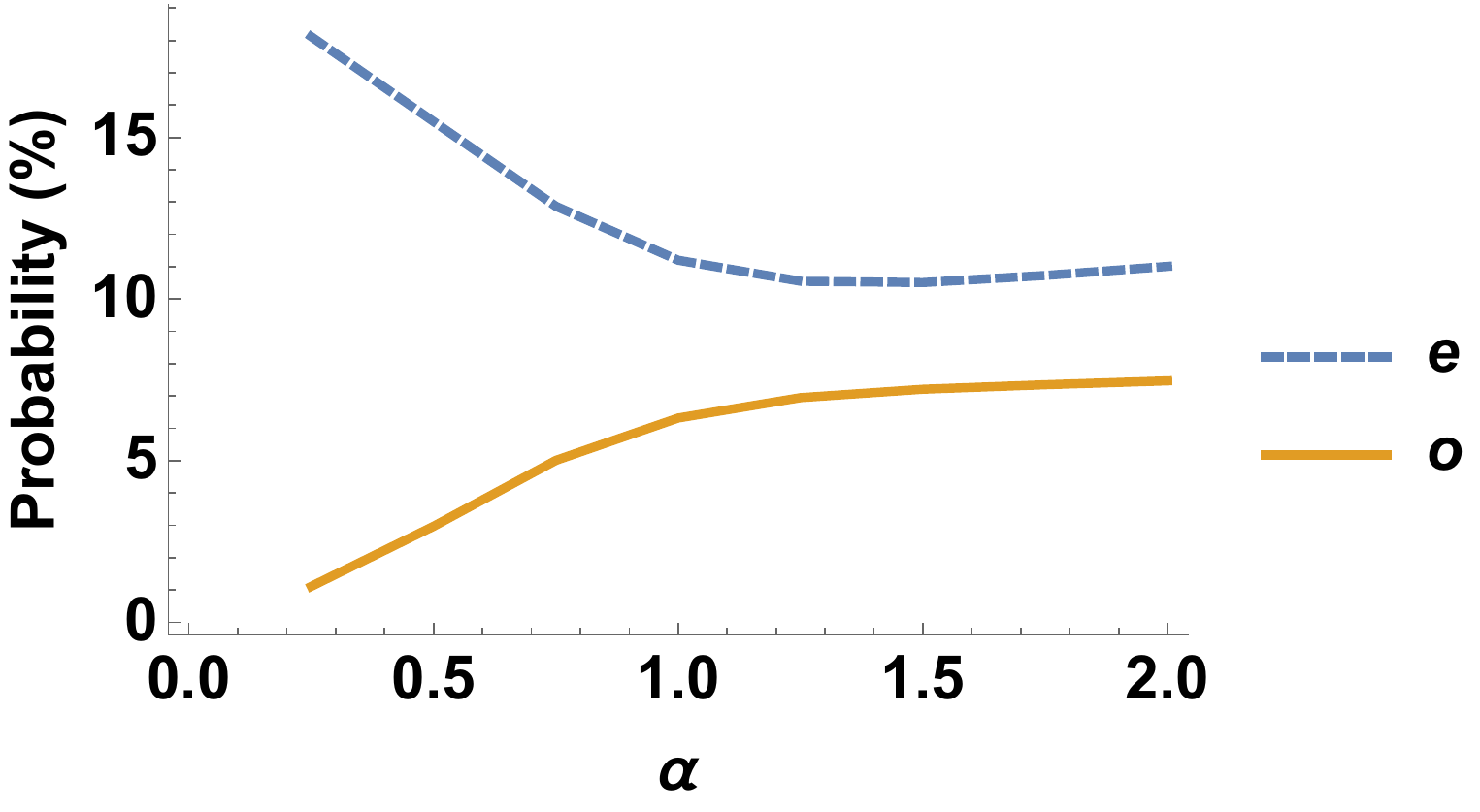}
\caption{Comparison of the maximum success probability for even and odd cat states as a function of the amplitude $\alpha$. We find that when using two-mode Gaussian input states, even cat states (e) are prepared with a higher success probability than the corresponding  (same amplitude) odd cat state (o).  }
\label{probcat}
\end{figure}

To obtain a better approximation for an odd cat state with a larger $\alpha$, we can replace the squeezed single-photon state by 
%Similarly, an odd cat state $\ket{\text{cat}_o}$ can be well approximated by $c_1 \ket{1} + c_3 \ket{3}$ for a small $\alpha$ and by 
$\hat{S}(\zeta_1)(c_1 \ket{1} + c_3 \ket{3})$. Again, we assume $\alpha$ is real, so $\zeta_1$, $c_1$ and $c_3$ are also real. 
To get a superposition of Fock states up to $\ket{3}$, one needs to detect three photons in the second mode. From Eq.~\eqref{eq:CE3photon-2mode} 
we find that $\mu_2 = 0$ and $f_{22} = \sqrt{2/3} \, c_1/c_3$. Therefore, the matrix ${\bf B}$ can be written as
\begin{eqnarray}\label{eq:BmatrixOddCat13}
{\bf B} =  
\begin{pmatrix}
	\tanh \zeta_1 & \kappa_2 \sech \zeta_1 \\
	\kappa_2 \sech \zeta_1 & \kappa_2^2 \big(\sqrt{2/3} \, c_{13}- \tanh \zeta_1 \big) 
	\end{pmatrix},
\end{eqnarray}
where we have defined $c_{13} = c_1/c_3$. Similarly, The parameter $\kappa_2$ has to be chosen to correspond to a physical two-mode Gaussian state. 
Provided this is true, the success probability of detecting three photons in the second mode is
\begin{eqnarray}\label{eq:ProbOddCat13}
%P(n=3) = \big(1+c_{13}^2 \big) \kappa_2^6 \sqrt{1-2 \, \kappa_2^2 + \bigg(1- \frac{2}{3} \, c_{02}^2 \bigg)\kappa_2^4}. 
P(3) = \big(1+c_{13}^2 \big) \kappa_2^6 \sqrt{1-2 \, \kappa_2^2 + \bigg(1- \frac{2}{3} \, c_{13}^2 \bigg)\kappa_2^4}. 
\end{eqnarray}
The free parameter $\kappa_2$ is further chosen to optimize the success probability, after which we substitute it back into Eq.~\eqref{eq:BmatrixOddCat13} to 
determine the optimal input squeezed states and the circuit. The results are summarized in Table~\ref{tab:ApproxOddCat}. We can see that a higher fidelity is 
obtained for a given $\alpha$, at the expense of a reduced success probability. To compare the generation of even and odd cat states we plot the maximum success probability as a function of the cat amplitude $\alpha$ for both the even and odd cat generation in Fig.~\ref{probcat}.

\subsection{Examples of detecting three-mode pure Gaussian states}

We now use our formalism to study the generation of single-mode non-Gaussian states via detecting two modes of a pure three-mode Gaussian state. Conjecture~\ref{conj} implies that increasing the number of modes should allow us to
target a larger region of state space. In particular, measuring a three-mode Gaussian state can generate an arbitrary superposition of Fock states up to $\ket{5}$,
followed by a Gaussian unitary operation. This means we can improve the fidelity and success probability for certain target states produced using the two-mode circuit.  We can also target more complex states, such as the ON states, GKP code states, and weak cubic phase states. We focus on searching for the best interferometer, input states and measurement schemes that give the highest success probability and fidelity, for a given target non-Gaussian state.

\subsubsection{GKP states}

The GKP code states were proposed in Ref.~\cite{PhysRevA.64.012310} to encode qubits in CV quantum modes, that would also protect against small quadrature shifts in phase space. 
It was recently shown that the GKP codes can also protect against excitation loss extremely well~\cite{PhysRevA.97.032346}. 
Although numerous methods have been proposed~\cite{pirandola2004constructing, pirandola2006generating, vasconcelos2010all, PhysRevA.97.022341, PhysRevA.95.053819}, 
generating optical GKP states remains very challenging. Here, we use our formalism to conditionally generate the GKP states.  The ideal GKP states are superpositions
of infinitely squeezed vacuum states, which are unphysical because they require infinite energy. In reality, one replaces the infinitely squeezed states by finitely squeezed states to 
construct approximate GKP states. The two codewords that represent the logical basis states $\ket{\tilde 0}$ and $\ket{\tilde 1}$ can be written in the position basis as~\cite{PhysRevA.64.012310}
\begin{eqnarray}\label{eq:GKPwavefunction}
\psi_{\tilde 0}(q) &=& \frac{N_0}{(\pi \Delta^2)^{1/4}} \sum_{s=-\infty}^{+\infty} e^{-2 \pi \Delta^2 s^2 - (q-2 s \sqrt{\pi})^2/(2 \Delta^2)}, \nonumber\\
\psi_{\tilde 1}(q) &=& \frac{N_1}{(\pi \Delta^2)^{1/4}} \sum_{s=-\infty}^{+\infty} \exp\bigg\{- \frac{1}{2} \pi \Delta^2 (2s+1)^2 
\nonumber\\
&& 
 - \frac{[q-(2s+1)\sqrt{\pi} \, ]^2}{2 \Delta^2} \bigg\},
\end{eqnarray}
where $\Delta$ is the standard deviation and characterizes the amount of squeezing of the codewords, and $N_0$ and $N_1$ are normalization factors. 

\begin{figure}
\includegraphics[width=\columnwidth]{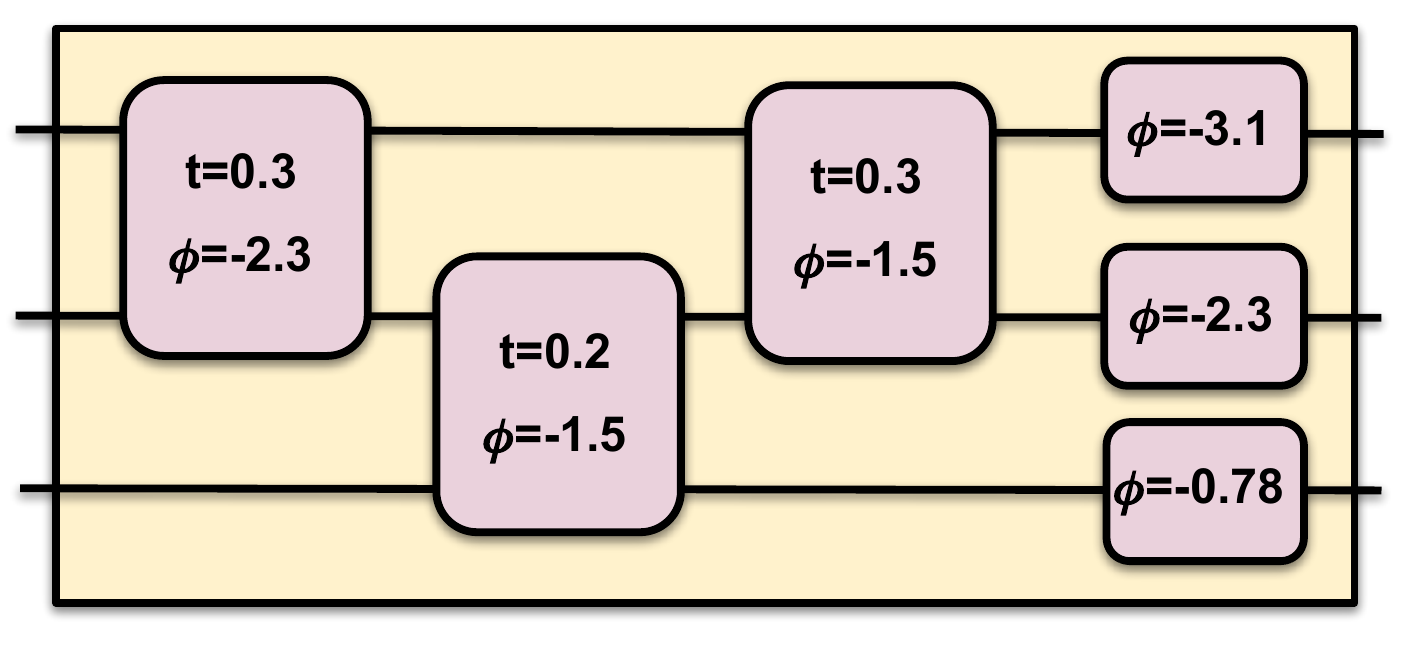}
\caption{Square decomposition of the unitary operator given in Eq.~\eqref{eq:weak-uni}. The first operator depicts a beam splitter of transmission $t$ preceded by a phase rotation by $\phi$ in the first mode alone. The operator denoted by only a phase angle is the standard phase rotation gate.  }
\label{fig:sq-gkp}
\end{figure}

\begin{figure}
\includegraphics[width=\columnwidth]{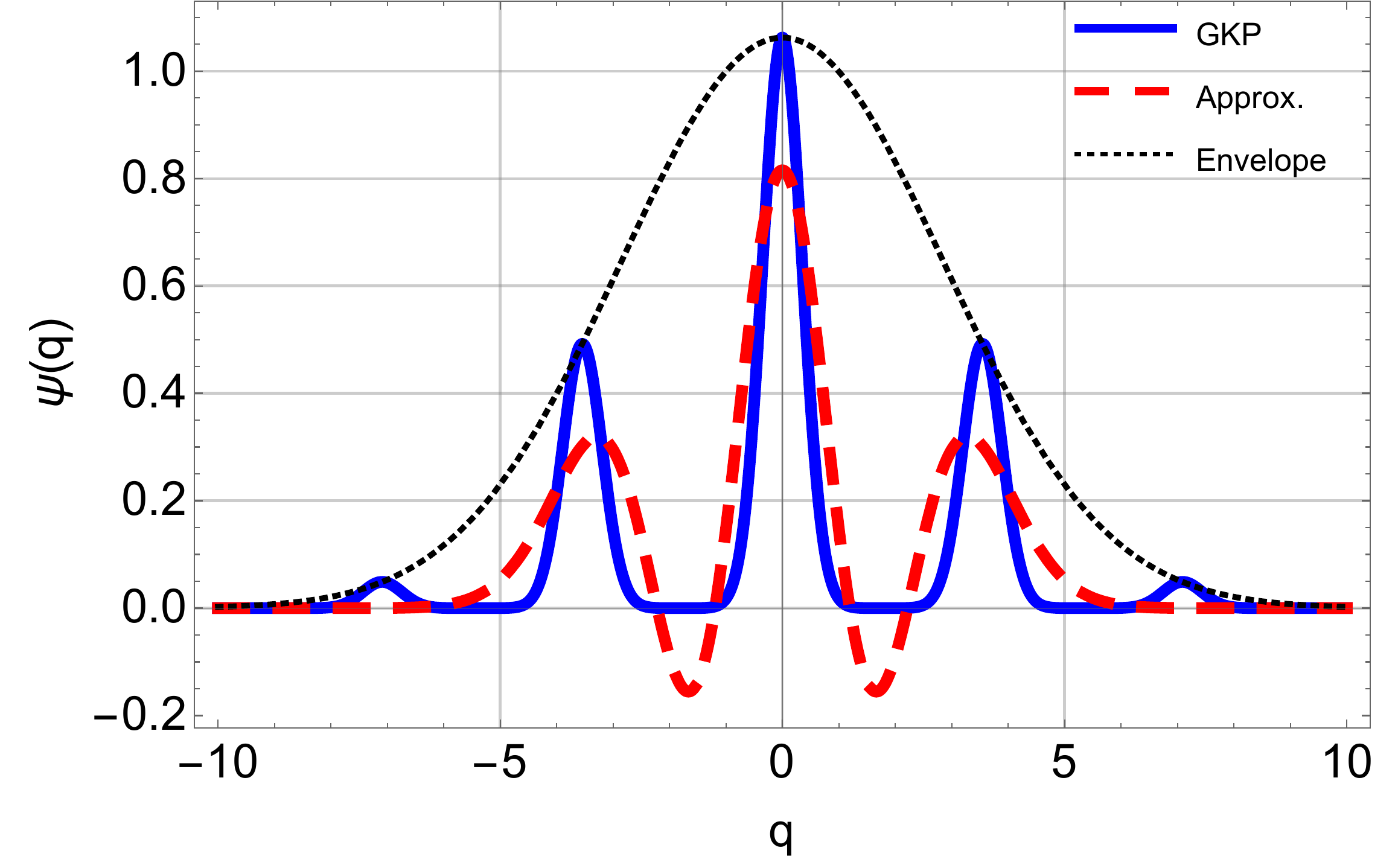}
\caption{The wave functions for the GKP state $\psi_{\tilde 0}(q)$ (blue solid line) from Eq.~(\ref{eq:GKPwavefunction}) and the approximate four-photon GKP state from Eq.~(\ref{eq:GKP024Approx}) (red dashed line) with $\zeta_1 = 0.294, c_0 = 0.669, c_2 = -0.216$ and  $ c_4 = 0.711$. The fidelity between these two states is $81.8\%$. The GKP Gaussian envelope is also shown (black dotted line).}
\label{fig:PlotGKPwavefun}
\end{figure}

It is evident that the wave functions in Eq.~\eqref{eq:GKPwavefunction} for the codewords $\ket{\tilde 0}$ and $\ket{\tilde 1}$ are even, therefore 
they should be expanded using only even Fock states. As an example, we approximate $\ket{\psi_{\tilde 0}}$ by 
\begin{eqnarray}\label{eq:GKP024Approx}
\hat{S}(\zeta_1) (c_0 \ket{0} + c_2 \ket{2} + c_4 \ket{4}),
\end{eqnarray}
which is in the form of Eq.~\eqref{eq:StateSingle}. Specifically, we choose $\Delta = 0.35$, corresponding to $9.12$ dB of squeezing. 
The highest fidelity between $\ket{\psi_{\tilde 0}}$ and the state \eqref{eq:GKP024Approx} is $81.8\%$ and is achieved when $\zeta_1 = 0.294, c_0 = 0.669, c_2 = -0.216$ and  $ c_4 = 0.711$. The wave functions for the GKP state $\psi_{\tilde 0}(q)$ from Eq.~(\ref{eq:GKPwavefunction}) and the approximate state in Eq.~(\ref{eq:GKP024Approx}) are shown in Fig.~\ref{fig:PlotGKPwavefun}. We generate the state \eqref{eq:GKP024Approx} by measuring two modes of a three-mode Gaussian state with
measurement outcome $\bar{\vt{n}} = (2, 2)$. The best success probability we obtained was approximately $1.1\%$.  
The three input squeezing parameters are $(r_1,r_2,r_3) = $ $(1.33803,0.101223,0.0994552)$ and the unitary corresponding to the interferometer is given by 
\begin{align}\label{eq:weak-uni}
    {\bf U} &= \begin{pmatrix}
        0& - 0.704006 i& -0.710195 \\ 0.707107 &
u_{22}&
 u_{33} \\
 -0.707107  &
 u_{22}& u_{33}
    \end{pmatrix},\nonumber\\
    u_{22} &=  0.355097 - 0.355098 i, \nonumber\\ 
    u_{33} &=  0.352003 + 0.352002 i.
\end{align}

One can perform a square decomposition~\cite{clements} of this interferometer as depicted in Fig.~\ref{fig:sq-gkp} using a python library\,\cite{web-interferometer}. The decomposition is made into two operators, the first is a beam splitter preceded by a phase rotation in the first mode, and the second only a phase rotation. The first operator has two parameters, a transmissivity $t =\cos^2{\theta}$ and a rotation angle $\phi$ that together induce the following unitary on the mode operators 
\begin{align}
    U(t,\phi) = \begin{pmatrix}
        e^{i \phi} \cos{\theta} & -\sin{\theta}\\
        e^{i \phi} \sin{\theta} & \cos{\theta}
    \end{pmatrix}.
\end{align}
The operator depicted with only a phase angle $\phi$ induces the transformation $\hat{a} \to e^{i\phi} \hat{a}$.

\subsubsection{Weak cubic phase states}

\begin{figure}
\includegraphics[width=\columnwidth]{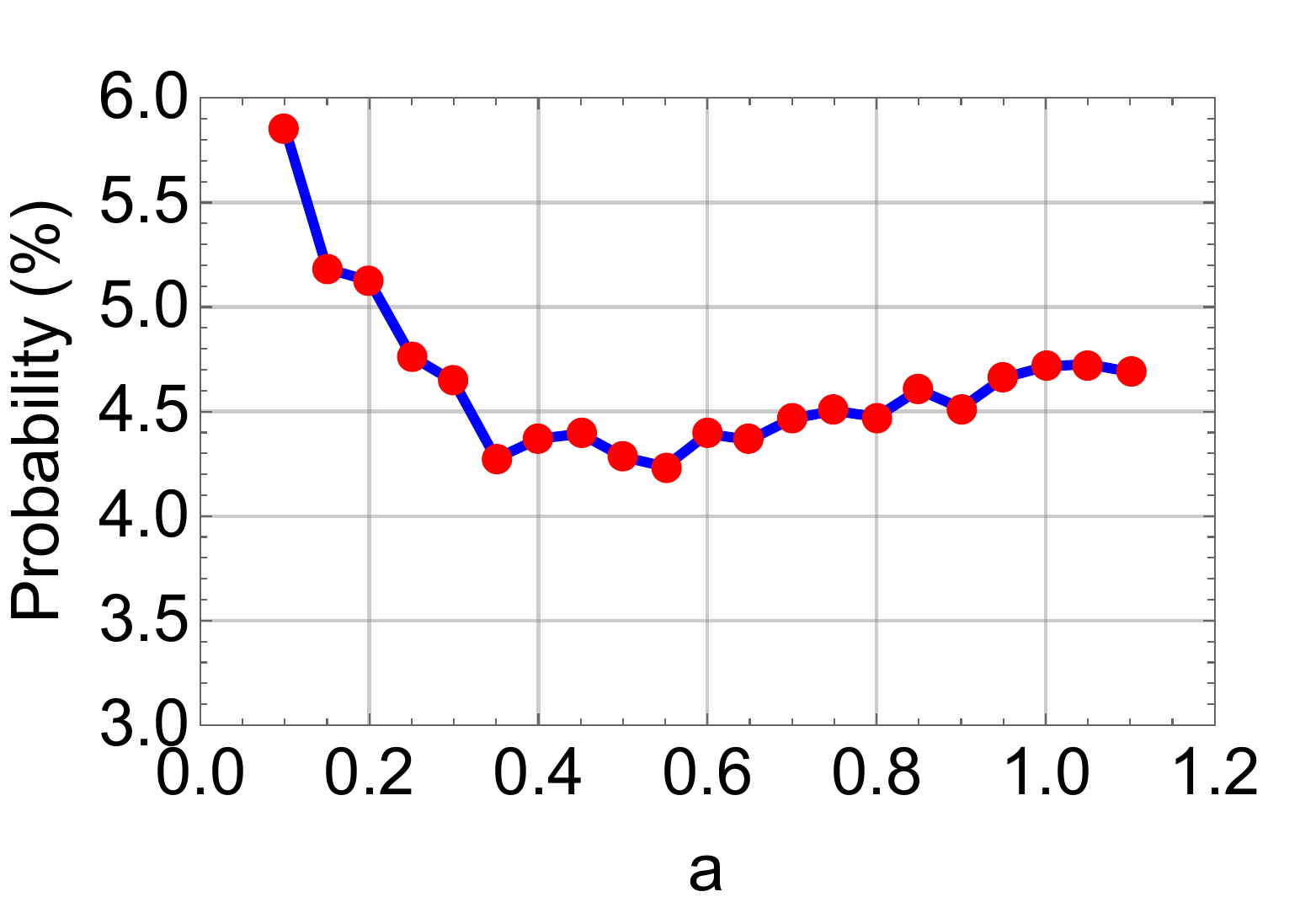}
\caption{Graph showing the probability of producing the state $\ket{\chi_a}$ in Eq.~\eqref{eq:WCBapprox} 
with $100\%$ fidelity. A three-mode circuit is used and the state is conditioned on detecting a photon number pattern $\bar{\vt{n}} = (1, 2)$. } 
\label{fig:PlotONstate}
\end{figure}

\begin{figure}
\includegraphics[width=\columnwidth]{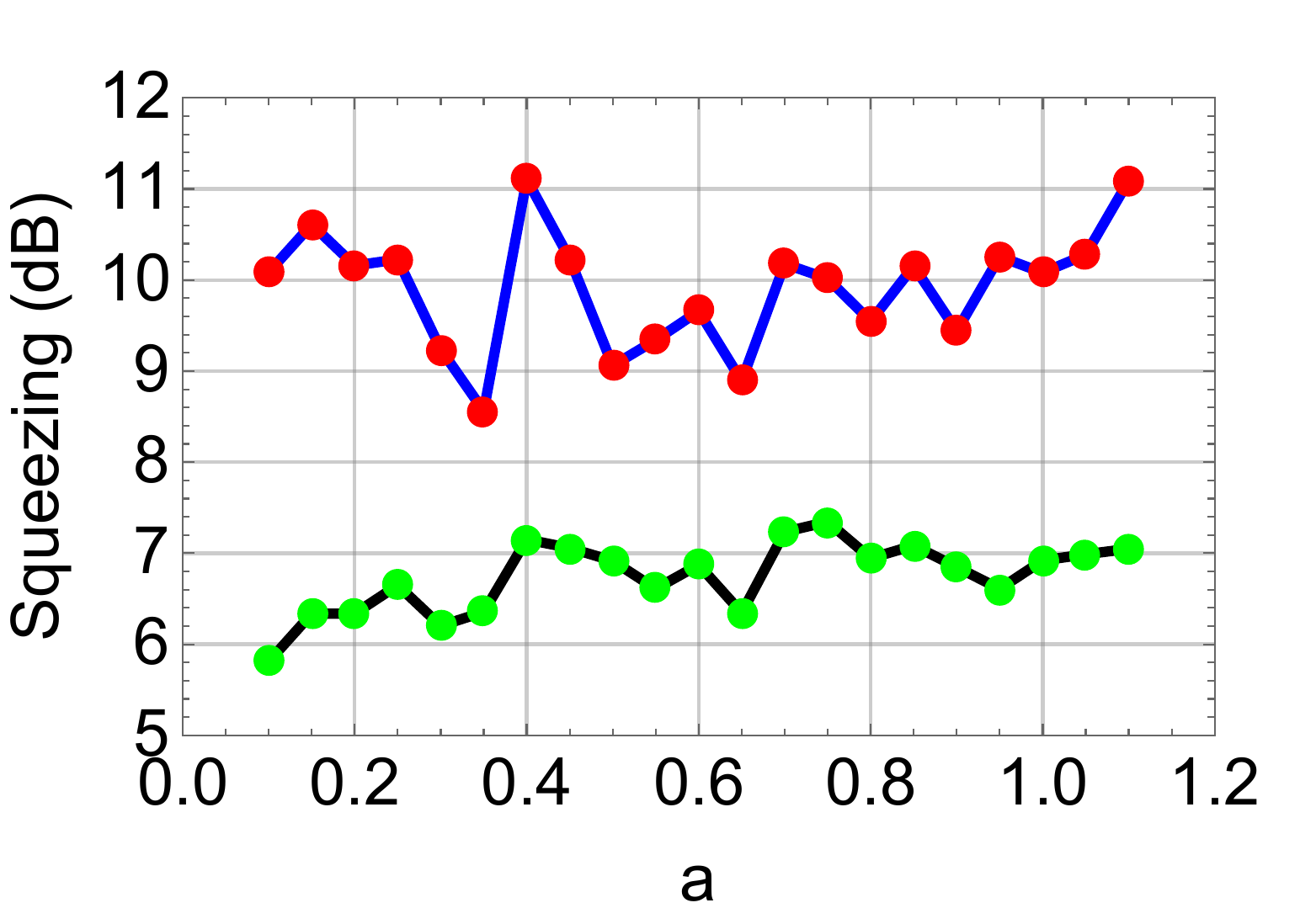}
\caption{The squeezing required to produce  the state $\ket{\chi_a}$. 
The top curve depicts the maximum squeezing that is required in any mode and the bottom curve provides the average squeezing per mode used to prepare the state. We observe that a higher success probability is achieved as compared to Ref. \cite{sabapathy2018near} at the cost of higher squeezing requirements.}
\label{fig:PlotONstate-sq}
\end{figure}

The cubic phase state is essential in CV quantum computation~\cite{PhysRevLett.82.1784}, e.g., it can be used as a resource state to implement a 
cubic phase gate through gate teleportation~\cite{PhysRevA.64.012310}. A recent proposal has also extended this notion to a two-mode gate that is non-Gaussian \cite{sefi2019}. 
A cubic phase state with a large phase parameter is usually difficult to generate, however, it 
can be generated by concatenating a sequence of weak cubic phase gates. Here, we focus on conditionally generating weak cubic phase states. 
In the weak coupling strength limit, the cubic phase states can be well approximated by superpositions of Fock states up to $\ket{3}$~\cite{ONstate}.
Specifically, we approximate the weak cubic phase state by~\cite{sabapathy2018near}
\begin{eqnarray}\label{eq:WCBapprox}
\ket{\chi_a} = \frac{1}{\sqrt{1+5|a|^2/2}} \bigg[ \ket{0} + i a \sqrt{\frac{3}{2}} \, \ket{1} + i a \ket{3} \bigg],
\end{eqnarray}
where $a \in \mathbb{R}$. 
A machine learning method was used to search a circuit and input states that can generate $\ket{\chi_a}$ with near perfect fidelity  and high probability~\cite{sabapathy2018near} ($1\% - 2\%$). We have shown that $\ket{\chi_a}$ can be generated with fidelity one by measuring two modes
of a three-mode Gaussian state, and use our formalism to optimize the success probability as well. 
As compared to Ref.~\cite{sabapathy2018near}, we obtained higher success probability of $4\% - 6\%$,
as shown in Fig.~\ref{fig:PlotONstate}. We also plot the maximum required squeezing and the average squeezing per mode in Fig.~\ref{fig:PlotONstate-sq}. 

\section{General formalism for multimode output states}\label{sec:MM-GeneralFormalism}

\begin{figure}
    \centering
    \scalebox{0.7}{\includegraphics{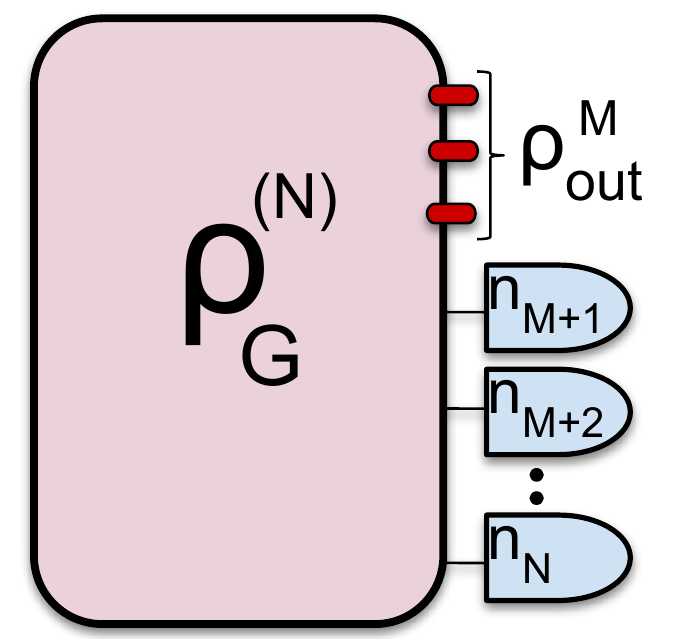}}
    \caption{Conditional generation of multimode non-Gaussian states. Here, we consider a general multimode Gaussian state $\rho^{(N)}_G$ of $N$-modes. $(N-M)$ modes are measured using PNRDs, giving values $n_k$ ($k = M+1, \cdots, N$) and resulting in a conditional output state $\rho_{\rm out}^M$ in the remaining $M$ modes. }
    \label{fig:ng-GBS-multimode-mixedM}
\end{figure}

We now derive a general formalism for generating multimode non-Gaussian states by detecting subsystems of multimode Gaussian states using PNRDs, 
as depicted in Fig.~\ref{fig:ng-GBS-multimode-mixedM}. It is a natural generalization of the formalism for generating single-mode non-Gaussian states. 
Most derivations carry over from the single-mode  case. The multimode formalism allows us to produce more complex non-Gaussian states, e.g., NOON states.

\subsection{Multimode output Wigner function}

Suppose we detect the last $(N-M)$ modes using PNRDs and obtain a photon number pattern $\bar{\vt{n}} = (n_{M+1}, n_{M+2}, \cdots, n_N)$, namely, the projected state in the 
detected modes is $| \bar{\vt{n}} \rangle = | n_{M+1}, n_{M+2}, \cdots, n_N \rangle$. By using Eqs.~\eqref{eq:DMcoherent} and \eqref{eq:DOcoherent-Matrix}
we find the unnormalized density matrix of the heralded modes to be
\begin{eqnarray}\label{eq:DM-M}
\tilde \rho_M 
&=& 
\frac{\mathcal{P}_0}{\pi^{2N}} \int \mathrm d^2 \vt{\alpha}_M \int \mathrm d^2 \vt{\beta}_M \int \mathrm d^2 \bar{\vt{\alpha}} \int \mathrm d^2 \bar{\vt{\beta}} ~
| \vt{\beta}_M \rangle \langle \vt{\alpha}_M | 
\nonumber\\
&&
\times \langle \bar{\vt{n}} | \bar{\vt{\beta}} \rangle \langle \bar{\vt{\alpha}} | \bar{\vt{n}} \rangle 
\exp\bigg(-\frac{| \tilde{\vt{\gamma}} |^2}{2} + \frac{1}{2} \tilde{\vt{\gamma}}^{\top} \tilde{\bf R} \tilde{\vt{\gamma}} + \tilde{\vt{\gamma}}^{\top} \tilde{\vt{y}} \bigg), \nonumber\\
\end{eqnarray}
where we have defined two $M$-component vectors $\vt{\alpha}_M = (\alpha_1, \cdots, \alpha_M)^{\top}$, $\vt{\beta}_M = (\beta_1, \cdots, \beta_M)^{\top}$,
and denoted the coherent states as
 $\ket{\vt{\alpha}_M} = \ket{ \alpha_1, \cdots, \alpha_M }$, $\ket{\vt{\beta}_M} = \ket{\beta_1, \cdots, \beta_M}$, 
$| \bar{\vt{\alpha}} \rangle = | \alpha_{M+1}, \cdots, \alpha_N \rangle$ and $| \bar{\vt{\beta}} \rangle = | \beta_{M+1}, \cdots, \beta_N \rangle$. 
By using the Fock state expansion of a coherent state from Eq.~\eqref{eq:CoherentState}, it is straightforward to find 
\begin{eqnarray}\label{eq:CoherentFock-Mmode}
\langle \bar{\vt{n}} | \bar{\vt{\beta}} \rangle \langle \bar{\vt{\alpha}} | \bar{\vt{n}} \rangle 
= \frac{1}{\bar{\vt{n}}!} \, e^{-(|\bar{\vt{\alpha}}|^2 + |\bar{\vt{\beta}}|^2 )/2} \prod_{k=M+1}^{N} \big(\alpha_k^* \beta_k \big)^{n_k}. %\nonumber\\
\end{eqnarray}

Similar to the single-mode output case, we define a $2N$-component vector $\vt{\gamma}$ by permuting the components of $\tilde{\vt{\gamma}}$ such that 
$\vt{\gamma} = (\vt{\beta}_M^*, \vt{\alpha}_M, \bar{\vt{\beta}}^*, \bar{\vt{\alpha}})^{\top}$. $\vt{\beta}_M^*$ and $\vt{\alpha}_M$ are collected to form a $2M$-component vector
$\vt{\gamma}_h =  (\vt{\beta}_M^*, \vt{\alpha}_M)^{\top}$, and $\bar{\vt{\beta}}^*$ and $ \bar{\vt{\alpha}}$ are collected to form a $2(N-M)$-component vector
$\vt{\gamma}_d = (\bar{\vt{\beta}}^*, \bar{\vt{\alpha}})^{\top}$. $\vt{\gamma}_h$ and $\vt{\gamma}_d$ correspond to the heralded and detected modes, respectively.
The vector $\vt{\gamma}$ and $\tilde{\vt{\gamma}}$ are related by a permutation matrix ${\bf P}$, namely, $\vt{\gamma} = {\bf P} \tilde{\vt{\gamma}}$. Correspondingly, 
we can define ${\bf R}$ and $\vt{y}$ as ${\bf R} = {\bf P} \tilde{\bf R} {\bf P}^{\top}$,  ${\vt{y}} = {\bf P}\tilde{\vt{y}}$. The matrix ${\bf R}$ can be partitioned as
\begin{eqnarray}
{\bf R} =
	\begin{pmatrix}
	{\bf R}_{hh} & {\bf R}_{hd} \\
	{\bf R}_{dh} & {\bf R}_{dd}
	\end{pmatrix},
\end{eqnarray}
where ${\bf R}_{hh}$ is now a $2M \times 2M$ symmetric matrix corresponding to the heralded modes, ${\bf R}_{dd}$ is a $(2N-2M) \times (2N-2M)$ symmetric matrix
corresponding to the detected modes and ${\bf R}_{hd}$ is a $2M \times (2N-2M)$ matrix that represents the connections between the heralded modes and 
detected modes. Since ${\bf R}$ is symmetric,  ${\bf R}_{dh} = {\bf R}_{hd}^{\top}$. Similarly, the vector $\vt{y}$ is partitioned into $(\vt{y}_h, \vt{y}_d)^{\top}$, 
where $\vt{y}_h$ has $2M$ components and corresponds to the heralded modes, and $\vt{y}_d$ has $2(N-M)$ components and corresponds to the detected modes. 
The three terms in the exponential in Eq.~\eqref{eq:DM-M} become
\begin{eqnarray}\label{eq:ModeSaperation-Mmode}
| \tilde{\vt{\gamma}} |^2 &=& %| \vt{\gamma} |^2 = 
| \vt{\gamma}_h |^2 + | \vt{\gamma}_d |^2, 
\nonumber\\
\tilde{\vt{\gamma}}^{\top} \tilde{\vt{y}} &=& %{\vt{\gamma}}^{\top} {\vt{y}} = 
\vt{\gamma}_h^{\top} \vt{y}_h + \vt{\gamma}_d^{\top} \vt{y}_d 
\\
\tilde{\vt{\gamma}}^{\top} \tilde{\bf R} \tilde{\vt{\gamma}} &=& %\vt{\gamma}^{\top} {\bf R} \vt{\gamma} =
	\vt{\gamma}_h^{\top} {\bf R}_{hh} \vt{\gamma}_h + \vt{\gamma}_d^{\top} {\bf R}_{dd} \vt{\gamma}_d 
	+ 2 \, \vt{\gamma}_h^{\top} {\bf R}_{hd} \vt{\gamma}_d.\nonumber 
\end{eqnarray}
Substituting Eqs.~\eqref{eq:CoherentFock-Mmode} and \eqref{eq:ModeSaperation-Mmode} into Eq.~\eqref{eq:DM-M}, 
we find that the unnormalized density matrix can be written as
\begin{eqnarray}\label{eq:DM-Mmode-2}
\tilde \rho_M = \frac{1}{\pi^{2M}} \int \mathrm d^2 \vt{\alpha}_M \int \mathrm d^2 \vt{\beta}_M ~ | \vt{\beta}_M \rangle \langle \vt{\alpha}_M | F(\vt{\alpha}_M, \vt{\beta}_M), \nonumber\\
\end{eqnarray}
where
\begin{align}\label{eq:Ffunction-Mmode-1}
&F(\vt{\alpha}_M, \vt{\beta}_M) =
\frac{\mathcal{P}_0}{\pi^{2(N-M)}\bar{\vt{n}}!}
\, \exp(L_2) \nonumber\\
&
~~~~~~~~~~\times \int \mathrm d^2 \bar{\vt{\alpha}} \int \mathrm d^2 \bar{\vt{\beta}} ~
\prod_{k=M+1}^{N} \big(\alpha_k^* \beta_k \big)^{n_k}
\exp(L_3)
\nonumber\\
&=
\frac{\mathcal{P}_0}{\bar{\vt{n}}!}
\, \exp(L_2) 
\prod_{k=M+1}^{N} \bigg(\frac{\partial^2}{\partial \alpha_k \partial \beta_k^*} \bigg)^{n_k}
\exp(L_3)\bigg|_{\vt{\gamma}_d = {\bf 0}},
\end{align}
where the expressions for $L_2$ and $L_3$ are in the same forms as the ones given in Eq.~\eqref{eq:Ffunction}. 
In the second equality of Eq.~\eqref{eq:Ffunction-Mmode-1}, %we have changed the integration over $\bar{\vt{\alpha}}$ and $\bar{\vt{\beta}}$ into partial derivative 
we have performed integration by parts over $\bar{\vt{\alpha}}$ and $\bar{\vt{\beta}}$, the detail of which 
is given by Eq.~\eqref{eq:IntegralDerivative} in Appendix \ref{appedix:IntegralDerivative}. 

From the unnormalized density matrix $\tilde \rho_M$ one can calculate the unnormalized characteristic function $ \chi(\vt{\beta};\tilde \rho_M)$
and the unnormalized Wigner function $ W(\vt{\alpha};\tilde \rho_M)$. By substituting  $\tilde \rho_M$ into Eq.~\eqref{eq:CharacteristicF-single} we have
\begin{align}\label{eq:CharacteristicF-1-M}
&\chi(\vt{\beta};\tilde \rho_M) = e^{-|\vt{\beta}|^2/2} \text{Tr}\big( e^{ -\vt{\beta}^{*\top} \hat{\vt a}} \tilde \rho_M e^{ \vt{\beta} \hat{\vt a}^{\dag}}  \big)
\nonumber\\
&=
\frac{1}{\pi^{2M}}   \int \mathrm d^2 \vt{\alpha}_M \int \mathrm d^2 \vt{\beta}_M ~ 
e^{-|\vt{\beta}|^2/2} e^{ -\vt{\beta}^{*\top} \vt{\beta}_M + \vt{\beta}^{\top} \vt{\beta}_M^*} 
\nonumber\\
&
\times \langle \vt{\alpha}_M  | \vt{\beta}_M \rangle F(\vt{\alpha}_M, \vt{\beta}_M),
\end{align}
where we have used the fact that the coherent state is the eigenstate of the annihilation operator, $\hat a \ket{\alpha} = \alpha \ket{\alpha}$.
Substituting  $ \chi(\vt{\beta};\tilde \rho_M)$ into Eq.~\eqref{eq:WignerCoherentBasis} we find the unnormalized Wigner function as
\begin{align}\label{eq:Wigner-Mmode-2}
&W(\vt{\alpha};\tilde \rho_M) = \frac{1}{\pi^{4M}}   \int \mathrm d^2 \vt{\alpha}_M \int \mathrm d^2 \vt{\beta}_M ~ 
\langle \vt{\alpha}_M  | \vt{\beta}_M \rangle 
\nonumber\\
&
\times 
F(\vt{\alpha}_M, \vt{\beta}_M) \int \mathrm d^2 \vt{\beta} ~ e^{-|\vt{\beta}|^2/2} e^{ i \vt{\beta}^{*\top} (\vt{\beta}_M-\vt{\alpha}) + i \vt{\beta}^{\top} (\vt{\alpha}_M^*-\vt{\alpha}^*)} \nonumber\\
&=
\frac{2^M}{\pi^{3M}}  e^{-2|\vt{\alpha}|^2} \int \mathrm d^2 \vt{\alpha}_M \int \mathrm d^2 \vt{\beta}_M ~ F(\vt{\alpha}_M, \vt{\beta}_M) \,
\nonumber\\
&
\times \, e^{-|\vt{\alpha}_M|^2/2 - |\vt{\beta}_M|^2/2 - \vt{\alpha}_M^{*\top} \vt{\beta}_M + 2(\vt{\alpha}^{\top} \vt{\alpha}_M^*+\vt{\alpha}^{*\top} \vt{\beta}_M)},
\end{align}
where in the last equality we have performed the integration over ${\vt{\beta}}$ and used the relation 
$\langle \vt{\alpha}_M  | \vt{\beta}_M \rangle = e^{-|\vt{\alpha}_M|^2/2 - | \vt{\beta}_M|^2/2 + \vt{\alpha}_M^{*\top} \vt{\beta}_M}$. 
By substituting the function $F(\vt{\alpha}_M, \vt{\beta}_M)$ of Eq.~\eqref{eq:Ffunction-Mmode-1} into Eq.~\eqref{eq:Wigner-Mmode-2}, interchanging the order of partial derivatives
and integration, and then performing the integration over $\vt{\alpha}_M$ and $\vt{\beta}_M$ (which is a Gaussian integration), 
we arrive at the final expression for the unnormalized Wigner function (see Appendix \ref{app:Wigner-Function-Mmode} for more details) given by
\begin{align}\label{eq:Wigner-Mmode-general}
& W(\vt{\alpha};\tilde \rho_M) = \frac{2^M \mathcal{ P}_0}{\pi^{M} \, \bar{\vt{n}}!} \exp (- \vt{v}^{\dag} {\bf L}_6 \vt{v} ) \nonumber\\ 
&~~~~~~~~\times  \frac{\exp \bigg\{ \frac{1}{2} \vt{y}_h^{\top} ({\bf I}_2 - {\bf X}_{2M} {\bf R}_{hh} )^{-1} {\bf X}_{2M} \vt{y}_h \bigg\}}{\sqrt{\text{det} ({\bf I}_{2M} + {\bf X}_{2M} {\bf R}_{hh} )}}
 \nonumber\\
& \times \prod_{k=M+1}^{N} \bigg(\frac{\partial^2}{\partial \alpha_k \partial \beta_k^*} \bigg)^{n_k}
\exp\bigg( \frac{1}{2} \vt{\gamma}_d^{\top} {\bf A} \vt{\gamma}_d + \vt{z}^{\top} \vt{\gamma}_d \bigg) \bigg|_{\vt{\gamma}_d = {\bf 0}},\nonumber\\
&{\bf L}_6 = ({\bf I}_{2M} + {\bf X}_{2M} {\bf R}_{hh})^{-1} ({\bf I}_{2M} - {\bf X}_{2M} {\bf R}_{hh}),
\end{align}
where
\begin{eqnarray}\label{eq:Az-Mmode-general}
\vt{v} &=& (\vt{\alpha}^*, \vt{\alpha})^{\top}
- ({\bf I}_{2M} - {\bf X}_{2M} {\bf R}_{hh})^{-1} {\bf X}_{2M} \vt{y}_h, \nonumber\\
{\bf A} &=& {\bf R}_{dd} - {\bf R}_{dh} ({\bf I}_{2M} + {\bf X}_{2M} {\bf R}_{hh} )^{-1} {\bf X}_{2M} {\bf R}_{hd}, \nonumber\\
\vt{z} &=& \vt{Y} + 2\, {\bf R}_{dh} ({\bf I}_{2M} + {\bf X}_{2M} {\bf R}_{hh})^{-1} \vt{v}, \nonumber\\
\vt{Y} &=& \vt{y}_d + {\bf R}_{dh} ({\bf I}_{2M} - {\bf X}_{2M} {\bf R}_{hh})^{-1} {\bf X}_{2M} \vt{y}_h. 
\end{eqnarray}

Similar to the single-mode output case, the unnormalized Wigner function in Eq.~\eqref{eq:Wigner-Mmode-general} is also factorized into two parts: the first part is  a 
Gaussian function of $\vt{v}$; the second part involving the partial derivatives  is  a polynomial in $\vt{v}$. The maximal order of the 
polynomial depends on the detected photon number $\{n_k\}$. If $n_k = 0$ for all $k$, namely, all PNRDs register no photons, then the polynomial  is  a constant. 
The output state is then a Gaussian state in the first $M$ modes. 
By comparing Eq.~\eqref{eq:Wigner-Mmode-general} with Eq.~\eqref{eq:WignerCoherent}, 
we can identify the displacement of the heralded Gaussian state as 
\begin{eqnarray}\label{eq:DisNoPhoton-Mmode}
\vt{d} = ({\bf I}_{2M} - {\bf X}_{2M} {\bf R}_{hh})^{-1} {\bf X}_{2M} \vt{y}_h
\end{eqnarray}
and the covariance matrix as 
\begin{eqnarray}\label{eq:CMnoPhoton-Mmode}
{\bf V}_M^{(c)} (\bar{\vt{n}} = {\bf 0}) =  \frac{1}{2} ({\bf I}_{2M} + {\bf X}_{2M} {\bf R}_{hh}) ({\bf I}_{2M} - {\bf X}_{2M} {\bf R}_{hh})^{-1}. \nonumber\\
\end{eqnarray}
To generate a non-Gaussian state, the polynomial should be nontrivial. Two conditions need to be satisfied to guarantee
a non-Gaussian state at the output\,: (1) the PNRDs must register photons; (2) the matrix ${\bf R}_{hd} \ne {\bf 0}$, which means the heralded modes must have 
some connections with the detected modes when viewed through the ${\bf R}$ matrix.

\subsection{Measurement probability}
 
The measurement probability $P(\bar{\vt{n}})$ can be obtained by tracing the unnormalized density operator \eqref{eq:DM-Mmode-2}, 
which corresponds to integrating the arguments ($\vt{\alpha}$) of the unnormalized Wigner function in Eq.~\eqref{eq:Wigner-Mmode-general}, and we get
\begin{eqnarray}
P(\bar{\vt{n}}) = \text{Tr}(\tilde \rho_M) = \int \mathrm d^2 \vt{\alpha} \,  W(\vt{\alpha};\tilde \rho_M). 
\end{eqnarray}
It is evident from Eq.~\eqref{eq:Wigner-Mmode-general} that the integration over $\vt{\alpha}$ is a Gaussian integration and can be performed in a direct manner.
Using the relation 
\begin{eqnarray}
\int \mathrm d^2 \vt{\alpha} \,\exp (-\vt{v}^{\dag} {\bf L}_6 \vt{v} )
= \frac{\pi^M }{2^M \, \sqrt{\text{det}\,{\bf L}_6}} \nonumber
\end{eqnarray}
with ${\bf L}_6$ a $2M \times 2M$ symmetric matrix,
we obtain the measurement probability
\begin{align}\label{eq:ProbabilityMmode}
&P(\bar{\vt{n}}) = \frac{\mathcal{ P}_0}{\bar{\vt{n}}!}
\frac{1}{\sqrt{\text{det} ({\bf I}_{2M} - {\bf X}_{2M} {\bf R}_{hh} )}}
\nonumber\\
&
\times \, \exp \bigg\{ \frac{1}{2} \vt{y}_h^{\top} ({\bf I}_{2M} - {\bf X}_{2M} {\bf R}_{hh} )^{-1} {\bf X}_{2M} \vt{y}_h \bigg\} 
\nonumber\\
&
\times 
\prod_{k=M+1}^{N} \bigg(\frac{\partial^2}{\partial \alpha_k \partial \beta_k^*} \bigg)^{n_k}
\exp\bigg( \frac{1}{2} \vt{\gamma}_d^{\top} {\bf A}_p \vt{\gamma}_d + \vt{z}_p^{\top} \vt{\gamma}_d \bigg) \bigg|_{\vt{\gamma}_d = {\bf 0}}, \nonumber\\
\end{align}
where
\begin{eqnarray}\label{eq:ProbabilityMAMmode}
{\bf A}_p &=& {\bf R}_{dd} + {\bf R}_{dh} ({\bf I}_{2M} - {\bf X}_{2M} {\bf R}_{hh} )^{-1} {\bf X}_{2M} {\bf R}_{hd}, \nonumber\\
\vt{z}_p &=& \vt{y}_d + {\bf R}_{dh} ({\bf I}_{2M} - {\bf X}_{2M} {\bf R}_{hh})^{-1} {\bf X}_{2M} \vt{y}_h.
\end{eqnarray}

\section{Multimode output states by measuring pure Gaussian states}\label{sec:MM-PureFormalism}

\begin{figure}
    \centering
    \includegraphics[width=0.95 \columnwidth]{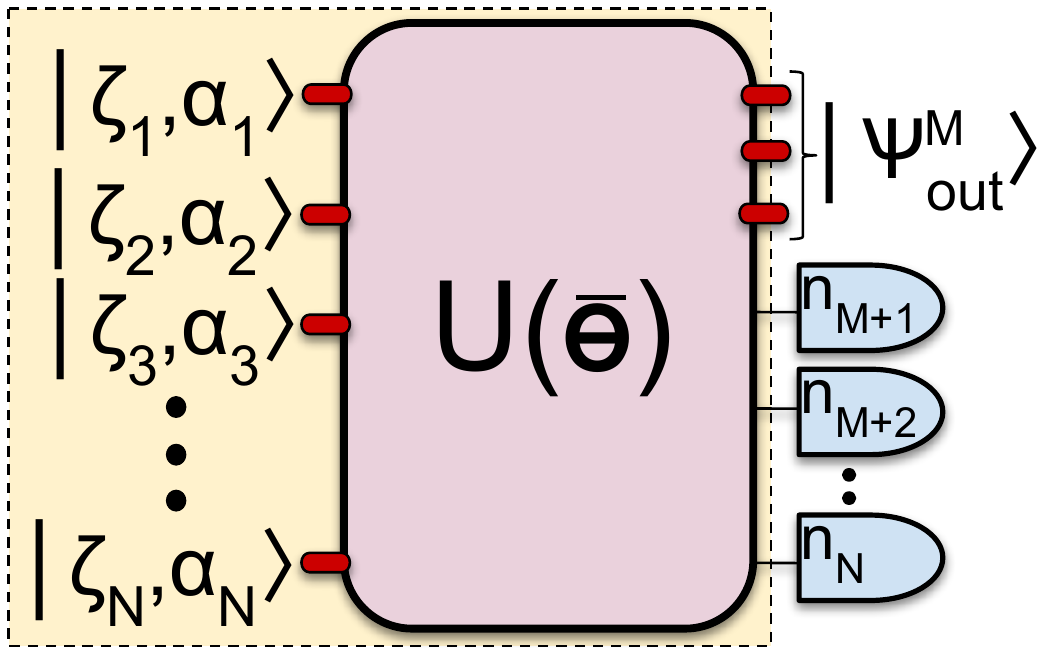}
    \caption{Conditional generation of multimode non-Gaussian states. Here, we consider a general pure multimode Gaussian state that can be decomposed into squeezed displaced states, $\ket{\zeta,\alpha} = \hat{S}(\zeta) \hat{D}(\mathsf{\alpha}) \ket{0}$, followed by an interferometer $\textsf{U}(\bar{\theta})$. The last $(N-M)$ modes are measured using PNRDs,
    giving values $n_k$ ($k=M+1, \cdots, N$) and resulting in a conditional output state $\ket{\psi_{\rm out}^M}$ in the remaining modes. }
    \label{fig:ng-GBS-M-mode-pure}
\end{figure}

We consider the case when $(N-M)$ modes of an $N$-mode pure Gaussian state are measured using 
PNRDs, as depicted in Fig.~\ref{fig:ng-GBS-M-mode-pure}. The heralded non-Gaussian state is a superposition of a finite number of Fock states, 
acted on by a multimode Gaussian unitary.

\subsection{Output Wigner function}

For multimode pure Gaussian states, the matrix $\tilde {\bf R}$ can be written in a block diagonal form: $\tilde {\bf R} = {\bf B} \oplus {\bf B}^*$. It is convenient to partition the 
matrix ${\bf B}$ as
\begin{eqnarray}
{\bf B} =
	\begin{pmatrix}
	{\bf B}_{hh} & {\bf B}_{hd} \\
	{\bf B}_{dh} & {\bf B}_{dd}
	\end{pmatrix},
\end{eqnarray}
where ${\bf B}_{hh}$ is an $M \times M$ symmetric matrix corresponding to the heralded modes, ${\bf B}_{dd}$ is an $(N-M) \times (N-M)$ symmetric matrix
corresponding to the detected modes, and ${\bf B}_{hd}$ is an $M \times (N-M)$ matrix that represents the connections between the detected modes and 
heralded mode. Then the matrices ${\bf R}_{hh}$, ${\bf R}_{hd}$ and ${\bf R}_{dd}$ can be written as: ${\bf R}_{hh} = {\bf B}_{hh} \oplus {\bf B}_{hh}^*$, 
${\bf R}_{hd} = {\bf B}_{hd} \oplus {\bf B}_{hd}^*$ and ${\bf R}_{dd} = {\bf B}_{dd} \oplus {\bf B}_{dd}^*$. 

If all PNRDs detect no photons, it is evident from Eq.~\eqref{eq:Wigner-Mmode-general} that the Wigner function is a Gaussian 
function, so the output state is an $M$-mode Gaussian state. The covariance matrix of the heralded Gaussian state is
\begin{align}\label{eq:CM0photon-Mmode}
&{\bf V}^{(c)}_M(\bar{\vt n}= {\bf 0}) %&=& \frac{1}{2} ({\bf I}_{2M} - {\bf X}_{2M} {\bf R}_{hh})^{-1} ({\bf I}_{2M} + {\bf X}_{2M} {\bf R}_{hh}) \nonumber\\
= \frac{1}{2} \begin{pmatrix}
    {\bf V}_{11} & {\bf V}_{12} \\ {\bf V}_{12}^{*} & {\bf V}_{22}
\end{pmatrix}, \nonumber\\
{\bf V}_{11} &= ({\bf I}_M - {\bf B}_{hh}^*{\bf B}_{hh})^{-1} ({\bf I}_M + {\bf B}_{hh}^*{\bf B}_{hh}),  \nonumber\\
{\bf V}_{12} &= 2 \, ({\bf I}_M - {\bf B}_{hh}^*{\bf B}_{hh})^{-1} {\bf B}_{hh}^*, \nonumber\\
{\bf V}_{22} &= ({\bf I}_M - {\bf B}_{hh} {\bf B}_{hh}^* )^{-1} ({\bf I}_M + {\bf B}_{hh} {\bf B}_{hh}^*).
\end{align}
It can be shown that the determinant of the covariance matrix ${\bf V}^{(c)}_M(\bar{\vt n}= {\bf 0})$ is one, indicating the output state is pure. 
Note that the matrix $({\bf I}_M - {\bf B}_{hh}^*{\bf B}_{hh})$ is Hermitian and we further require that it is positive definite to correspond to a valid quantum state.
If we define a Hermitian matrix ${\bf T}_{2M}$ as
\begin{eqnarray}
{\bf T}_{2M} = 
	\begin{pmatrix}
	\sqrt{{\bf I}_M - {\bf B}_{hh}^*{\bf B}_{hh}} & {\bf 0} \\
	{\bf 0} & \sqrt{{\bf I}_M - {\bf B}_{hh}{\bf B}_{hh}^*}
	\end{pmatrix}
\end{eqnarray}
and a vector $\vt{w} = (\vt{\delta}^*, \vt{\delta})^{\top}$ as
\begin{eqnarray}\label{eq:VW-Mmode}
\vt{w} = {\bf T}_{2M} ({\bf I}_{2M} + {\bf X}_{2M} {\bf R}_{hh})^{-1} \vt{v}, %~~~~~~ \text{or} ~~~~~~
%\vt{v} = ({\bf I}_{2M} + {\bf X}_{2M} {\bf R}_{hh}) \, {\bf T}_{2M}^{-1}  \, \vt{w}
\end{eqnarray}
%so that
%\begin{eqnarray}
%- \vt{v}^{\top} {\bf X}_{2M}({\bf I}_{2M} + {\bf X}_{2M} {\bf R}_{hh})^{-1} ({\bf I}_{2M} - {\bf X}_{2M} {\bf R}_{hh}) \vt{v} 
%\nonumber\\
%&=&
%- \vt{w}^{\top} \, {\bf T}_{2M}^{-\top} ({\bf I}_{2M} + {\bf R}_{hh} {\bf X}_{2M} ) {\bf X}_{2M}({\bf I}_{2M} + {\bf X}_{2M} {\bf R}_{hh})^{-1} ({\bf I}_{2M} - {\bf X}_{2M} {\bf R}_{hh}) ({\bf I}_{2M} + {\bf X}_{2M} {\bf R}_{hh}) \, {\bf T}_{2M}^{-1} \vt{w}
%\nonumber\\
%&=&
%- \vt{w}^{\top} {\bf X}_{2M} \, {\bf T}_{2M}^{-1} ({\bf I}_{2M} - {\bf X}_{2M} {\bf R}_{hh}) ({\bf I}_{2M} + {\bf X}_{2M} {\bf R}_{hh}) \, {\bf T}_{2M}^{-1} \vt{w}
%\nonumber\\
%&=&
%- \vt{w}^{\top} {\bf X}_{2M} \vt{w}. 
%\end{eqnarray}
then the Wigner function becomes
\begin{align}\label{eq:Wigner-Mmode-general-pure}
W(\vt{\alpha}; \rho_M) 
&\propto 
e^{- \vt{w}^{\dag} \vt{w}}
\prod_{k=M+1}^{N} \bigg(\frac{\partial^2}{\partial \alpha_k \partial \beta_k^*} \bigg)^{n_k} \nonumber\\ 
&\times \exp\bigg( \frac{1}{2} \vt{\gamma}_d^{\top} {\bf A} \vt{\gamma}_d + \vt{z}^{\top} \vt{\gamma}_d \bigg) \bigg|_{\vt{\gamma}_d = {\bf 0}},
\end{align}
where $\vt{z} = \vt{Y} + 2\, {\bf R}_{dh} {\bf T}^{-1}_{2M} \vt{w}$ and ${\bf A}$ is given by Eq.~\eqref{eq:Az-Mmode-general}. 

The transformation in Eq.~\eqref{eq:VW-Mmode} is  a symplectic transformation. To see that we define a matrix ${\bf S}_{2M}$ as
$\vt{v} = {\bf S}_{2M} \vt{w}$, and write it in terms of the matrix ${\bf B}_{hh}$ as 
\begin{align}\label{eq:Symplectic}
&{\bf S}_{2M} = ({\bf I}_{2M} + {\bf X}_{2M} {\bf R}_{hh}) {\bf T}_{2M}^{-1}
\nonumber\\
&=
	\begin{pmatrix}
	({\bf I}_{M} - {\bf B}_{hh}^* {\bf B}_{hh} )^{-1/2} & {\bf B}_{hh}^* ({\bf I}_{M} - {\bf B}_{hh} {\bf B}_{hh}^* )^{-1/2} \\
	{\bf B}_{hh} ({\bf I}_{M} - {\bf B}_{hh}^* {\bf B}_{hh} )^{-1/2} & ({\bf I}_{M} - {\bf B}_{hh} {\bf B}_{hh}^* )^{-1/2}
	\end{pmatrix}. 
\end{align}
According to the Autonne-Takagi factorization (see Corollary 4.4.4 in \cite{horn1990matrix}), the complex symmetric matrix ${\bf B}_{hh}$ can be decomposed as
${\bf B}_{hh} = {\bf K} {\bf \Lambda} {\bf K}^{\top}$, where ${\bf K}$ is a unitary matrix and ${\bf \Lambda}$ is a complex diagonal matrix defined as
${\bf \Lambda} = \text{diag}(\lambda_1, \lambda_2, \cdots, \lambda_M)$. By substituting the decomposition of ${\bf B}_{hh}$ into Eq.~\eqref{eq:Symplectic}, 
we find ${\bf S}_{2M} = {\bf K}_{2M} {\bf S}_{\text{sq}} {\bf K}_{2M}^{\dag}$, where
\begin{align}
&{\bf S}_{\text{sq}} =
	\begin{pmatrix}
	({\bf I}_{M} - {\bf \Lambda}^* {\bf \Lambda} )^{-1/2} & {\bf \Lambda}^* ({\bf I}_{M} - {\bf \Lambda} {\bf \Lambda}^* )^{-1/2} \\
	{\bf \Lambda} ({\bf I}_{M} - {\bf \Lambda}^* {\bf \Lambda} )^{-1/2} & ({\bf I}_{M} - {\bf \Lambda} {\bf \Lambda}^* )^{-1/2}
	\end{pmatrix},
\nonumber\\
&{\bf K}_{2M} =
	\begin{pmatrix}
	{\bf K}^* & {\bf 0} \\
	{\bf 0} & {\bf K}
	\end{pmatrix}.
\end{align}
It is evident that ${\bf K}_{2M}$ represents a transformation of a linear interferometer and ${\bf S}_{\text{sq}}$ represents $M$ independent single-mode squeezing
transformations. The matrix ${\bf S}_{\text{sq}}$ transforms the annihilation operators $\{ \hat{a}_k\}$ as
\begin{eqnarray}
\hat{a}_k \rightarrow \frac{1}{\sqrt{1 - |\lambda_k|^2}} \hat{a}_k + \frac{\lambda_k}{\sqrt{1 - |\lambda_k|^2}} \hat{a}_k^{\dag}. 
\end{eqnarray}
Therefore, the squeezing amplitude of the $k$-th mode is $\zeta_k = r_k e^{i \varphi_k}$, with $r_k = \tanh^{-1}(|\lambda_k|)$ and $\varphi_k = \text{Arg}(\lambda_k)+\pi$. Collecting the above facts together, we have that the multimode output state can be written in the form 
\begin{align}\label{eq:mmode-output}
    \ket{\psi} = \hat{U}_{{\bf M}} \sum_{\vt{\ell}= {\bf 0}}^{\infty} c_{\vt{\ell}} \ket{\vt{\ell}},  
\end{align}
$\hat{U}_{{\bf M}}$ is an $M$-mode Gaussian gate and $\{c_{\vt{\ell}} \}$ are Fock basis coefficients with $\vt{\ell} = (\ell_1, \ell_2,\cdots \ell_M)^{\top}$ are the Fock basis elements of the $M$-mode system.

\subsection{Coefficients $\{c_{\boldsymbol{\ell}}\}$ in the Fock state superposition}

The coefficients of the superposition of Fock states remain to be determined. 
Let us suppose that the position space wave function of an $M$-mode quantum state $\ket{\psi}$ is $\psi(\vt{q})$, where $\vt{q} = (q_1, q_2, \cdots, q_M)^{\top}$ is a real vector
with $M$ components. The wave function $\psi(\vt{q})$ can be expanded in the Fock basis as 
\begin{eqnarray}
\psi(\vt{q}) = \sum_{\ell_1=0}^{\infty}  \sum_{\ell_2=0}^{\infty} \cdots \sum_{\ell_M=0}^{\infty} c_{\vt{\ell}} \psi_{\vt{\ell}} (\vt{q}),
\end{eqnarray}
where $\vt{\ell} = (\ell_1, \ell_2, \cdots, \ell_M)^{\top}$, 
$c_{\vt{\ell}}$ is the coefficient, and $\psi_{\vt{\ell}} (\vt{q})$ is the wave function of the Fock state $\ket{\vt{\ell}}$ given by
\begin{eqnarray}\label{eq:FockWF-Mmode}
\psi_{\vt{\ell}} (\vt{q}) = \frac{1}{\pi^{M/4}} \prod_{k=1}^M \frac{1}{\sqrt{2^{\ell_k} \ell_k !}} e^{-q_k^2/2} H_{\ell_k} (q_k),
\end{eqnarray}
with $H_{\ell_k} (q_k)$ the corresponding Hermite polynomial. From Eq.~\eqref{eq:WignerXPgeneral}, the $M$-mode Wigner function is 
\begin{eqnarray}\label{eq:WignerFockSuper-Mmode}
\xoverline{W}(\vt{p}, \vt{q}) &=& \frac{1}{\pi^M} \int \mathrm d \vt y \, e^{- 2i \vt{p}^{\top} \vt{y}} \langle \vt q - \vt y \ket{\psi} \langle \psi | \vt q + \vt y \rangle
\nonumber\\
&=&
\frac{1}{\pi^M} \sum_{{\vt{\ell}} = \vt{0}}^{\vt{\infty}}  \sum_{{\vt{m}} = \vt{0}}^{\vt{\infty}} c_{\vt{\ell}} c_{\vt{m}}^* \, W_{\vt{\ell} \vt{m}}(\vt{p}, \vt{q}),
\end{eqnarray}
where we have introduced the notation $\sum_{{\vt{\ell}} = \vt{0}}^{\vt{\infty}} = \sum_{\ell_1=0}^{\infty} \cdots \sum_{\ell_M=0}^{\infty}$ to simplify the expression and 
$W_{\vt{\ell} \vt{m}}(\vt{p}, \vt{q})$ is defined as 
\begin{eqnarray}
W_{\vt{\ell} \vt{m}} (\vt{p}, \vt{q}) &=& \int \mathrm d \vt y \, e^{- 2i \vt{p}^{\top} \vt{y}} \langle \vt q - \vt y \ket{\psi_{\vt{\ell}}} \langle \psi_{\vt{m}} | \vt q + \vt y \rangle
\nonumber\\
&=&
e^{-\vt{q}^{\top} \vt{q} - \vt{p}^{\top} \vt{p}} \prod_{k=1}^M \frac{1}{\sqrt{\ell_k! \, m_k!}}  H_{\ell_k m_k}(2 \alpha_k, 2 \alpha_k^*).  \nonumber
\end{eqnarray}
By using the orthogonality relation of Ito's 2D-Hermite polynomials Eq.~\eqref{eq:HermiteOrthogonality-alpha}, we can write down the coefficient $c_{\vt{\ell}}$ as an 
overlap integral of the Wigner function and Ito's 2D-Hermite polynomials,
\begin{eqnarray}\label{eq:CoeffWignerHermite-Mmode-1}
c_{\vt{\ell}} \, c_{\vt{m}}^* =  \int \frac{\mathrm d^2 \vt{\alpha}}{\sqrt{\vt{\ell}\,! \, \vt{m}!}} \, W(\vt{\alpha}) e^{-2|\vt{\alpha}|^2} \prod_{k=1}^M H_{\ell_k m_k}(2 \alpha_k, 2 \alpha_k^*), \nonumber\\
\end{eqnarray}
where $\vt{\ell} \,! = \ell_1! \, \ell_2! \cdots \ell_N!$, $\vt{m} \,! = m_1! \, m_2! \cdots m_N!$, and we have used the convention that $W(\vt p, \vt q) = 2^M \xoverline{W}(\vt p, \vt q)$. 

If the quantum state $\ket{\psi}$ is acted upon by a Gaussian unitary, according to the transformation rule of the Wigner function and from Eq.~\eqref{eq:WignerFockSuper-Mmode} we see that
the coefficients $\{ c_n\}$ are unchanged while the arguments of the Wigner function are changed. This change can be taken into account by replacing $\vt{\alpha}$ by $\vt{\delta}$,
where $\vt{\delta}$ contains the information of the Gaussian unitary. Now by substituting the Wigner function of Eq.~\eqref{eq:Wigner-Mmode-general-pure} 
into Eq.~\eqref{eq:CoeffWignerHermite-Mmode-1} and performing the integration over $\vt{\delta}$, 
we find (see Appendix \ref{app:Fock-Coefficients-Mmode} for more details) that the coefficients $\{c_{\boldsymbol \ell}\}$ satisfy
\begin{align}\label{eq:CoeffWignerHermite-Mmode-2}
&c_{\vt{\ell}} \, c_{\vt{m}}^* = \frac{1}{\sqrt{\vt{\ell}\,! \, \vt{m}!}} \int \mathrm d^2 \vt{\delta} \, W(\vt{\alpha}) e^{-2|\vt{\delta}|^2} 
\prod_{k=1}^M H_{\ell_k m_k}(2 \delta_k, 2 \delta_k^*) \nonumber\\
&=
%\bigg( \frac{\pi}{4}\bigg)^M 
\frac{\mathcal{N} \pi^M}{ 4^M \sqrt{\vt{\ell}\,! \vt{m}!}} \, 
\prod_{k=1}^M \bigg( \frac{\partial^{\ell_k}}{\partial t_k^{\ell_k}}  \frac{\partial^{m_k}}{\partial s_k^{m_k}} \bigg)
\prod_{k=M+1}^{N} \bigg(\frac{\partial^2}{\partial \alpha_k \partial \beta_k^*} \bigg)^{n_k}
\nonumber\\
%\bigg\{
& \times \exp\bigg[ \frac{1}{2} (\vt{u}^{\top}, \vt{\gamma}_d^{\top}) \, {\bf M} \begin{pmatrix} \vt{u} \\ \vt{\gamma}_d \end{pmatrix} + \vt{Y}^{\top} \vt{\gamma}_d \bigg] 
%\bigg\}
\bigg|_{\vt{\gamma}_d = \vt{0}, \vt{u}=\vt{0}}, 
\end{align}
%\end{eqnarray}
%\end{widetext}
where $\mathcal{N}$ is the normalization factor of the Wigner function, $\vt{u} = \vt{t} \oplus \vt{s}$ with $\vt{t} = (t_1, t_2, \cdots, t_M)$ and $\vt{s} = (s_1, s_2, \cdots, s_M)$,
and the matrix ${\bf M}$ is defined as
%\begin{eqnarray}
%{\bf M} = 
%	\begin{pmatrix}
%	{\bf 0} & {\bf X}_{2M} {\bf T}_{2M}^{-\top}{\bf R}_{hd}  \\
%	%\\
%	{\bf R}_{dh} {\bf T}_{2M}^{-1} {\bf X}_{2M}  & {\bf C}
%	\end{pmatrix}
%\end{eqnarray}
\begin{eqnarray}
{\bf M} = 
	\begin{pmatrix}
	{\bf 0} & {\bf 0} & {\bf 0} & {\bf C}_1  \\
	%\\
	{\bf 0} & {\bf 0} & {\bf C}_1^* & {\bf 0} \\
	%\\
	{\bf 0} & {\bf C}_1^{\dag} & {\bf C}_2 & \vt{0} \\
	%\\
	{\bf C}_1^{\top} & {\bf 0} & {\bf 0} & {\bf C}_2^*
	\end{pmatrix}
\end{eqnarray}
with ${\bf C}_1$ and ${\bf C}_2$ given by 
\begin{eqnarray}
{\bf C}_1 &=& ({\bf I}_M - {\bf B}_{hh}^*{\bf B}_{hh})^{-1/2} {\bf B}_{hd}^*, \nonumber\\
{\bf C}_2 &=& {\bf B}_{dd} + {\bf B}_{dh} ({\bf I}_M - {\bf B}_{hh}^*{\bf B}_{hh})^{-1} {\bf B}_{hh}^* {\bf B}_{hd}. 
\end{eqnarray}

\section{Examples of generating multimode non-Gaussian states}\label{sec:MM-Example}

We now consider several examples of generating multimode non-Gaussian states via measuring pure multimode Gaussian states. We focus on the W state and NOON states.

\subsection{W states}

Let us measure one mode (the $N$-th mode without loss of generality) of an $N$-mode pure Gaussian state and postselect the measurement outcome with one photon detected. 
From Eq.~\eqref{eq:Wigner-Mmode-general-pure} it is clear that the heralded state is a superposition of Fock states with the total photon number to be at most one,
followed by a Gaussian operation. For simplicity, we choose Gaussian states such that the Gaussian operation is an identity in Eq.~\eqref{eq:mmode-output}, namely, the heralded state is only a superposition of Fock states. To simplify the notation, we define $\ket{{\bf 0}}$ as the vacuum state, and $\ket{{\bf 1}_k}$ as the state with one photon in the $k$-th mode
and zero photons in other modes. The heralded state  can thus be written as
\begin{eqnarray}\label{eq:Wstate}
%\ket{\psi_1} = 
c_{\bf 0} \ket{{\bf 0}} + \sum_{k=1}^{N-1} c_{{\bf 1}_k} \ket{{\bf 1}_k},
\end{eqnarray}
where $c_{\bf 0}$ and $c_{{\bf 1}_k}$ are coefficients that are determined by Eq.~\eqref{eq:CoeffWignerHermite-Mmode-2}. 
To guarantee that the heralded state is in the form of Eq.~\eqref{eq:Wstate}, we choose $\vt{y}_h = \vt{0}$ and ${\bf B}_{hh} = {\bf 0}$, 
then ${\bf C}_1 = {\bf B}_{hd}^* = (b_{1N}, b_{2N}, \cdots, b_{N-1,N})^{\dag}$ is a vector with $(N-1)$ components and ${\bf C}_2 = {\bf B}_{dd} = b_{NN}$ 
is  a complex number. 

It is straightforward to calculate the coefficients from Eq.~\eqref{eq:CoeffWignerHermite-Mmode-2} and we have
\begin{eqnarray}
|c_{\bf 0}|^2 &\propto& |y_N|^2, \nonumber\\
c_{\bf 0} c_{{\bf 1}_k}^* &\propto& y_N b_{kN}, \nonumber\\
c_{{\bf 1}_{\ell}} c_{{\bf 1}_{k}}^* &\propto& b_{\ell N}^* b_{k N},
\end{eqnarray}
where we have used the fact that ${\vt Y} = \vt{y}_d = (y_N^*, y_N)^{\top}$. It is therefore evident that 
\begin{eqnarray}\label{eq:CoeffWstate-zero}
c_{\bf 0} \propto y_N, ~~~~~~  c_{{\bf 1}_{k}} \propto b_{k N}^*. 
\end{eqnarray}
Since $y_N$ and $b_{k N}$ are independent free parameters,  they can be chosen arbitrarily, provided that the corresponding detected Gaussian state is physical. 
This guarantees that one can generate an arbitrary superposition state of  $\ket{{\bf 0}}$ and $\ket{{\bf 1}_k}$. 
Of the particular interest are the states which do not contain the vacuum state. They can be obtained by setting $y_N = 0$, namely, 
the mean of the detected $N$-mode Gaussian state is zero. From Eq.~\eqref{eq:CoeffWstate-zero}, the normalized state can be written as
\begin{eqnarray}
\ket{\psi} = \frac{1}{\sqrt{\mathcal{N}_w}} \sum_{k=1}^{N-1}  b_{k N}^* \ket{{\bf 1}_k},
\end{eqnarray}
where $\mathcal{N}_w = \sum_{k=1}^{N-1}  |b_{k N}|^2$. 

From the above constraints, the matrix ${\bf B}$ can be written as
\begin{eqnarray}\label{eq:BmatrixWstate}
{\bf B} = 
	\begin{pmatrix}
	{\bf 0} & {\bf B}_{hd} \\
	{\bf B}_{dh} & {b}_{NN}
	\end{pmatrix},
\end{eqnarray}
from which we can calculate the matrix ${\bf A}_p$ by using Eq.~\eqref{eq:ProbabilityMAMmode}, 
\begin{eqnarray}\label{eq:ApWstate}
{\bf A}_p = 
	\begin{pmatrix}
	{\bf B}_{dd} & {\bf B}_{dh} {\bf B}_{hd}^* \\
	{\bf B}_{dh} {\bf B}_{hd}^* & {\bf B}_{dd}^*
	\end{pmatrix}
=
	\begin{pmatrix}
	{b}_{NN} & \mathcal{N}_w \\
	\mathcal{N}_w & {b}_{NN}^*
	\end{pmatrix}. 
\end{eqnarray}
Substituting Eq.~\eqref{eq:ApWstate} into Eq.~\eqref{eq:ProbabilityMmode} and taking into account the fact that the mean of the detected Gaussian state is zero,
the success probability is 
\begin{eqnarray}\label{eq:ProbabilityWstate}
P(1) = \mathcal{N}_w \sqrt{(1 - \mathcal{N}_w)^2 - | {b}_{NN}|^2},
\end{eqnarray}
where we have used the result
\begin{eqnarray}
\mathcal{ P}_0 &=& \sqrt{\text{det} ({\bf I}_{2N} - {\bf X}_{2N} \tilde{\bf R})} = \sqrt{\text{det} ({\bf I}_{N} - {\bf B}^* {\bf B})} \nonumber\\
&=&\sqrt{(1 - \mathcal{N}_w)^2 - | {b}_{NN}|^2}. 
\end{eqnarray}
It is evident from Eq.~\eqref{eq:ProbabilityWstate} that the maximum success probability is $1/4$ when ${b}_{NN} = 0$ and $\mathcal{N}_w = 1/2$. 

The input squeezed states and the interferometer that are used to produce the measured Gaussian states can be extracted from the matrix ${\bf B}$. 
According to Eq.~\eqref{eq:Bmatrix-Signle} or the Autonne-Takagi decomposition \cite{horn1990matrix}, $r_j$ determines the squeezing parameter of the 
input squeezed state at the $j$-th mode and ${\bf U}$ represents the interferometer transformation. The unitary matrix ${\bf U}$ also diagonalizes
${\bf B}^*{\bf B}$ with eigenvalues $\tanh^2 r_j$. From the matrix ${\bf B}$ given by Eq.~\eqref{eq:BmatrixWstate}, we find that ${\bf B}^*{\bf B}$ has only two nonzero eigenvalues:
\begin{eqnarray*}
\mathcal{N}_w + \frac{1}{2} |{b}_{NN}|^2 \pm  \frac{1}{2} |{b}_{NN}| \sqrt{4 \mathcal{N}_w + |{b}_{NN}|^2}.
\end{eqnarray*}
This implies that there are two input squeezed states and all other inputs are vacuum states. 
Note that when determining the actual value of $r_j$, there might be a negative sign indicating the phase of the input squeezed states. 
%The interferometer is determined by the unitary matrix that diagonalizes ${\bf B}^*{\bf B}$. 

In the case where the success probability is optimal, the two nonzero eigenvalues of ${\bf B}^*{\bf B}$ are the same: $\tanh^2 r_1 = \tanh^2 r_2 = 1/2$. This corresponds to 
$r_1 = - r_2 \approx - 0.88$, or about $7.66$ dB of input squeezing. In the special case where all $b_{k N}$ are the same for $k<N$, the heralded state is an 
equal superposition of all $\ket{{\bf 1}_k}$, known as the W state. The unitary matrix that diagonalizes ${\bf B}$ when the success probability is maximum is given by
\begin{eqnarray}
{\bf U}_5 = \frac{1}{2\sqrt{2}}
	\begin{pmatrix}
	- 1 & 1 & - 2 & ~~0 & - \sqrt{2} \\
	%\\
	- 1 & 1 &  ~~0 & - 2 & ~~\sqrt{2} \\
	%\\
	- 1 & 1 & ~~0 & ~~2 &  ~~\sqrt{2} \\
	%\\
	- 1 & 1 & ~~2 & ~~0 & - \sqrt{2} \\
	%\\
	~~2 & 2 & ~~0 & ~~0 & ~~~0 
	\end{pmatrix}
\end{eqnarray}
for $N=5$.

\subsection{Generation of NOON states $\ket{\eta_N}$}

\begin{figure}[ht!]
\includegraphics[width=8.cm]{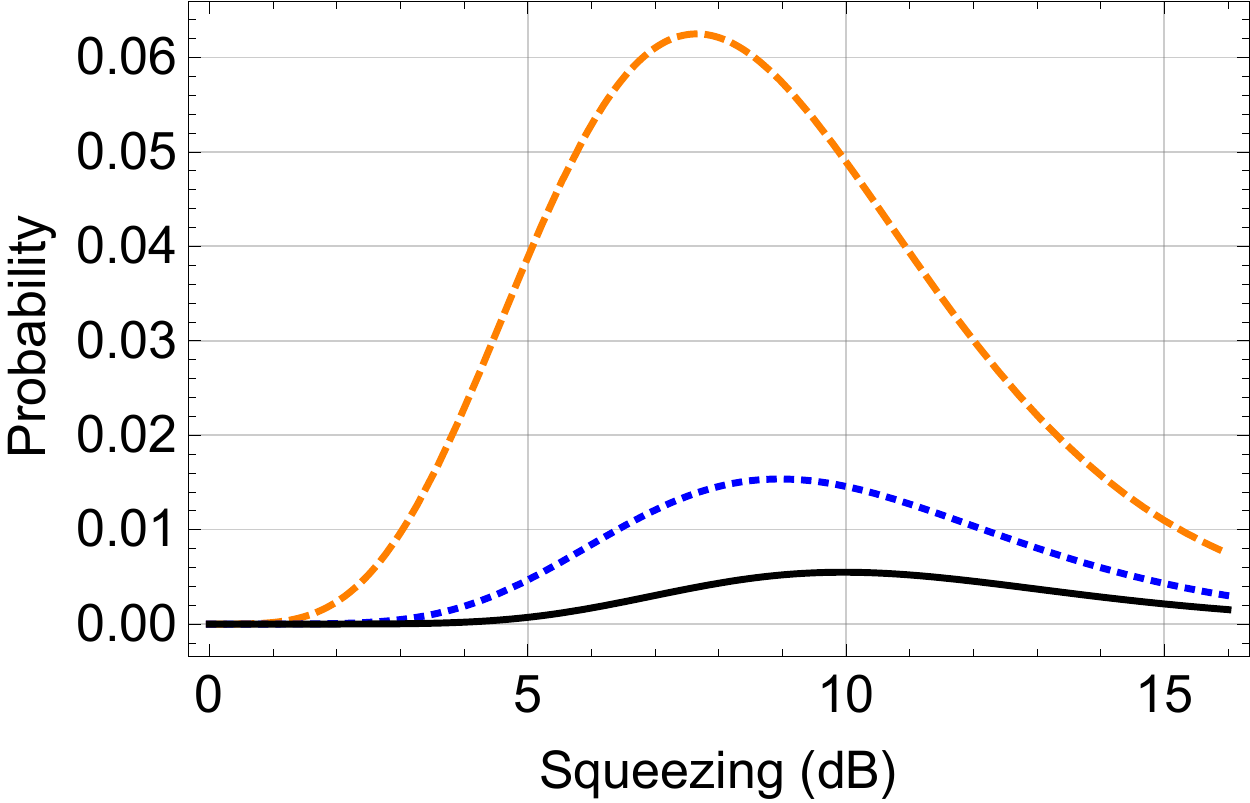}
\caption{ Success probability versus input squeezing for NOON state $\ket{\eta_2}$ (orange dashed), $\ket{\eta_3}$ (blue dotted) and $\ket{\eta_4}$ (black solid). Here, we 
consider the case where squeezing parameters of the input squeezed states are the same. }
\label{fig:ProbabilitySqueezing}
\end{figure}

An important class of two-mode non-Gaussian states is the NOON states, which is defined as 
\begin{align} \label{eq:noon}
\ket{\eta_N} = \frac{1}{\sqrt{2}} (\ket{N0} + \ket{0N}),
\end{align}
with $N$ a positive integer. It has applications both in quantum metrology and quantum computation, in particular, the error-correcting bosonic codes \cite{PhysRevA.56.1114, PhysRevA.94.012311, michael2016new}.  
The NOON state can be generated by the method of photon subtraction \cite{yoshikawa2018heralded}. 
Here, we generate NOON states up to $N=4$ using our formalism and optimize the success probability. 
The results are summarized in Table \ref{tab:NOONstate} and Fig. \ref{fig:ProbabilitySqueezing}. 
Note that the maximal success probabilities are substantially bigger than those found in Ref.~\cite{yoshikawa2018heralded}. 
We discuss in detail how to generate the NOON states in the following subsections.

\begin{table}[tp]
\caption{ Generating NOON states (Eq.~\eqref{eq:noon}) by detecting multimode Gaussian states using PNRDs. In the title row, Modes represents the number of modes of the initial Gaussian state, followed by the detection pattern, followed by the success probability of producing the particular NOON state, and the final column lists the required amount of squeezing in the input modes.}  
\label{tab:NOONstate}
\centering
\begin{center}
\resizebox{0.48\textwidth}{!}{
    \begin{tabular}{| c | c | c | c |c|}
    \hline  \hline
    {\bf State} & {\bf Modes} & {\bf Detection} & {\bf Probability} &{\bf Req. sq.} \\ \hline 
    $\ket{\eta_2}$ & 4 & (1, 1) & $1/16 = 6.25 \%$ & $7.66$\,\textsf{dB}\\ \hline 
    $\ket{\eta_3}$ & 5 & (1, 1, 1) & $48/3125 \approx 1.54 \%$ & $8.96$\,\textsf{dB} \\ \hline
    $\ket{\eta_4}$ & 6 & (1, 1, 1, 1) & $4/729 \approx 0.55 \% $  & $9.96$\,\textsf{dB}\\ \hline
    \end{tabular}
   }
\end{center}
\end{table}

\subsubsection{Generation of $\ket{\eta_2}$}

To generate $\ket{\eta_2}$, the PNRDs should register at least 2 photons in total. We find that detecting one mode of a three-mode 
Gaussian state cannot generate the desired NOON state. Specifically, one cannot generate an arbitrary state in the Hilbert space expanded by two-photon Fock bases:
$\ket{20}, \ket{02}$ and $\ket{11}$. This issue can be resolved by detecting two modes of a four-mode Gaussian state and postselecting the photon measurement pattern
$\bar{\vt{n}} = (1, 1)$. 

Since the target state is a superposition of a finite number of Fock states, there should  be no final displacement or squeezing operator applied to the heralded state.
These conditions can be satisfied by choosing $\vt{y}_h = \vt{0}$ and ${\bf B}_{hh} = {\bf 0}$. The vector $\vt{Y}$ is thus equal to $\vt{y}_d = (y_3^*, y_4^*, y_3, y_4)^{\top}$ 
and the matrix ${\bf B}$ becomes
\begin{eqnarray}\label{eq:Bmatrix2002}
{\bf B} = 
	\begin{pmatrix}
	0 & 0 & b_{13} & b_{14}  \\
	0 & 0 & b_{23} & b_{24}  \\
	b_{13} & b_{23} & b_{33} & b_{34}  \\
	b_{14} & b_{24} & b_{34} & b_{44}  \\
	\end{pmatrix}. 
\end{eqnarray}
From Eq.~\eqref{eq:CoeffWignerHermite-Mmode-2}, we can calculate the coefficients of all Fock basis states up to 2 photons: $\ket{00}$, $\ket{10}$, $\ket{01}$, $\ket{20}$, $\ket{02}$
and $\ket{11}$. These coefficients are explicitly given by Eq.~\eqref{eq:Coef11-4mode} in Appendix \ref{appedix:CoeffFourStates}. It can be shown that any state in the Hilbert
space expanded by the Fock bases up to 2 photons can be generated by appropriately tuning the matrix ${\bf B}$ and vector $\vt{y}_d$. In particular, to obtain $\ket{\eta_2}$, one requires that $c_{00} = c_{10} = c_{01} = c_{11} = 0$ and $c_{20} = c_{02}$. These constraints result in $b_{23} = \pm i b_{13}$,
$b_{24} = \mp i b_{14}$ and $b_{34} = y_3 = y_4 = 0$. 

The success probability can be calculated from Eq.~\eqref{eq:ProbabilityMmode} by taking into account the above constraints. We find 
\begin{eqnarray}
P(1, 1) &=& 4 \, |b_{13}|^2 |b_{14}|^2 \sqrt{\big[(1-2 \,  |b_{13}|^2)^2 -  |b_{33}|^2 \big]} \nonumber\\
&&
\times \sqrt{\big[(1-2 \,  |b_{14}|^2)^2 -  |b_{44}|^2 \big]}. 
\end{eqnarray}
It is evident that the presence of $b_{33}$ and $b_{44}$ reduces the success probability. To maximize the success probability, we thus assume that $b_{33} = b_{44} = 0$.
Under this condition, the success probability is optimized when $|b_{13}| = |b_{14}| = 1/2$, and the maximal success probability is $1/16 = 6.25 \%$. 
We want to emphasize that this is the best success probability one can achieve by measuring four-mode Gaussian states. 

\begin{figure}
\includegraphics[width=\columnwidth]{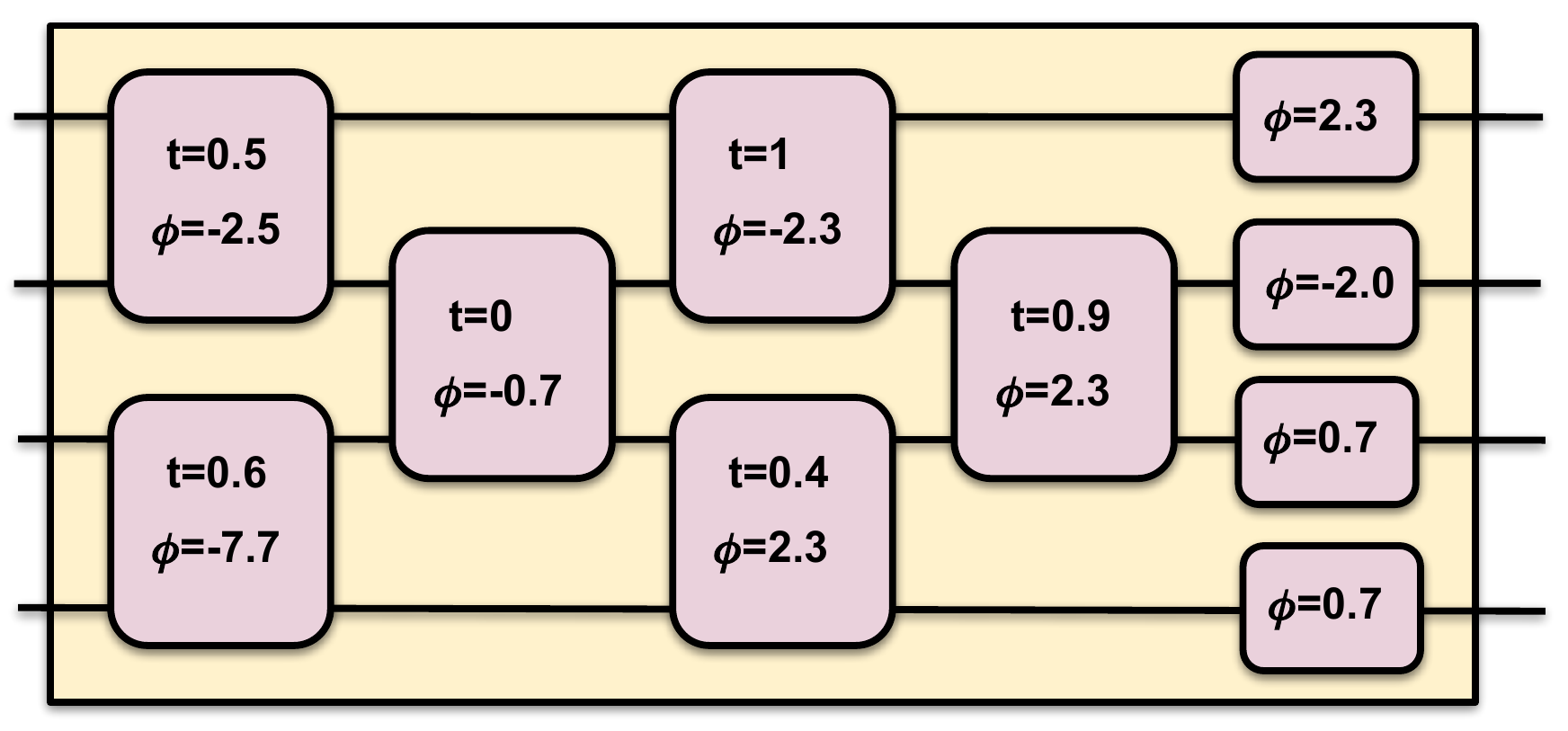}
\caption{Square decomposition \cite{clements} of the unitary in Eq.~\eqref{eq:Umatrix2002-optimal}. The gate denoted by the pair $(t,\phi)$ denotes a beamsplitter with transmissivity $t=\cos^2{\theta}$ preceded by a phase rotation by angle $\phi$ in the first mode alone. The gate denoted by just $\phi$ is a single-mode phase rotation.  }
\label{fig:2002}
\end{figure}

One of the possible options for the matrix ${\bf B}$ to achieve the maximal success probability is
\begin{eqnarray}\label{eq:Bmatrix2002-optimal}
{\bf B}_{2002}^{\text{max}} = \frac{1}{2}
	\begin{pmatrix}
	0 & 0 & 1 & 1 \\
	0 & 0 & -i & i  \\
	1 & -i & 0 & 0  \\
	1 & i & 0 & 0  \\
	\end{pmatrix}. 
\end{eqnarray}
The input states and linear interferometer that produce the detected four-mode Gaussian state are fully determined by the matrix in Eq.~\eqref{eq:Bmatrix2002-optimal}. 
It is found that the input squeezing parameters in the input modes are $r_1 = - r_2 = r_3 = - r_4 = \tanh^{-1}(1/2)$, corresponding to about $7.66$ dB of squeezing. 
The corresponding unitary is 
\begin{eqnarray}\label{eq:Umatrix2002-optimal}
{\bf U}_{2002}^{\text{max}} = \frac{1}{2}
	\begin{pmatrix}
	\sqrt{2} & -\sqrt{2} & 0 & 0 \\
	0 & 0 &  \sqrt{2}e^{i \pi/4} & -\sqrt{2}e^{i \pi/4}  \\
	1 & 1 & -e^{i \pi/4} & -e^{i \pi/4}  \\
	1 & 1 & e^{i \pi/4} & e^{i \pi/4}  \\
	\end{pmatrix}. 
\end{eqnarray}
As before we provide the square decomposition of the unitary in Eq.~\eqref{eq:Umatrix2002-optimal} schematically in Fig.~\ref{fig:2002}. 

\subsubsection{Generation of $\ket{\eta_3}$}

Similarly, generating $\ket{\eta_3}$ requires the PNRDs to detect at least three photons in total. We find that detecting three photons 
in two modes of a four-mode Gaussian state cannot generate the $\ket{\eta_3}$. This can be seen from the coefficients of the Fock basis 
states up to three photons given by Eqs.~\eqref{eq:Coef12-4mode} and \eqref{eq:Coef21-4mode} in Appendix \ref{appedix:CoeffFourStates}. The coefficients 
in the subspace with three photons are not independent. To resolve this issue, we have to detect three modes of a 5-mode Gaussian state with a photon measurement pattern
$\bar{\vt{n}} = (1, 1, 1)$. 

Again, we choose $\vt{y}_h = \vt{0}$ and ${\bf B}_{hh} = {\bf 0}$. The vector $\vt{Y}$ is now equal to $\vt{y}_d = (y_3^*, y_4^*, y_5^*, y_3, y_4, y_5)^{\top}$ 
and the matrix ${\bf B}$ becomes
\begin{eqnarray}\label{eq:Bmatrix3003}
{\bf B} = 
	\begin{pmatrix}
	0 & 0 & b_{13} & b_{14} & b_{15}  \\
	0 & 0 & b_{23} & b_{24} & b_{25} \\
	b_{13} & b_{23} & b_{33} & b_{34} & b_{35} \\
	b_{14} & b_{24} & b_{34} & b_{44}  & b_{45} \\
	b_{15} & b_{25} & b_{35} & b_{45}  & b_{55} \\
	\end{pmatrix}. 
\end{eqnarray}
From Eq.~\eqref{eq:CoeffWignerHermite-Mmode-2}, we can calculate the coefficients of all Fock basis states up to 3 photons, which are 
given by Eq.~\eqref{eq:Coef111-5mode} in Appendix \ref{appedix:CoeffFiveStates}. To obtain $\ket{\eta_3}$, one requires that 
$c_{30} = c_{03}$ and all other coefficients are zero. The above constraints imply that 
$b_{34} = b_{35} = b_{45} = y_3 = y_4 = y_5 = 0$ and $(b_{23}, b_{24}, b_{25}) = (\tau_1 b_{13}, \tau_2 b_{14}, \tau_3 b_{15})$, where the three parameters
$(\tau_1, \tau_2, \tau_3)$ have solutions:
\begin{eqnarray}\label{eq:6solution3003}
\big( 1, ~ e^{2i\pi/3}, ~ e^{4i\pi/3} \big)
\end{eqnarray}
and all possible permutations of these three numbers.

We expect that the maximal probability is obtained when $b_{33} = b_{44} = b_{55} = 0$. 
Under this condition, all solutions in 
Eq.~\eqref{eq:6solution3003} lead to the same success probability expression, 
\begin{eqnarray}
&& P(1, 1, 1) \nonumber\\
&= &12 \, |b_{13}|^2 |b_{14}|^2 |b_{15}|^2 \bigg[1 - 2 \, (|b_{13}|^2 + |b_{14}|^2 + |b_{15}|^2) 
\nonumber\\
&&
+ 3 \, |b_{13}|^2 |b_{14}|^2 + 3 \, |b_{13}|^2 |b_{15}|^2 + 3 \, |b_{14}|^2 |b_{15}|^2
\bigg]. 
\end{eqnarray}
The success probability is maximized when $|b_{13}|^2 = |b_{14}|^2 = |b_{15}|^2 = 1/5$, and the maximal success probability is $48/5^5 = 1.536 \%$. 
A representative matrix ${\bf B}$ that gives the maximal success probability is 
\begin{eqnarray}\label{eq:Bmatrix3003-optimal}
{\bf B}_{3003}^{\text{max}} = \frac{1}{\sqrt{5}}
	\begin{pmatrix}
	0 & 0 & 1 & 1 & 1 \\
	0 & 0 & 1 & e^{2i\pi/3}  & e^{4i\pi/3}   \\
	1 & 1 & 0 & 0 & 0  \\
	1 & e^{2i\pi/3} & 0 & 0 & 0 \\
	1 & e^{4i\pi/3} & 0 & 0 & 0 \\
	\end{pmatrix}.
\end{eqnarray}
The input states and linear interferometer that produce the detected 5-mode Gaussian state are fully determined by the matrix in Eq.~\eqref{eq:Bmatrix3003-optimal}. 
It is found that one of the inputs is vacuum and other inputs are squeezed vacuum states with the same squeezing parameter, $\tanh^2 r = 3/5$, 
corresponding to about $8.96$ dB of squeezing.

\subsubsection{Generation of $\ket{\eta_4}$}

The NOON state $\ket{\eta_4}$ has to be generated by detecting four modes of a 6-mode Gaussian state with a photon measurement pattern
$\bar{\vt{n}} = (1, 1, 1, 1)$. Again, we choose $\vt{y}_h = \vt{0}$ and ${\bf B}_{hh} = {\bf 0}$. 
The vector $\vt{Y}$ is now equal to $\vt{y}_d = (y_3^*, y_4^*, y_5^*, y_6^*, y_3, y_4, y_5, y_6)^{\top}$ 
and the matrix ${\bf B}$ becomes
\begin{eqnarray}\label{eq:Bmatrix4004}
{\bf B} = 
	\begin{pmatrix}
	0 & 0 & b_{13} & b_{14} & b_{15}  & b_{16} \\
	0 & 0 & b_{23} & b_{24} & b_{25} & b_{26} \\
	b_{13} & b_{23} & b_{33} & b_{34} & b_{35} & b_{36}\\
	b_{14} & b_{24} & b_{34} & b_{44}  & b_{45} & b_{46} \\
	b_{15} & b_{25} & b_{35} & b_{45}  & b_{55} & b_{56}\\
	b_{16} & b_{26} & b_{36} & b_{46}  & b_{56} & b_{66}\\
	\end{pmatrix}. 
\end{eqnarray}
From Eq.~\eqref{eq:CoeffWignerHermite-Mmode-2}, we can calculate the coefficients of all Fock basis states up to four photons, which are 
given by Eqs.~\eqref{eq:Coef1111-6mode-01}-\eqref{eq:Coef1111-6mode-4} in Appendix \ref{appedix:CoeffSixStates}. 
To obtain $\ket{\eta_4}$, one requires that 
$c_{40} = c_{04}$ and all other coefficients are zero. The above constraints imply that %(\textcolor{red}{\bf prove this!})
$b_{34} = b_{35} = b_{36} = b_{45} = b_{46} = b_{56} = 0$, $y_3 = y_4 = y_5 = y_6 = 0$ and 
$(b_{23}, b_{24}, b_{25}, b_{26}) = (\tau_1 b_{13}, \tau_2 b_{14}, \tau_3 b_{15}, \tau_4 b_{16})$, where the four parameters $(\tau_1, \tau_2, \tau_3, \tau_4)$ have solutions:
\begin{eqnarray}\label{eq:3solution4004}
&&\big( e^{i \pi/4}, ~e^{3 i \pi/4}, ~e^{5 i \pi/4}, ~e^{7 i \pi/4} \big)
\end{eqnarray}
and all possible permutations of these four numbers.

We expect that the maximal probability is obtained when $b_{33} = b_{44} = b_{55} = b_{6} = 0$. % (\textcolor{red}{\bf check}). 
Under this condition, 
all possible solutions given by Eq.~\eqref{eq:3solution4004} lead to the same success probability expression,
\begin{eqnarray}
&& P(1, 1, 1, 1) \nonumber\\
&=& 64 \, |b_{13}|^2 |b_{14}|^2 |b_{15}|^2  |b_{16}|^2 \big[1 - 2 \, (|b_{13}|^2 + |b_{15}|^2) \big]
\nonumber\\
&&
\times \big[1 - 2 \, (|b_{14}|^2 + |b_{16}|^2) \big]. 
\end{eqnarray}
The success probability is maximized when $|b_{13}|^2 = |b_{14}|^2 = |b_{15}|^2 = |b_{16}|^2 = 1/6$, and the maximal success probability is $4/3^9 \approx 0.55 \%$. 
A representative matrix ${\bf B}$ that gives the maximal success probability is 
\begin{eqnarray}\label{eq:Bmatrix4004-optimal}
{\bf B}_{4004}^{\text{max}} = \frac{1}{\sqrt{6}}
	\begin{pmatrix}
	0 & 0 & 1 & 1 & 1 & 1\\
	0 & 0 & e^{i \pi/4} & e^{3 i \pi/4} & e^{5 i \pi/4} & e^{7 i \pi/4}  \\
	1 & e^{i \pi/4} & 0 & 0 & 0  & 0 \\
	1 & e^{3 i \pi/4} & 0 & 0 & 0 & 0 \\
	1 & e^{5 i \pi/4}  & 0 & 0 & 0 & 0 \\
	1 & e^{7 i \pi/4} & 0 & 0 & 0 & 0 \\
	\end{pmatrix}. \nonumber\\
\end{eqnarray}
The input states and linear interferometer that produce the detected 6-mode Gaussian state are fully determined by the matrix in Eq.~\eqref{eq:Bmatrix4004-optimal}. 
It is found that two of the inputs are vacuum states and other inputs are squeezed vacuum states with the same squeezing parameter, $\tanh^2 r = 2/3$, 
corresponding to about $9.96$ dB of squeezing.

\section{Conclusion}\label{sec:conclusion}

We develop a detailed analytic framework for the study of probabilistic generation of non-Gaussian states by measuring multimode Gaussian states via PNR detectors. 
We derive explicit expressions for the output Wigner function and the measurement success probability, which show clearly the mapping between the properties of the multimode
Gaussian states whose subsystems are being measured and the measurement outcome, and that of the heralded non-Gaussian output states. The framework unifies many state preparation schemes, and more importantly, it provides a procedure to optimize the fidelity and success probability of the target state. 

We demarcate the analysis into single mode and multimode mode cases and focus on measuring  pure Gaussian states to obtain pure non-Gaussian outputs. For the single-mode case we consider the generation of GKP states, cat states, ON states, and weak cubic phase states. For the multimode case we consider illustrative examples such as the W and NOON states. In all the cases, we find that both the fidelity
and success probability are improved as compared to previous schemes. 

The formalism can also deal with the case when the initial Gaussian state that is being measured  is mixed. This is an important point when one is dealing with realistic experimental setups. A common noise model is that of photon loss that is modeled using lossy channels. The Gaussian state that one obtains just before the photon detection is then mixed. One way to look at these mixed states is to purify the resulting Gaussian state and ignoring a few modes to obtain the mixed state. 

It is also expected that increasing the number of modes and choosing a particular photon detection pattern may not scale favorably with the number of input modes. One possible way to get around this could be to coarse-grain the output detection. In this case we would end up in a scenario where there is a natural trade-off between the fidelity to a given target state and its success probability. 

Our general framework is closely related to a sampling algorithm called \textsf{Gaussian BosonSampling} (GBS) \cite{PhysRevLett.119.170501,kruse2018detailed} which is a variant of the famous \textsf{BosonSampling} problem, as was briefly mentioned earlier. In GBS, one has the same state preparation scheme, namely, squeezed displaced vacuum states are input to a multimode interferometer. Then all the output modes of the interferometer are detected using PNR detectors to generate samples of photon detection on multimode Gaussian states. It is believed that GBS is one route to demonstrating quantum advantage, and has received much attention in the quantum optics community, with various groups around the world pushing experimental boundaries of the number of modes GBS is executed in. While much effort has been dedicated to the statistical behaviour and its implications to computational complexity, very little attention was diverted to the study of the non-Gaussian states that are generated from the GBS device when only few modes are detected. Our framework proposes the use of these GBS device for the purpose of non-Gaussian state preparation, which has been a challenge from an experimental point of view.

One final application for our framework that we envisage is to the quantum resource theory of non-Gaussianity \cite{quntao2018,shor2018,albarelli2018resource,lami2018all}. In this language, our Gaussian state preparation would fall under the class of free-operations. The only non-Gaussian resources that we use is that of PNR measurements. Using this resource, we convert Gaussian to non-Gaussian states. It would be fruitful to quantify these conversions from the resource perspective of non-Gaussianity. For example, can the non-Gaussianity of the output states be quantified by the parameters of the Gaussian state that is being measured and the post-selected photon-detection pattern?

With steady improvements in the optical technology of PNR detectors \cite{magana2019multiphoton, tiedau2019scalability}, our framework would be a promising candidate to generate non-Gaussianity that is essential in applications such as quantum metrology and quantum computing, in particular, 
the generation of fault-tolerant error-correcting codes.

\acknowledgements 
We thank Haoyu Qi, Kamil Br\'adler, Christian Weedbrook, Saikat Guha and Christos Gagatsos for insightful discussions.

\clearpage

\appendix

\begin{widetext}

\section{From integration to derivative}\label{appedix:IntegralDerivative}

Suppose $f(z)$ is an analytic function in the complex $z$ plane and $e^{-z z^*} f(z) \rightarrow 0$ when $z \rightarrow \infty$. We want to evaluate the following integral,
\begin{eqnarray}
\mathcal{I}_n = \int \mathrm d z \, (z^*)^{n} e^{-z z^*} f(z). 
\end{eqnarray}
We start from the simplest case where $n=1$ and assume that $z = x + i y$ with $x$ and $y$ real numbers. We find
\begin{eqnarray}
\mathcal{I}_1 &=& \int \mathrm d z \, z^* e^{-z z^*} f(z) = \int \mathrm dx \int \mathrm d y \, (x - i y) e^{- x^2 - y^2} f(z) 
= - \int \mathrm dx \int \mathrm d y \bigg[ \bigg(\frac{1}{2} \frac{\partial}{\partial x} - \frac{i}{2} \frac{\partial}{\partial y} \bigg) e^{- x^2 - y^2} \bigg] f(z)
\nonumber\\
&=&
-  \frac{1}{2} \int \mathrm d y \, e^{-y^2} \bigg[ e^{-x^2} f(z) \bigg|_{-\infty}^{+\infty} - \int \mathrm dx \, e^{-x^2} \frac{\partial}{\partial x} f(z) \bigg]
+  \frac{i}{2} \int \mathrm d x \, e^{-x^2} \bigg[ e^{-y^2} f(z) \bigg|_{-\infty}^{+\infty} - \int \mathrm dy \, e^{-y^2} \frac{\partial}{\partial y} f(z) \bigg]
\nonumber\\
&=&
\int \mathrm dx \int \mathrm d y \, e^{- x^2 - y^2}  \bigg[ \bigg(\frac{1}{2} \frac{\partial}{\partial x} - \frac{i}{2} \frac{\partial}{\partial y} \bigg) f(z) \bigg]
\nonumber\\
&=&
\int \mathrm d z \, e^{- z z^*}  \frac{\partial}{\partial z} f(z).
\end{eqnarray}
By using the relation
\begin{eqnarray}
\int \mathrm dz \, z^m e^{- z z^*} = 0 
\end{eqnarray}
for any positive integer $m$, we have 
\begin{eqnarray}
\mathcal{I}_1 &=& 
\int \mathrm d z \, e^{- z z^*}  \frac{\partial}{\partial z} f(z)
=
\frac{\partial}{\partial z} f(z) \bigg|_{z=0} \times \int \mathrm d z \, e^{- z z^*}
%\nonumber\\
%&=&
=
\pi \frac{\partial}{\partial z} f(z) \bigg|_{z=0}. 
\end{eqnarray}
By repeatedly performing the integration in part, we find
\begin{eqnarray}\label{eq:IntegralDerivative}
\mathcal{I}_n = \pi \frac{\partial^n }{\partial z^n} f(z) \bigg|_{z=0}. 
\end{eqnarray}

\section{Measuring subsystems of three-mode Gaussian states}\label{appedix:CoeffThreeStates}

In this appendix, we explicitly derive the coefficients of the superposition of Fock states when detecting two modes of a three-mode Gaussian states. 
Here, we assume that $\kappa_2 \ne 0$ and $\kappa_3 \ne 0$, implying that $n_{\text{max}} = n_T$. When $n_T = 0$, the heralded state is a Gaussian state,
which is not interesting in the perspective of non-Gaussian state generation. So in the following, we will consider cases where PNRDs register photons. 

{\bf Detecting one photon}. When the total number of detected photons is $n_T = 1$, there are two possible photon number patterns: $(n_2, n_3) = (1, 0)$ and $(n_2, n_3) = (0, 1)$. 
The heralded state is in the form
\begin{eqnarray}\label{eq:FormPN1}
\hat{D}(\alpha_1) \hat{S}(\zeta_1) (c_0 \ket{0} + c_1 \ket{1}), 
\end{eqnarray}
where $\alpha_1$ and $\zeta_1$ are the displacement and squeezing amplitudes, respectively. For the photon number pattern $(n_2, n_3) = (1, 0)$, $c_0$ and $c_1$ satisfy
\begin{eqnarray}\label{eq:CE10photon3mode}
\frac{c_0}{c_1} = \mu_2;
\end{eqnarray}
for the photon number pattern $(n_2, n_3) = (0, 1)$, $c_0$ and $c_1$ satisfy
\begin{eqnarray}\label{eq:CE01photon3mode}
\frac{c_0}{c_1} = \mu_3. 
\end{eqnarray}

{\bf Detecting two photons}. When the total number of detected photons is $n_T = 2$, there are three possible photon number patterns: 
$(n_2, n_3) = (1, 1)$, $(n_2, n_3) = (2, 0)$ and $(n_2, n_3) = (0, 2)$. The heralded state is in the form 
\begin{eqnarray}\label{eq:FormPN2}
\hat{D}(\alpha_1) \hat{S}(\zeta_1) (c_0 \ket{0} + c_1 \ket{1} + c_2 \ket{2}).
\end{eqnarray} 
For the photon number pattern $(n_2, n_3) = (1, 1)$, the coefficients satisfy
\begin{eqnarray}\label{eq:CE11photon3mode}
\frac{c_1}{c_2} = \frac{1}{\sqrt{2}} (\mu_2 + \mu_3), ~~~~~~ \frac{c_0}{c_2} = \frac{1}{\sqrt{2}} \big( \mu_2 \mu_3 + f_{23}^* \big). 
\end{eqnarray}
For the photon number pattern $(n_2, n_3) = (2, 0)$, the coefficients satisfy
\begin{eqnarray}\label{eq:CE20photon3mode}
\frac{c_1}{c_2} = \sqrt{2} \, \mu_2, ~~~~~~ \frac{c_0}{c_2} = \frac{1}{\sqrt{2}} \big( \mu_2^2 + f_{22}^* \big). 
\end{eqnarray}
For the photon number pattern $(n_2, n_3) = (0, 2)$, the coefficients satisfy
\begin{eqnarray}\label{eq:CE02photon3mode}
\frac{c_1}{c_2} = \sqrt{2} \, \mu_3, ~~~~~~ \frac{c_0}{c_2} = \frac{1}{\sqrt{2}} \big(\mu_3^2 + f_{33}^* \big). 
\end{eqnarray}

{\bf Detecting three photons}. When the total number of detected photons is $n_T = 3$, there are four possible photon number patterns: 
$(n_2, n_3) = (2, 1)$, $(n_2, n_3) = (1, 2)$, $(n_2, n_3) = (3, 0)$ and $(n_2, n_3) = (0, 3)$. The heralded state is in the form 
\begin{eqnarray}\label{eq:FormPN3}
\hat{D}(\alpha_1) \hat{S}(\zeta_1) (c_0 \ket{0} + c_1 \ket{1} + c_2 \ket{2} + c_3 \ket{3}).
\end{eqnarray}
For the photon number pattern $(n_2, n_3) = (2, 1)$, the coefficients satisfy
\begin{eqnarray}\label{eq:CE21photon3mode}
\frac{c_2}{c_3} &=& \frac{1}{\sqrt{3}} (2 \mu_2 + \mu_3), ~~~~~~ 
\frac{c_1}{c_3} = \frac{1}{\sqrt{6}} \big[ \mu_2 (\mu_2 + 2 \mu_3) + f_{22}^* + 2\, f_{23}^* \big], ~~~~~~
\frac{c_0}{c_3} = \frac{1}{\sqrt{6}} \big[ \mu_2^2 \, \mu_3 + \mu_3 f_{22}^* + 2\, \mu_2 f_{23}^* \big]. 
\end{eqnarray}
For the photon number pattern $(n_2, n_3) = (1, 2)$, the coefficients satisfy
\begin{eqnarray}\label{eq:CE12photon3mode}
\frac{c_2}{c_3} &=& \frac{1}{\sqrt{3}} (\mu_2 + 2 \mu_3), ~~~~~~ 
\frac{c_1}{c_3} = \frac{1}{\sqrt{6}} \big[ \mu_3 (2 \mu_2 + \mu_3) + f_{33}^* + 2\, f_{23}^* \big], ~~~~~~
\frac{c_0}{c_3} = \frac{1}{\sqrt{6}} \big[ \mu_2 \, \mu_3^2 + \mu_2 f_{33}^* + 2\, \mu_3 f_{23}^* \big]. 
\end{eqnarray}
For the photon number pattern $(n_2, n_3) = (3, 0)$, the coefficients satisfy
\begin{eqnarray}\label{eq:CE30photon3mode}
\frac{c_2}{c_3} = \sqrt{3} \, \mu_2, ~~~~~~ \frac{c_1}{c_3} = \sqrt{\frac{3}{2}} \big( \mu_2^2 + f_{22}^* \big), ~~~~~~
\frac{c_0}{c_3} = \frac{1}{\sqrt{6}} \mu_2 \big( \mu_2^2 + 3 f_{22}^* \big). 
\end{eqnarray}
For the photon number pattern $(n_2, n_3) = (0, 3)$, the coefficients satisfy
\begin{eqnarray}\label{eq:CE03photon3mode}
\frac{c_2}{c_3} = \sqrt{3} \, \mu_3, ~~~~~~ \frac{c_1}{c_3} = \sqrt{\frac{3}{2}} \big( \mu_3^2 + f_{33}^* \big), ~~~~~~
\frac{c_0}{c_3} = \frac{1}{\sqrt{6}} \mu_3 \big( \mu_3^2 + 3 f_{33}^* \big). 
\end{eqnarray}

{\bf Detecting four photons}. When the total number of detected photons is $n_T = 4$, there are five possible photon number patterns: 
$(n_2, n_3) = (2, 2)$, $(n_2, n_3) = (3, 1)$, $(n_2, n_3) = (1, 3)$, $(n_2, n_3) = (4, 0)$ and $(n_2, n_3) = (0, 4)$. The heralded state is in the form 
\begin{eqnarray}\label{eq:FormPN4}
\hat{D}(\alpha_1) \hat{S}(\zeta_1) (c_0 \ket{0} + c_1 \ket{1} + c_2 \ket{2} + c_3 \ket{3} + c_4 \ket{4}).
\end{eqnarray}
For the photon number pattern $(n_2, n_3) = (2, 2)$, the coefficients satisfy
\begin{eqnarray}\label{eq:CE22photon3mode}
\frac{c_3}{c_4} &=& \mu_2 + \mu_3, ~~~~~~ \frac{c_2}{c_4} = \frac{1}{2\sqrt{3}} \bigg( \mu_2^2 +4  \mu_2  \mu_3 +  \mu_3^2 + f_{22}^* + 4 f_{23}^* + f_{33}^* \bigg), \nonumber\\
\frac{c_1}{c_4} &=& \frac{1}{\sqrt{6}} \bigg[ \mu_2^2\mu_3 + \mu_2 \mu_3^2 + \mu_3 f_{22}^* + 2(\mu_2 + \mu_3)f_{23}^* + \mu_2 f_{33}^* \bigg], \nonumber\\
\frac{c_0}{c_4} &=& \frac{1}{2\sqrt{6}} \bigg( \mu_2^2\mu_3^2 +  \mu_3^2 f_{22}^* + 4 \mu_2 \mu_3 f_{23}^* + \mu_2^2 f_{33}^* + f_{22}^*f_{33}^* + 2 f_{23}^{*2} \bigg). 
\end{eqnarray}
For the photon number pattern $(n_2, n_3) = (3, 1)$, the coefficients satisfy
\begin{eqnarray}\label{eq:CE31photon3mode}
\frac{c_3}{c_4} &=& \frac{1}{2} (3 \mu_2 + \mu_3), ~~~~~~ \frac{c_2}{c_4} = \frac{\sqrt{3}}{2} \bigg( \mu_2^2 + \mu_2  \mu_3 + f_{22}^* + f_{23}^*\bigg), \nonumber\\
\frac{c_1}{c_4} &=& \frac{1}{2\sqrt{6}} \bigg[ \mu_2^3 + 3 \mu_2^2 \mu_3 + 3(\mu_2 + \mu_3) f_{22}^* + 6 \mu_2 f_{23}^* \bigg], \nonumber\\
\frac{c_0}{c_4} &=& \frac{1}{2\sqrt{6}} \bigg( \mu_2^3\mu_3 +  3 \mu_2 \mu_3 f_{22}^* + 3 \mu_2^2 f_{23}^* + 3 f_{22}^*f_{23}^* \bigg). 
\end{eqnarray}
For the photon number pattern $(n_2, n_3) = (1, 3)$, the coefficients satisfy
\begin{eqnarray}\label{eq:CE13photon3mode}
\frac{c_3}{c_4} &=& \frac{1}{2} (\mu_2 + 3 \mu_3), ~~~~~~ \frac{c_2}{c_4} = \frac{\sqrt{3}}{2} \bigg( \mu_2  \mu_3 + \mu_3^2 + f_{23}^* + f_{33}^*\bigg), \nonumber\\
\frac{c_1}{c_4} &=& \frac{1}{2\sqrt{6}} \bigg[ 3 \mu_2 \mu_3^2 + \mu_3^3 + 3(\mu_2 + \mu_3) f_{33}^* + 6 \mu_3 f_{23}^* \bigg], \nonumber\\
\frac{c_0}{c_4} &=& \frac{1}{2\sqrt{6}} \bigg( \mu_2\mu_3^3 +  3 \mu_2 \mu_3 f_{33}^* + 3 \mu_3^2 f_{23}^* + 3 f_{23}^*f_{33}^* \bigg). 
\end{eqnarray}
For the photon number pattern $(n_2, n_3) = (4, 0)$, the coefficients satisfy
\begin{eqnarray}\label{eq:CE40photon3mode}
\frac{c_3}{c_4} = 2  \mu_2, ~~~~~~ \frac{c_2}{c_4} = \sqrt{3} \big( \mu_2^2 + f_{22}^* \big), ~~~~~~
\frac{c_1}{c_4} = \sqrt{\frac{2}{3}} \mu_2 \big( \mu_2^2 + 3 f_{22}^* \big), ~~~~~~
\frac{c_0}{c_4} = \frac{1}{2\sqrt{6}} \big( \mu_2^4 + 6 \mu_2^2 f_{22}^* + 3 f_{22}^{*2} \big). 
\end{eqnarray}
For the photon number pattern $(n_2, n_3) = (0, 4)$, the coefficients satisfy
\begin{eqnarray}\label{eq:CE04photon3mode}
\frac{c_3}{c_4} = 2  \mu_3, ~~~~~~ \frac{c_2}{c_4} = \sqrt{3} \big( \mu_3^2 + f_{33}^* \big), ~~~~~~
\frac{c_1}{c_4} = \sqrt{\frac{2}{3}} \mu_3 \big( \mu_3^2 + 3 f_{33}^* \big), ~~~~~~
\frac{c_0}{c_4} = \frac{1}{2\sqrt{6}} \big( \mu_3^4 + 6 \mu_3^2 f_{33}^* + 3 f_{33}^{*2} \big). 
\end{eqnarray}

{\bf Detecting five photons}. When the total number of detected photons is $n_T = 5$, there are six possible photon number patterns: 
$(n_2, n_3) = (3, 2)$, $(n_2, n_3) = (2, 3)$, $(n_2, n_3) = (4, 1)$, $(n_2, n_3) = (1, 4)$, $(n_2, n_3) = (5, 0)$ and $(n_2, n_3) = (0, 5)$. The heralded state is in the form 
\begin{eqnarray}\label{eq:FormPN5}
\hat{D}(\alpha_1) \hat{S}(\zeta_1) (c_0 \ket{0} + c_1 \ket{1} + c_2 \ket{2} + c_3 \ket{3} + c_4 \ket{4} +  c_5 \ket{5}).
\end{eqnarray}
For the photon number pattern $(n_2, n_3) = (3, 2)$, the coefficients satisfy
\begin{eqnarray}\label{eq:CE32photon3mode}
\frac{c_4}{c_5} &=& \frac{1}{\sqrt{5}} (3 \mu_2 + 2 \mu_3), \nonumber\\
\frac{c_3}{c_5} &=& \frac{1}{2 \sqrt{5}} \bigg[ (3 \mu_2^2 + 6 \mu_2 \mu_3 + \mu_3^2) + 3\, f_{22}^* + f_{33}^* + 6 \, f_{23}^* \bigg], \nonumber\\
\frac{c_2}{c_5} &=& \frac{1}{ 2 \sqrt{15}} \bigg[ (\mu_2^3 + 6 \mu_2^2 \mu_3 + 3 \mu_2 \mu_3^2) 
+ 3(\mu_2 + 2 \mu_3) f_{22}^* + 3 \mu_2 f_{33}^* + 6 (2 \mu_2 + \mu_3) f_{23}^* \bigg], \nonumber\\
\frac{c_1}{c_5} &=& \frac{1}{2 \sqrt{30}} \bigg[ (2 \mu_2^3 \, \mu_3 + 3 \mu_2^2 \, \mu_3^2) + 3(2 \mu_2 \mu_3+\mu_3^2) f_{22}^* 
+ 3 \mu_2^2 f_{33}^* +  6( \mu_2^2 + 2\mu_2 \mu_3) f_{23}^* + 6\, f_{22}^* f_{23}^*
+ 3\, (f_{22}^* f_{33}^* + 2 \, f_{23}^{*2}) \bigg], \nonumber\\
\frac{c_0}{c_5} &=& \frac{1}{2 \sqrt{30}} \bigg[ \mu_2^3 \, \mu_3^2 + 3 \mu_2 \mu_3^2 f_{22}^* 
+  \mu_2^3 f_{33}^* +  6 \mu_2^2 \mu_3 f_{23}^* + 6 \mu_3 f_{22}^* f_{23}^*
+ 3 \mu_2 (f_{22}^* f_{33}^* + 2 \, f_{23}^{*2}) \bigg]. 
\end{eqnarray}

\section{Measuring subsystems of four-mode Gaussian states}\label{appedix:CoeffFourStates}

In this appendix, we list the coefficients of a two-mode non-Gaussian states by detecting two modes of a four-mode Gaussian state. 

When the detection event is $\bar{\vt{n}} = (1, 1)$, we find:
\begin{eqnarray}\label{eq:Coef11-4mode}
c_{00} &\propto& b_{34}^* + y_3 y_4, ~~~~~~ c_{10} \propto b_{14}^* y_3 + b_{13}^* y_4, ~~~~~~ c_{01} \propto b_{24}^* y_3 + b_{23}^* y_4
\nonumber\\
c_{20} &\propto& \sqrt{2} \, b_{13}^*  b_{14}^*, ~~~~~~ c_{02} \propto \sqrt{2} \, b_{23}^*  b_{24}^*,  ~~~~~~
c_{11} \propto b_{13}^*  b_{24}^* + b_{23}^*  b_{14}^*. 
\end{eqnarray}

When the detection event is $\bar{\vt{n}} = (1, 2)$, we find:
\begin{eqnarray}\label{eq:Coef12-4mode}
c_{00} &\propto& b_{44}^* y_3 + y_4 (2 b_{34}^* + y_3 y_4), 
\nonumber\\
 c_{10} &\propto& 2 \, b_{14}^* ( b_{34}^* + y_3 y_4 ) + b_{13}^* (b_{44}^* + y_4^2), ~~~~~~ 
 c_{01} \propto 2 \, b_{24}^* ( b_{34}^* + y_3 y_4 ) + b_{23}^* (b_{44}^* + y_4^2),
\nonumber\\
c_{20} &\propto& \sqrt{2} \, b_{14}^* (b_{14}^* y_3 + 2 \, b_{13}^* y_4), ~~~~~~ c_{02} \propto \sqrt{2} \, b_{24}^* (b_{24}^* y_3 + 2 \, b_{23}^* y_4), 
\nonumber\\
 c_{11} &\propto& 2 \, b_{14}^*  b_{24}^* y_3 + 2(b_{13}^*  b_{24}^* + b_{23}^*  b_{14}^*)y_4,
 \nonumber\\
 c_{30} &\propto& \sqrt{6} \, b_{13}^* b_{14}^{*2}, ~~~~~~ c_{03} \propto \sqrt{6} \, b_{23}^* b_{24}^{*2},
 \nonumber\\
 c_{21} &\propto& \sqrt{2} \, b_{14}^* (2 \, b_{13}^*  b_{24}^* + b_{23}^*  b_{14}^*), ~~~~~~ c_{12} \propto \sqrt{2} \, b_{24}^* (2 \, b_{23}^*  b_{14}^* + b_{13}^*  b_{24}^*). 
\end{eqnarray}

When the detection event is $\bar{\vt{n}} = (2, 1)$, we find:
\begin{eqnarray}\label{eq:Coef21-4mode}
c_{00} &\propto& b_{33}^* y_4 + y_3 (2 \, b_{34}^* + y_3 y_4), 
\nonumber\\
 c_{10} &\propto& 2 \, b_{13}^* ( b_{34}^* + y_3 y_4 ) + b_{14}^* (b_{33}^* + y_3^2), ~~~~~~ 
 c_{01} \propto 2 \, b_{23}^* ( b_{34}^* + y_3 y_4 ) + b_{24}^* (b_{33}^* + y_3^2),
\nonumber\\
c_{20} &\propto& \sqrt{2} \, b_{13}^* (b_{13}^* y_4 + 2 \, b_{14}^* y_3), ~~~~~~ c_{02} \propto \sqrt{2} \, b_{23}^* (b_{23}^* y_4 + 2 \, b_{24}^* y_3), 
\nonumber\\
 c_{11} &\propto& 2 \, b_{13}^*  b_{23}^* y_4 + 2(b_{23}^*  b_{14}^* + b_{13}^*  b_{24}^*)y_3,
 \nonumber\\
 c_{30} &\propto& \sqrt{6} \, b_{13}^{*2} b_{14}^{*}, ~~~~~~ c_{03} \propto \sqrt{6} \, b_{23}^{*2} b_{24}^{*},
 \nonumber\\
 c_{21} &\propto& \sqrt{2} \, b_{13}^* (2 \, b_{23}^*  b_{14}^* + b_{13}^*  b_{24}^*), ~~~~~~ c_{12} \propto \sqrt{2} \, b_{23}^* (2 \, b_{13}^*  b_{24}^* + b_{23}^*  b_{14}^*). 
\end{eqnarray}

\section{Measuring subsystems of five-mode Gaussian states}\label{appedix:CoeffFiveStates}

In this appendix, we list the coefficients of a two-mode non-Gaussian states by detecting three modes of a five-mode Gaussian state.

When the detection event is $\bar{\vt{n}} = (1, 1, 1)$, we find:
\begin{eqnarray}\label{eq:Coef111-5mode}
c_{00} &\propto& b_{34}^* y_5 + b_{35}^* y_4 + b_{45}^* y_3 + y_3 y_4 y_5, \nonumber\\
c_{10} &\propto& (b_{13}^* b_{45}^* + b_{14}^* b_{35}^* + b_{15}^* b_{34}^*) + (b_{13}^* y_4 y_5 + b_{14}^* y_3 y_5 + b_{15}^* y_3 y_4), \nonumber\\
c_{01} &\propto& (b_{23}^* b_{45}^* + b_{24}^* b_{35}^* + b_{25}^* b_{34}^*) + (b_{23}^* y_4 y_5 + b_{24}^* y_3 y_5 + b_{25}^* y_3 y_4), \nonumber\\
c_{20} &\propto& \sqrt{2} \, (b_{13}^*  b_{14}^* y_5 + b_{13}^*  b_{15}^* y_4 + b_{14}^*  b_{15}^* y_3), \nonumber\\
c_{02} &\propto& \sqrt{2} \, (b_{23}^*  b_{24}^* y_5 + b_{23}^*  b_{25}^* y_4 + b_{24}^*  b_{25}^* y_3), \nonumber\\
c_{11} &\propto& (b_{13}^*  b_{24}^* + b_{23}^*  b_{14}^*) y_5 + (b_{13}^*  b_{25}^* + b_{23}^*  b_{15}^*) y_4 + (b_{14}^*  b_{25}^* + b_{24}^*  b_{15}^*) y_3, \nonumber\\
c_{30} &\propto& \sqrt{6} \, b_{13}^* b_{14}^* b_{15}^*, \nonumber\\
c_{03} &\propto& \sqrt{6} \, b_{23}^* b_{24}^* b_{25}^*, \nonumber\\
c_{21} &\propto& \sqrt{2} \, (b_{13}^* b_{14}^* b_{25}^* + b_{13}^* b_{15}^* b_{24}^* + b_{14}^* b_{15}^* b_{23}^*), \nonumber\\
c_{12} &\propto& \sqrt{2} \, (b_{13}^* b_{24}^* b_{25}^* + b_{14}^* b_{23}^* b_{25}^* + b_{15}^* b_{23}^* b_{24}^*). 
\end{eqnarray}

When the detection event is $\bar{\vt{n}} = (1, 1, 2)$, we find that in the zero and one photon subspace,
\begin{eqnarray}
c_{00} &\propto& b_{34}^* b_{55}^* + 2 \, b_{35}^* b_{45}^* + b_{34}^* y_5^2 + 2 \, b_{35}^* y_4 y_5 + 2 \, b_{45}^* y_3 y_5 + b_{55}^* y_3 y_4 + y_3 y_4 y_5^2, \nonumber\\
c_{10} &\propto& (2 \, b_{15}^* b_{45}^* + b_{14}^* b_{55}^*) y_3 + (2 \, b_{15}^* b_{35}^* + b_{13}^* b_{55}^*) y_4 + 2 (b_{15}^* b_{34}^* + b_{14}^* b_{35}^* + b_{13}^* b_{45}^* ) y_5
+ 2 \, b_{15}^* y_3 y_4 y_5 + b_{14}^* y_3 y_5^2 + b_{13}^* y_4 y_5^2, \nonumber\\
c_{01} &\propto& (2 \, b_{25}^* b_{45}^* + b_{24}^* b_{55}^*) y_3 + (2 \, b_{25}^* b_{35}^* + b_{23}^* b_{55}^*) y_4 + 2 (b_{25}^* b_{34}^* + b_{24}^* b_{35}^* + b_{23}^* b_{45}^* ) y_5
+ 2 \, b_{25}^* y_3 y_4 y_5 + b_{24}^* y_3 y_5^2 + b_{23}^* y_4 y_5^2, \nonumber\\
\end{eqnarray}
and in the two photon subspace,
\begin{eqnarray}
c_{20} &\propto& \sqrt{2} \, b_{15}^{*2} b_{34}^* + b_{13}^* b_{14}^* b_{55}^* + 2 \, b_{15}^* b_{14}^* b_{35}^* + 2 \, b_{15}^* b_{13}^* b_{45}^* 
+ b_{15}^{*2} y_3 y_4 + b_{13}^* b_{14}^* y_5^2 + 2 \, b_{15}^* b_{14}^* y_3 y_5 + 2 \, b_{15}^* b_{13}^* y_4 y_5, \nonumber\\
c_{02} &\propto& \sqrt{2} \, b_{25}^{*2} b_{34}^* + b_{23}^* b_{24}^* b_{55}^* + 2 \, b_{25}^* b_{24}^* b_{35}^* + 2 \, b_{25}^* b_{23}^* b_{45}^* 
+ b_{25}^{*2} y_3 y_4 + b_{23}^* b_{24}^* y_5^2 + 2 \, b_{25}^* b_{24}^* y_3 y_5 + 2 \, b_{25}^* b_{23}^* y_4 y_5, \nonumber\\
c_{11} &\propto& 
b_{13}^*(2 \, b_{25}^* b_{45}^* + b_{24}^* b_{55}^*) + b_{14}^*(2 \, b_{25}^* b_{35}^* + b_{23}^* b_{55}^*) 
+ 2 \, b_{15}^*(b_{24}^* b_{35}^* + b_{23}^* b_{45}^* + b_{25}^* b_{34}^*) + 2 \, b_{15}^* b_{25}^* y_3 y_4 
\nonumber\\
&&
+ 2 (b_{14}^* b_{25}^* + b_{15}^* b_{24}^*) y_3 y_5 + 2 (b_{13}^* b_{25}^* + b_{15}^* b_{23}^*) y_4 y_5 
+ (b_{14}^* b_{23}^* + b_{13}^* b_{24}^*) y_5^2 , %\nonumber
\end{eqnarray}
and in the three photon subspace,
\begin{eqnarray}
c_{30} &\propto& \sqrt{6} \, b_{15}^* (2 \, b_{13}^*b_{14}^* y_5 + b_{13}^* b_{15}^* y_4 + b_{14}^* b_{15}^* y_3 ), \nonumber\\
c_{03} &\propto& \sqrt{6} \, b_{25}^* (2 \, b_{23}^*b_{24}^* y_5 + b_{23}^* b_{25}^* y_4 + b_{24}^* b_{25}^* y_3 ), \nonumber\\
c_{21} &\propto& \sqrt{2} \, b_{15}^*(b_{15}^* b_{24}^* + 2 \, b_{14}^* b_{25}^*) y_3 + b_{15}^* (b_{15}^* b_{23}^* + 2 \, b_{13}^* b_{25}^*) y_4 
+ 2( b_{13}^* b_{14}^* b_{25}^* + b_{14}^* b_{15}^* b_{23}^* + b_{13}^* b_{15}^* b_{24}^* ) y_5, \nonumber\\
c_{12} &\propto& \sqrt{2} \, b_{25}^*(b_{14}^* b_{25}^* + 2 \, b_{15}^* b_{24}^*) y_3 + b_{25}^* (b_{13}^* b_{25}^* + 2 \, b_{15}^* b_{23}^*) y_4 
+ 2 ( b_{13}^* b_{24}^* b_{25}^* + b_{14}^* b_{23}^* b_{25}^* + b_{15}^* b_{23}^* b_{24}^* ) y_5, %\nonumber
\end{eqnarray}
and in the four photon subspace,
\begin{eqnarray}
c_{40} &\propto& 2 \sqrt{6} \, b_{13}^* b_{14}^* b_{15}^{*2}, \nonumber\\
c_{04} &\propto& 2 \sqrt{6} \, b_{23}^* b_{24}^* b_{25}^{*2}, \nonumber\\
c_{31} &\propto& \sqrt{6} \, b_{15}^*(2 \, b_{13}^* b_{14}^* b_{25}^* + b_{13}^* b_{15}^* b_{24}^*+ b_{14}^* b_{15}^* b_{23}^*), \nonumber\\
c_{13} &\propto& \sqrt{6} \, b_{25}^*(2 \, b_{15}^* b_{23}^* b_{24}^* + b_{14}^* b_{23}^* b_{25}^* + b_{13}^* b_{24}^* b_{25}^*), \nonumber\\
c_{22} &\propto& 2(b_{13}^* b_{14}^* b_{25}^{*2} + 2 \, b_{13}^* b_{15}^* b_{24}^*b_{25}^* + 2 \, b_{14}^* b_{15}^* b_{23}^* b_{25}^* + b_{15}^{*2} b_{23}^* b_{24}^*). 
\end{eqnarray}

\section{Measuring subsystems of six-mode Gaussian states}\label{appedix:CoeffSixStates}

In this appendix, we list the coefficients of a two-mode non-Gaussian states by detecting four modes of a six-mode Gaussian state.

When the detection event is $\bar{\vt{n}} = (1, 1, 1, 1)$, we find that in the zero and one photon subspace,
\begin{eqnarray}\label{eq:Coef1111-6mode-01}
c_{00} &\propto& % (cb34 cb56 + cb56 y3 y4 + cb46 y3 y5 + cb36 (cb45 + y4 y5) + cb45 y3 y6 + cb34 y5 y6 + y3 y4 y5 y6 + cb35 (cb46 + y4 y6))
(b_{34}^* b_{56}^* + b_{35}^* b_{46}^* + b_{36}^* b_{45}^*) + b_{34}^* y_5 y_6 + b_{35}^* y_4 y_6 + b_{36}^* y_4 y_5 + b_{45}^* y_3 y_6 + b_{46}^* y_3 y_5 + b_{56}^* y_3 y_4 
+ y_3 y_4 y_5 y_6,
\nonumber\\
c_{10} &\propto& %(cb14 cb56 y3 + cb13 cb56 y4 + cb14 cb36 y5 + cb13 cb46 y5 + cb16 (cb45 y3 + cb35 y4 + cb34 y5 + y3 y4 y5) + cb14 cb35 y6 + cb13 cb45 y6 + cb14 y3 y5 y6 + cb13 y4 y5 y6 + cb15 (cb46 y3 + cb36 y4 + cb34 y6 + y3 y4 y6))
(b_{14}^* b_{56}^* + b_{15}^* b_{46}^* + b_{16}^* b_{45}^*) y_3 + (b_{13}^* b_{56}^* + b_{15}^* b_{36}^* + b_{16}^* b_{35}^* ) y_4 
+ (b_{13}^* b_{46}^* + b_{14}^* b_{36}^* + b_{16}^* b_{34}^*) y_5 
\nonumber\\
&& + (b_{13}^* b_{45}^* + b_{14}^* b_{35}^* + b_{15}^* b_{34}^*) y_6 
+ b_{13}^* y_4 y_5 y_6 + b_{14}^* y_3 y_5 y_6 + b_{15}^* y_3 y_4 y_6 + b_{16}^* y_3 y_4 y_5, 
\nonumber\\
c_{01} &\propto& 
(b_{24}^* b_{56}^* + b_{25}^* b_{46}^* + b_{26}^* b_{45}^*) y_3 + (b_{23}^* b_{56}^* + b_{25}^* b_{36}^* + b_{26}^* b_{35}^* ) y_4 
+ (b_{23}^* b_{46}^* + b_{24}^* b_{36}^* + b_{26}^* b_{34}^*) y_5 
\nonumber\\
&& + (b_{23}^* b_{45}^* + b_{24}^* b_{35}^* + b_{25}^* b_{34}^*) y_6 
+ b_{23}^* y_4 y_5 y_6 + b_{24}^* y_3 y_5 y_6 + b_{25}^* y_3 y_4 y_6 + b_{26}^* y_3 y_4 y_5, 
\end{eqnarray}
and in the two photon subspace,
\begin{eqnarray}
c_{20} &\propto& \sqrt{2} \, \big[
%(cb13 cb16 (cb45 + y4 y5) + cb15 (cb14 cb36 + cb13 cb46 + cb16 (cb34 + y3 y4) + cb14 y3 y6 + cb13 y4 y6) + cb14 (cb16 cb35 + cb13 cb56 + cb16 y3 y5 + cb13 y5 y6))
( b_{13}^* b_{14}^* b_{56}^* +  b_{13}^* b_{15}^* b_{46}^* +  b_{13}^* b_{16}^* b_{45}^* + b_{14}^* b_{15}^* b_{36}^* + b_{14}^* b_{16}^* b_{35}^* + b_{15}^* b_{16}^* b_{34}^*)
\nonumber\\
&&
+ (b_{13}^* b_{14}^* y_5 y_6 + b_{13}^* b_{15}^* y_4 y_6 + b_{13}^* b_{16}^* y_4 y_5 + b_{14}^* b_{15}^* y_3 y_6 + b_{14}^* b_{16}^* y_3 y_5 + b_{15}^* b_{16}^* y_3 y_4)
\big], 
\nonumber\\
c_{02} &\propto& \sqrt{2} \, \big[
( b_{23}^* b_{24}^* b_{56}^* +  b_{23}^* b_{25}^* b_{46}^* +  b_{23}^* b_{26}^* b_{45}^* + b_{24}^* b_{25}^* b_{36}^* + b_{24}^* b_{26}^* b_{35}^* + b_{25}^* b_{26}^* b_{34}^*)
\nonumber\\
&&
+ (b_{23}^* b_{24}^* y_5 y_6 + b_{23}^* b_{25}^* y_4 y_6 + b_{23}^* b_{26}^* y_4 y_5 + b_{24}^* b_{25}^* y_3 y_6 + b_{24}^* b_{26}^* y_3 y_5 + b_{25}^* b_{26}^* y_3 y_4)
\big], 
\nonumber\\
c_{11} &\propto& 
b_{13}^*(b_{24}^* b_{56}^* + b_{25}^* b_{46}^* + b_{26}^* b_{45}^*) + b_{14}^* (b_{23}^* b_{56}^* + b_{25}^* b_{36}^* + b_{26}^* b_{35}^*) 
+ b_{15}^* (b_{23}^* b_{46}^* + b_{24}^* b_{36}^* + b_{26}^* b_{34}^*) 
\nonumber\\
&&
+ b_{16}^* (b_{23}^* b_{45}^* + b_{24}^* b_{35}^* + b_{25}^* b_{34}^*) 
+ b_{13}^*(b_{24}^* y_5 y_6 + b_{25}^* y_4 y_6 + b_{26}^* y_4 y_5) + b_{14}^* (b_{23}^* y_5 y_6 + b_{25}^* y_3 y_6 + b_{26}^* y_3 y_5) 
\nonumber\\
&&
+ b_{15}^* (b_{23}^* y_4 y_6 + b_{24}^* y_3 y_6 + b_{26}^* y_3 y_4) + b_{16}^* (b_{23}^* y_4 y_5 + b_{24}^* y_3 y_5 + b_{25}^* y_3 y_4),
\end{eqnarray}
and in the three photon subspace,
\begin{eqnarray}
c_{30} &\propto& \sqrt{6} \, (b_{14}^* b_{15}^* b_{16}^* y_3 + b_{13}^* b_{15}^* b_{16}^* y_4 + b_{13}^* b_{14}^* b_{16}^* y_5 + b_{13}^* b_{14}^* b_{15}^*  y_6), 
\nonumber\\
c_{03} &\propto& \sqrt{6} \, (b_{24}^* b_{25}^* b_{26}^* y_3 + b_{23}^* b_{25}^* b_{26}^* y_4 + b_{23}^* b_{24}^* b_{26}^* y_5 + b_{23}^* b_{24}^* b_{25}^*  y_6), 
\nonumber\\
c_{21} &\propto& \sqrt{2} \, \big[ 
(b_{14}^* b_{15}^* b_{26}^* + b_{14}^* b_{16}^* b_{25}^* + b_{15}^* b_{16}^* b_{24}^* ) y_3 
+ (b_{13}^* b_{15}^* b_{26}^* + b_{13}^* b_{16}^* b_{25}^* + b_{15}^* b_{16}^* b_{23}^*) y_4
\nonumber\\
&&
+ (b_{13}^* b_{14}^* b_{26}^* + b_{13}^* b_{16}^* b_{24}^* + b_{14}^* b_{16}^* b_{23}^*) y_5
+ (b_{13}^* b_{14}^* b_{25}^* + b_{13}^* b_{15}^* b_{24}^* + b_{14}^* b_{15}^* b_{23}^*) y_6
\big],
\nonumber\\
c_{12} &\propto& \sqrt{2} \, \big[
(b_{14}^* b_{25}^* b_{26}^* + + b_{15}^* b_{24}^* b_{26}^* + b_{16}^* b_{24}^* b_{25}^*  ) y_3 
+ (b_{13}^* b_{25}^* b_{26}^* + b_{15}^* b_{23}^* b_{26}^* + b_{16}^* b_{23}^* b_{25}^*) y_4
\nonumber\\
&&
+ (b_{13}^* b_{24}^* b_{26}^* + b_{14}^* b_{23}^* b_{26}^* +  b_{16}^* b_{23}^*b_{24}^*) y_5
+ (b_{13}^* b_{24}^* b_{25}^* + b_{14}^* b_{23}^* b_{25}^* + b_{15}^* b_{23}^*b_{24}^*) y_6
\big],
\end{eqnarray}
and in the four photon subspace,
\begin{eqnarray}\label{eq:Coef1111-6mode-4}
c_{40} &\propto& 2 \sqrt{6} \, b_{13}^* b_{14}^* b_{15}^* b_{16}^*, \nonumber\\
c_{04} &\propto& 2 \sqrt{6} \, b_{23}^* b_{24}^* b_{25}^* b_{26}^*, \nonumber\\
c_{31} &\propto& \sqrt{6} \, ( b_{13}^* b_{14}^* b_{15}^* b_{26}^* + b_{13}^* b_{14}^* b_{16}^* b_{25}^* + b_{13}^* b_{15}^* b_{16}^* b_{24}^* + b_{14}^* b_{15}^* b_{16}^* b_{23}^*),
\nonumber\\
c_{13} &\propto& \sqrt{6} \, ( b_{13}^* b_{24}^* b_{25}^* b_{26}^* + b_{14}^* b_{23}^* b_{25}^* b_{26}^* + b_{15}^* b_{23}^* b_{24}^* b_{26}^* + b_{16}^* b_{23}^* b_{24}^* b_{25}^*),
 \nonumber\\
c_{22} &\propto& 2 \,
(b_{13}^* b_{14}^* b_{25}^* b_{26}^* + b_{13}^* b_{15}^* b_{24}^* b_{26}^* + b_{13}^* b_{16}^* b_{24}^* b_{25}^* 
+ b_{14}^* b_{15}^* b_{23}^* b_{26}^* + b_{14}^* b_{16}^* b_{23}^* b_{25}^* + b_{15}^* b_{16}^* b_{23}^* b_{24}^*).
\end{eqnarray}

\section{Derivation of Eq. \eqref{eq:Wigner-Ito}}\label{app:Wigner-Ito-Hermite}

In this appendix, we explain how to express the Wigner function in terms of the Ito's 2D-Hermite polynomials, in particular, we derive Eq. \eqref{eq:Wigner-Ito} in detail. 
The 2D-Hermite polynomials are defined as \cite{ismail2017review}
\begin{eqnarray}\label{eq:}
H_{mn}(z, z^*) = \frac{\partial^m}{\partial t_1^m} \frac{\partial^n}{\partial t_2^n} e^{- t_1t_2 + z t_1 + z^* t_2 } \bigg|_{t_1 = t_2 =0},
\end{eqnarray}
where $z$ is a complex number. 

Equation \eqref{eq:FockWF} shows that the wave function of a Fock state $\ket{n}$ is related to the Hermite polynomial $H_n (q)$. 
The generating function of Hermite polynomials is 
\begin{eqnarray}
%\sum_{n=0}^{\infty} \frac{H_n(q)}{n!} t^n = \exp(-t^2 + 2 q t),
\sum_{n=0}^{\infty} \frac{H_n(q)}{n!} t^n = e^{-t^2 + 2 q t},
\end{eqnarray}
so we have 
\begin{eqnarray}
%H_n(q)= \frac{\mathrm{d}^n}{\mathrm d t^n} \exp(-t^2 + 2 q t) \bigg|_{t=0}. 
H_n(q)= \frac{\mathrm{d}^n}{\mathrm d t^n} e^{-t^2 + 2 q t} \bigg|_{t=0}.
\end{eqnarray}
Therefore, the wave function of the Fock state $\ket{n}$ can be written as 
\begin{eqnarray}\label{eq:Fock-Hermite-Generating}
\psi_n(q) = \frac{1}{\pi^{1/4}\sqrt{ 2^n \, n!}} e^{-q^2/2} \frac{\mathrm{d}^n}{\mathrm d t^n} e^{-t^2 + 2 q t} \bigg|_{t=0}.
\end{eqnarray}
By substituting Eqs. \eqref{eq:FockWF} and \eqref{eq:Fock-Hermite-Generating} into Eq. \eqref{eq:Wigner-Ito}, we can calculate $W_{mn}(p, q)$ straightforwardly:
\begin{eqnarray}
W_{mn}(p, q) &=& 
\frac{1}{\sqrt{\pi} \sqrt{ 2^{n+m} \, n! \, m!}} \int \mathrm d y \, e^{- 2i py} \exp\bigg\{ - \frac{1}{2} \big[(q-y)^2 + (q+y)^2\big] \bigg\} H_m(q-y) H_n(q+y) \nonumber\\
&=&
\frac{1}{\sqrt{\pi} \sqrt{ 2^{n+m} \, n! \, m!}} \frac{\partial^m}{\partial t_1^m} \frac{\partial^n}{\partial t_2^n} e^{-t_1^2 - t_2^2}
 \int \mathrm d y \, e^{- 2i py} e^{-q^2 - y^2 + 2 (q-y) t_1 + 2 (q+y) t_2} \bigg|_{t_1 = t_2 =0} \nonumber\\
&=&
\frac{1}{\sqrt{\pi} \sqrt{ 2^{n+m} \, n! \, m!}} \frac{\partial^m}{\partial t_1^m} \frac{\partial^n}{\partial t_2^n} e^{-t_1^2 - t_2^2 + 2 (t_1+t_2) q - q^2}
 \int \mathrm d y \, e^{- y^2 - 2 (t_1-t_2) y - 2i py} \bigg|_{t_1 = t_2 =0} \nonumber\\
&=&
\frac{1}{\sqrt{ 2^{n+m} \, n! \, m!}} e^{-q^2 - p^2} \frac{\partial^m}{\partial t_1^m} \frac{\partial^n}{\partial t_2^n} e^{-2 t_1t_2 + 2 (t_1+t_2) q + 2i(t_1-t_2)p} \bigg|_{t_1 = t_2 =0}
\nonumber\\
&=&
\frac{1}{\sqrt{n! \, m!}} e^{-q^2 - p^2} \frac{\partial^m}{\partial t_1^m} \frac{\partial^n}{\partial t_2^n} e^{- t_1t_2 + \sqrt{2}(q + ip) t_1 + \sqrt{2} (q - ip) t_2 } \bigg|_{t_1 = t_2 =0}
\nonumber\\
&=&
\frac{1}{\sqrt{n! \, m!}} e^{-q^2 - p^2} \frac{\partial^m}{\partial t_1^m} \frac{\partial^n}{\partial t_2^n} e^{- t_1t_2 + 2 \alpha t_1 + 2 \alpha^* t_2 } \bigg|_{t_1 = t_2 =0}
\nonumber\\
&=&
\frac{1}{\sqrt{n! \, m!}} e^{-q^2 - p^2} H_{mn}(2 \alpha, 2 \alpha^*).
\end{eqnarray}
where we have defined $\alpha = (q + ip)/\sqrt{2}$.

\section{Derivation of Eq. \eqref{eq:CmCn-1-single} }\label{app:Fock-Coefficients}

To clarify the calculation, we rewrite the Wigner function \eqref{eq:WignerSignlePure} as 
\begin{eqnarray}\label{eq:WignerSinglePure-full}
W(\alpha; \rho_1) = \mathcal{N}_1 \, e^{- 2|\delta|^2}
\prod_{k=2}^{N} \bigg(\frac{\partial^2}{\partial \alpha_k \partial \beta_k^*} \bigg)^{n_k}
\exp\bigg( \frac{1}{2} \vt{\gamma}_d^{\top} {\bf A} \vt{\gamma}_d + \vt{z}^{\top} \vt{\gamma}_d \bigg) \bigg|_{\vt{\gamma}_d = 0},
\end{eqnarray}
where $\mathcal{N}_1$ is a normalization factor and 
\begin{eqnarray}
\vt{z} &=& {\vt Y} + \frac{2}{\sqrt{1-|b_{11}|^2}} \, {\bf R}_{dh} \vt{w}, \nonumber\\
\vt{w} &=& \begin{pmatrix} \delta^* \\ \delta \end{pmatrix} = \sqrt{1 - |b_{11}|^2} \, ({\bf I}_2 + {\bf X}_2 {\bf R}_{hh})^{-1} 
\bigg[ \begin{pmatrix} \alpha^* \\ \alpha \end{pmatrix} - ({\bf I}_2 - {\bf X}_2 {\bf R}_{hh})^{-1} {\bf X}_2 \vt{y}_h \bigg]. 
\end{eqnarray}
The Fock-state coefficients now can be written as
\begin{eqnarray}
c_m c_n^* &=& \frac{1}{\sqrt{m! n!}} \int \mathrm d^2 \delta \, W(\alpha; \rho_1) H^*_{m n}(2\delta, 2\delta^*) e^{- 2 |\delta|^2}
\nonumber\\
&=&
\frac{\mathcal{N}_1}{\sqrt{m! n!}} %\, e^{- 2|\delta|^2}
\prod_{k=2}^{N} \bigg(\frac{\partial^2}{\partial \alpha_k \partial \beta_k^*} \bigg)^{n_k}
\exp\bigg( \frac{1}{2} \vt{\gamma}_d^{\top} {\bf A} \vt{\gamma}_d + \vt{Y}^{\top} \vt{\gamma}_d \bigg) 
\nonumber\\
&&
\times
\int \mathrm d^2 \delta \, \exp\bigg( \frac{2}{\sqrt{1 - |b_{11}|^2}} \vt{w}^{\top} {\bf R}_{hd} \vt{\gamma}_d \bigg)  H^*_{m n}(2\delta, 2\delta^*) e^{- 4 |\delta|^2}
\bigg|_{\vt{\gamma}_d = 0}
\nonumber\\
&=&
\frac{\mathcal{N}_1}{\sqrt{m! n!}} %\, e^{- 2|\delta|^2}
\prod_{k=2}^{N} \bigg(\frac{\partial^2}{\partial \alpha_k \partial \beta_k^*} \bigg)^{n_k}
\exp\bigg( \frac{1}{2} \vt{\gamma}_d^{\top} {\bf A} \vt{\gamma}_d + \vt{Y}^{\top} \vt{\gamma}_d \bigg) 
\frac{\partial^m}{\partial t_1^m}  \frac{\partial^n}{\partial s_1^n} \, e^{-t_1 s_1} 
\nonumber\\
&&
\times
\int \mathrm d^2 \delta \, \exp\bigg( \frac{2}{\sqrt{1 - |b_{11}|^2}} \vt{w}^{\top} {\bf R}_{hd} \vt{\gamma}_d + 2 \, \delta^* t_1 + 2 \, \delta s_1 \bigg) e^{- 4 |\delta|^2}
\bigg|_{\vt{\gamma}_d = 0, \, t_1=s_1 = 0} 
\nonumber\\
&=&
\frac{\mathcal{N}_1}{\sqrt{m! n!}} %\, e^{- 2|\delta|^2}
\prod_{k=2}^{N} \bigg(\frac{\partial^2}{\partial \alpha_k \partial \beta_k^*} \bigg)^{n_k}
\exp\bigg( \frac{1}{2} \vt{\gamma}_d^{\top} {\bf A} \vt{\gamma}_d + \vt{Y}^{\top} \vt{\gamma}_d \bigg) 
\frac{\partial^m}{\partial t_1^m}  \frac{\partial^n}{\partial s_1^n} \, e^{-t_1 s_1} 
\nonumber\\
&&
\times
\int \mathrm d^2 \delta \, \exp\bigg[ 2 \, \vt{w}^{\top} \bigg( \vt{t} + \frac{1}{\sqrt{1 - |b_{11}|^2}}  {\bf R}_{hd} \vt{\gamma}_d \bigg) \bigg] e^{- 2 \, \vt{w}^{\top} {\bf X}_2 \vt{w} }
\bigg|_{\vt{\gamma}_d = 0, \, t_1=s_1 = 0},
\end{eqnarray}
where in the last equality we have defined a vector $\vt{t} = (t_1, s_1)^{\top}$. The integration over $\delta$ is a Gaussian integration and can be integrated
straightforwardly. We have
\begin{eqnarray}
c_m c_n^* &=&
\frac{\pi \, \mathcal{N}_1}{4 \sqrt{m! n!}} %\, e^{- 2|\delta|^2}
\prod_{k=2}^{N} \bigg(\frac{\partial^2}{\partial \alpha_k \partial \beta_k^*} \bigg)^{n_k}
\exp\bigg( \frac{1}{2} \vt{\gamma}_d^{\top} {\bf A} \vt{\gamma}_d + \vt{Y}^{\top} \vt{\gamma}_d \bigg) 
\frac{\partial^m}{\partial t_1^m}  \frac{\partial^n}{\partial s_1^n} \, e^{-t_1 s_1} 
\nonumber\\
&&
\times
\exp\bigg[ \frac{1}{2} \bigg( \vt{t}^{\top} + \frac{1}{\sqrt{1 - |b_{11}|^2}} \vt{\gamma}_d^{\top}  {\bf R}_{dh} \bigg) {\bf X}_2  \bigg( \vt{t} + \frac{1}{\sqrt{1 - |b_{11}|^2}}  {\bf R}_{hd} \vt{\gamma}_d \bigg) \bigg] 
\bigg|_{\vt{\gamma}_d = 0, \, t_1=s_1 = 0}
\nonumber\\
&=&
\frac{\pi \, \mathcal{N}_1}{4 \sqrt{m! n!}} %\, e^{- 2|\delta|^2}
\prod_{k=2}^{N} \bigg(\frac{\partial^2}{\partial \alpha_k \partial \beta_k^*} \bigg)^{n_k}
\exp\bigg( \frac{1}{2} \vt{\gamma}_d^{\top} {\bf A} \vt{\gamma}_d + \vt{Y}^{\top} \vt{\gamma}_d \bigg) 
\nonumber\\
&&
\times
\frac{\partial^m}{\partial t_1^m} \frac{\partial^n}{\partial s_1^n}  
\exp\bigg[ \frac{1}{2(1 - |b_{11}|^2)} \vt{\gamma}_d^{\top}  {\bf R}_{dh} {\bf X}_2 {\bf R}_{hd} \vt{\gamma}_d + \frac{1}{\sqrt{1 - |b_{11}|^2}} \vt{\gamma}_d^{\top}  {\bf R}_{dh} {\bf X}_2 \vt{t} \bigg] 
\bigg|_{\vt{\gamma}_d = 0, \, t_1=s_1 = 0}
\nonumber\\
&=&
\frac{\pi \, \mathcal{N}_1}{4 \sqrt{m! n!}} %\, e^{- 2|\delta|^2}
\prod_{k=2}^{N} \bigg(\frac{\partial^2}{\partial \alpha_k \partial \beta_k^*} \bigg)^{n_k}
\exp\bigg( \frac{1}{2} \vt{\gamma}_d^{\top} {\bf C} \vt{\gamma}_d + \vt{Y}^{\top} \vt{\gamma}_d \bigg)
\bigg( \sum_{j=2}^N \kappa_j^* \alpha_j \bigg)^m \bigg(  \sum_{i=2}^N \kappa_i \beta_i^* \bigg)^n
\bigg|_{\vt{\gamma}_d = 0},
%\nonumber\\
\end{eqnarray}
where in the last equality we have used the relation 
\begin{eqnarray}
 \frac{1}{\sqrt{1 - |b_{11}|^2}} \vt{\gamma}_d^{\top}  {\bf R}_{dh} = \bigg(  \sum_{i=2}^N \kappa_i \beta_i^*, \, \sum_{j=2}^N \kappa_j^* \alpha_j \bigg)^{\top}
\end{eqnarray}
and defined a matrix ${\bf C}$ as
\begin{eqnarray}
{\bf C} &=& {\bf A} + \frac{1}{(1 - |b_{11}|^2)}{\bf R}_{dh} {\bf X}_2 {\bf R}_{hd}
\nonumber \\
&=&
{\bf R}_{dd} + \frac{1}{(1 - |b_{11}|^2)} {\bf R}_{dh} {\bf X}_2 {\bf R}_{hd} - {\bf R}_{dh} ({\bf I}_2 + {\bf X}_2 {\bf R}_{hh} )^{-1} {\bf X}_2 {\bf R}_{hd}
\nonumber\\
&=&
{\bf R}_{dd} + \frac{1}{(1 - |b_{11}|^2)} {\bf R}_{dh} 
	\begin{pmatrix}
	b_{11}^* & 0\\
	0 & b_{11}
	\end{pmatrix}
{\bf R}_{hd}. 
\end{eqnarray}

\section{Derivation of Eq.~\eqref{eq:CoeffWignerHermite-Mmode-2} }\label{app:Fock-Coefficients-Mmode}

We rewrite the Wigner function \eqref{eq:Wigner-Mmode-general-pure} as 
\begin{eqnarray}\label{eq:WignerMmodePure-full}
W(\vt{\alpha}; \rho_M) = \mathcal{N} \, e^{- 2|\vt{\delta}|^2}
\prod_{k=M+1}^{N} \bigg(\frac{\partial^2}{\partial \alpha_k \partial \beta_k^*} \bigg)^{n_k}
\exp\bigg( \frac{1}{2} \vt{\gamma}_d^{\top} {\bf A} \vt{\gamma}_d + \vt{z}^{\top} \vt{\gamma}_d \bigg) \bigg|_{\vt{\gamma}_d = {\bf 0}},
\end{eqnarray}
where $\mathcal{N}$ is a normalization factor and 
\begin{eqnarray}
\vt{z} &=& \vt{Y} + 2\, {\bf R}_{dh} {\bf T}^{-1}_{2M} \vt{w}, \nonumber\\
\vt{w} &=& \begin{pmatrix} \vt{\delta}^* \\ \vt{\delta} \end{pmatrix} = {\bf T}_{2M} ({\bf I}_{2M} + {\bf X}_{2M} {\bf R}_{hh})^{-1} 
\bigg[ \begin{pmatrix} \vt{\alpha}^* \\ \vt{\alpha} \end{pmatrix} - ({\bf I}_{2M} - {\bf X}_{2M} {\bf R}_{hh})^{-1} {\bf X}_{2M} \vt{y}_h \bigg]. 
\end{eqnarray}
The Fock-state coefficients now can be written as
\begin{eqnarray}
c_{\vt{\ell}} \, c_{\vt{m}}^* &=& \frac{1}{\sqrt{\vt{\ell}\,! \, \vt{m}!}} \int \mathrm d^2 \vt{\delta} \, W(\vt{\alpha}; \rho_M) e^{-2|\vt{\delta}|^2} 
\prod_{k=1}^M H_{\ell_k m_k}(2 \delta_k, 2 \delta_k^*) 
\nonumber\\
&=&
\frac{\mathcal{N}}{\sqrt{\vt{\ell}! \vt{m}!}} %\, e^{- 2|\delta|^2}
\prod_{k=M+1}^{N} \bigg(\frac{\partial^2}{\partial \alpha_k \partial \beta_k^*} \bigg)^{n_k}
\exp\bigg( \frac{1}{2} \vt{\gamma}_d^{\top} {\bf A} \vt{\gamma}_d + \vt{Y}^{\top} \vt{\gamma}_d \bigg) 
\nonumber\\
&&
\times
\int \mathrm d^2 \vt{\delta} \, \exp\bigg( 2 \, \vt{w}^{\top}  {\bf T}^{-\top}_{2M} {\bf R}_{hd} \vt{\gamma}_d \bigg)  
\prod_{k=1}^M H_{\ell_k m_k}(2 \delta_k, 2 \delta_k^*) e^{- 4 |\vt{\delta}|^2} \bigg|_{\vt{\gamma}_d = {\bf 0}}
\nonumber\\
&=&
\frac{\mathcal{N}}{\sqrt{\vt{\ell}! \vt{m}!}} %\, e^{- 2|\delta|^2}
\prod_{k=M+1}^{N} \bigg(\frac{\partial^2}{\partial \alpha_k \partial \beta_k^*} \bigg)^{n_k}
\exp\bigg( \frac{1}{2} \vt{\gamma}_d^{\top} {\bf A} \vt{\gamma}_d + \vt{Y}^{\top} \vt{\gamma}_d \bigg) 
\bigg( \prod_{k=1}^M \frac{\partial^{\ell_k}}{\partial t_k^{\ell_k}}  \frac{\partial^{m_k}}{\partial s_k^{m_k}} \bigg) \, e^{- \vt{t}^{\top} \vt{s}} 
\nonumber\\
&&
\times
\int \mathrm d^2 \vt{\delta} \, \exp\bigg[ 2 \, \vt{w}^{\top}  \bigg( \vt{u}  + {\bf T}^{-\top}_{2M} {\bf R}_{hd} \vt{\gamma}_d \bigg) \bigg] e^{- 2 \, \vt{w}^{\top} {\bf X}_{2M} \vt{w}}
\bigg|_{\vt{\gamma}_d = {\bf 0}, \, \vt{u} = {\bf 0}}. 
\end{eqnarray}
where in the last equality we have defined a vector $\vt{u} = (t_1, \cdots, t_M, s_1, \cdots, s_M)^{\top}$. The integration over $\vt{\delta}$ is a Gaussian integration and can be integrated
straightforwardly. We have
\begin{eqnarray}
c_{\vt{\ell}} \, c_{\vt{m}}^* &=&
\frac{\pi^M \, \mathcal{N}}{4^M \sqrt{\vt{\ell}! \vt{m}!}} %\, e^{- 2|\delta|^2}
\prod_{k=M+1}^{N} \bigg(\frac{\partial^2}{\partial \alpha_k \partial \beta_k^*} \bigg)^{n_k}
\exp\bigg( \frac{1}{2} \vt{\gamma}_d^{\top} {\bf A} \vt{\gamma}_d + \vt{Y}^{\top} \vt{\gamma}_d \bigg) 
\bigg( \prod_{k=1}^M \frac{\partial^{\ell_k}}{\partial t_k^{\ell_k}}  \frac{\partial^{m_k}}{\partial s_k^{m_k}} \bigg) \, e^{- \vt{t}^{\top} \vt{s}}
\nonumber\\
&&
\times
\exp\bigg[ \frac{1}{2} \bigg( \vt{u}^{\top} + \vt{\gamma}_d^{\top}  {\bf R}_{dh} {\bf T}^{-1}_{2M} \bigg) {\bf X}_{2M}  \bigg( \vt{u} +  {\bf T}^{-\top}_{2M} {\bf R}_{hd} \vt{\gamma}_d \bigg) \bigg] 
\bigg|_{\vt{\gamma}_d = {\bf 0}, \, \vt{u} = {\bf 0}}
\nonumber\\
&=&
\frac{\pi^M \, \mathcal{N}}{4^M \sqrt{\vt{\ell}! \vt{m}!}} %\, e^{- 2|\delta|^2}
\prod_{k=M+1}^{N} \bigg(\frac{\partial^2}{\partial \alpha_k \partial \beta_k^*} \bigg)^{n_k}
\exp\bigg( \frac{1}{2} \vt{\gamma}_d^{\top} {\bf A} \vt{\gamma}_d + \vt{Y}^{\top} \vt{\gamma}_d \bigg) 
\nonumber\\
&&
\times
\bigg( \prod_{k=1}^M \frac{\partial^{\ell_k}}{\partial t_k^{\ell_k}}  \frac{\partial^{m_k}}{\partial s_k^{m_k}} \bigg)
\exp\bigg( \vt{u}^{\top} {\bf X}_{2M} {\bf T}^{-\top}_{2M} {\bf R}_{hd} \vt{\gamma}_d 
%+ \frac{1}{2} \vt{\gamma}_d^{\top}  {\bf R}_{dh} {\bf T}^{-1}_{2M} {\bf X}_{2M} \vt{u}
+ \frac{1}{2} \vt{\gamma}_d^{\top}  {\bf R}_{dh} {\bf T}^{-1}_{2M} {\bf X}_{2M} {\bf T}^{-\top}_{2M} {\bf R}_{hd} \vt{\gamma}_d \bigg)
\bigg|_{\vt{\gamma}_d = {\bf 0}, \, \vt{u} = {\bf 0}}
\nonumber\\
&=&
\frac{\pi^M \mathcal{N} }{ 4^M \sqrt{\vt{\ell}\,! \vt{m}!}} \, 
\prod_{k=1}^M \bigg( \frac{\partial^{\ell_k}}{\partial t_k^{\ell_k}}  \frac{\partial^{m_k}}{\partial s_k^{m_k}} \bigg)
\prod_{k=M+1}^{N} \bigg(\frac{\partial^2}{\partial \alpha_k \partial \beta_k^*} \bigg)^{n_k}
\exp\bigg[ \frac{1}{2} (\vt{u}^{\top}, \vt{\gamma}_d^{\top}) \, {\bf M} \begin{pmatrix} \vt{u} \\ \vt{\gamma}_d \end{pmatrix} + \vt{Y}^{\top} \vt{\gamma}_d \bigg] 
\bigg|_{\vt{\gamma}_d = \vt{0}, \vt{u}=\vt{0}}, 
\end{eqnarray}
where we have defined a matrix ${\bf M}$ as
\begin{eqnarray}
{\bf M} = 
	\begin{pmatrix}
	{\bf 0} &  {\bf X}_{2M} {\bf T}^{-\top}_{2M} {\bf R}_{hd} \\
	\\
	{\bf R}_{dh} {\bf T}^{-1}_{2M} {\bf X}_{2M} & {\bf A} + \frac{1}{2} {\bf R}_{dh} {\bf T}^{-1}_{2M} {\bf X}_{2M} {\bf T}^{-\top}_{2M} {\bf R}_{hd}
	\end{pmatrix}
&\equiv&
	\begin{pmatrix}
	{\bf 0} & {\bf 0} & {\bf 0} & {\bf C}_1  \\
	%\\
	{\bf 0} & {\bf 0} & {\bf C}_1^* & {\bf 0} \\
	%\\
	{\bf 0} & {\bf C}_1^{\dag} & {\bf C}_2 & \vt{0} \\
	%\\
	{\bf C}_1^{\top} & {\bf 0} & {\bf 0} & {\bf C}_2^*
	\end{pmatrix}, 
\end{eqnarray}
with ${\bf C}_1$ and ${\bf C}_2$ given by 
\begin{eqnarray}
{\bf C}_1 &=& ({\bf I}_M - {\bf B}_{hh}^*{\bf B}_{hh})^{-1/2} {\bf B}_{hd}^*, \nonumber\\
{\bf C}_2 &=& {\bf B}_{dd} + {\bf B}_{dh} ({\bf I}_M - {\bf B}_{hh}^*{\bf B}_{hh})^{-1} {\bf B}_{hh}^* {\bf B}_{hd}. 
\end{eqnarray}

\section{Derivation of Wigner function}\label{app:Wigner-Function-Mmode}

In this appendix, we provide details of deriving the Wigner function for the multimode output case, namely, to derive Eq.~\eqref{eq:Wigner-Mmode-general} from Eq.~\eqref{eq:Wigner-Mmode-2}. The single-mode output case can be obtained by setting $M=1$. 
To perform the integration over $\vt{\alpha}_M$ and $\vt{\beta}_M$, we extract the part that is only relevant to $\vt{\alpha}_M$ and $\vt{\beta}_M$ in Eq. \eqref{eq:Wigner-Mmode-2}, 
which is basically a Gaussian function. If we define $\vt{v}_1 = \vt{y}_h + {\bf R}_{hd} \vt{\gamma}_d$ and $\vt{v}_2 = 2 (\vt{\alpha}^*,\vt{\alpha})^{\top}$, the exponential of the integrand becomes
\begin{eqnarray}
&&- |\vt{\gamma}_h|^2 + \frac{1}{2} \vt{\gamma}_h^{\top} {\bf R}_{hh} \vt{\gamma}_h - \vt{\alpha}_M^{*\top} \vt{\beta}_M + 2 \, (\vt{\alpha}^{\top} \vt{\alpha}_M^*+\vt{\alpha}^{*\top} \vt{\beta}_M)
+ \vt{\gamma}_h^{\top} \vt{y}_h + \vt{\gamma}_h^{\top} {\bf R}_{hd} \vt{\gamma}_d  \nonumber\\
&=& - \frac{1}{2} (\vt{\gamma}_h^{\top}, \vt{\gamma}_h^{\dag})
	\begin{pmatrix}
	-{\bf R}_{hh} & {\bf I}_{2M} \\
	{\bf I}_{2M} & {\bf X}_{2M}
	\end{pmatrix}
	\begin{pmatrix}
	\vt{\gamma}_h \\
	\vt{\gamma}_h^*
	\end{pmatrix}
+ (\vt{\gamma}_h^{\top}, \vt{\gamma}_h^{\dag})
	\begin{pmatrix}
	\vt{v}_1 \\
	\vt{v}_2
	\end{pmatrix} \nonumber\\
&=& - \frac{1}{2} \vt{\Gamma}^{\top}_h
	\begin{pmatrix}
	- {\bf R}_{hh} & {\bf I}_{2M} \\
	{\bf I}_{2M} & {\bf X}_{2M}
	\end{pmatrix}
	\vt{\Gamma}_h 
+ \frac{1}{2} (\vt{v}_1^{\top}, \vt{v}_2^{\top})
\begin{pmatrix}
	- {\bf R}_{hh} & {\bf I}_{2M} \\
	{\bf I}_{2M} & {\bf X}_{2M}
	\end{pmatrix}^{-1}
	\begin{pmatrix}
	\vt{v}_1 \\
	\vt{v}_2
	\end{pmatrix} \nonumber\\
&=&  - \frac{1}{2} \vt{\Gamma}^{\top}_h
	\begin{pmatrix}
	-{\bf R}_{hh} & {\bf I}_{2M} \\
	{\bf I}_{2M} & {\bf X}_{2M}
	\end{pmatrix}
	\vt{\Gamma}_h 
+ \frac{1}{2} (\vt{v}_1^{\top}, \vt{v}_2^{\top})
\begin{pmatrix}
	-({\bf R}_{hh} + {\bf X}_{2M} )^{-1}  &  ({\bf I}_{2M} + {\bf X}_{2M}{\bf R}_{hh})^{-1} \\
	%({\bf I}_{2M} + {\bf R}_{hh}{\bf X}_{2M} )^{-1} & {\bf R}_{hh} ({\bf I}_{2M} + {\bf X}_{2M}{\bf R}_{hh})^{-1}
	({\bf I}_{2M} + {\bf R}_{hh}{\bf X}_{2M} )^{-1} & {\bf X}_{2M} - {\bf X}_{2M} ({\bf I}_{2M} + {\bf X}_{2M}{\bf R}_{hh})^{-1}
	\end{pmatrix}
	\begin{pmatrix}
	\vt{v}_1 \\
	\vt{v}_2
	\end{pmatrix}, 
\end{eqnarray}
where we have introduced
\begin{eqnarray}
\vt{\Gamma}_h =
	\begin{pmatrix}
	\vt{\gamma}_h \\
	\vt{\gamma}_h^*
	\end{pmatrix} - 
	\begin{pmatrix}
	- {\bf R}_{hh} & {\bf I}_{2M} \\
	{\bf I}_{2M} & {\bf X}_{2M}
	\end{pmatrix}^{-1}
	\begin{pmatrix}
	\vt{v}_1 \\
	\vt{v}_2
	\end{pmatrix}.
\end{eqnarray}
Accprding to the Gaussian integration formula, the integration over $\vt{\alpha}_M$ and $\vt{\beta}_M$ gives
\begin{eqnarray}
\pi^{2M} \bigg[\text{det}
\begin{pmatrix}
	-{\bf R}_{hh} & {\bf I}_{2M} \\
	{\bf I}_{2M} & {\bf X}_{2M}
	\end{pmatrix} \bigg]^{-1/2}
\exp\bigg\{ 
\frac{1}{2} (\vt{v}_1^{\top}, \vt{v}_2^{\top})
\begin{pmatrix}
	-({\bf R}_{hh} + {\bf X}_{2M} )^{-1}  &  ({\bf I}_{2M} + {\bf X}_{2M} {\bf R}_{hh})^{-1} \\
	({\bf I}_{2M} + {\bf R}_{hh}{\bf X}_{2M} )^{-1} & {\bf X}_{2M} - {\bf X}_{2M} ({\bf I}_{2M} + {\bf X}_{2M} {\bf R}_{hh})^{-1}
	\end{pmatrix}
	\begin{pmatrix}
	\vt{v}_1 \\
	\vt{v}_2
	\end{pmatrix}
\bigg\}. \nonumber\\
\end{eqnarray}

\begin{eqnarray}\label{eq:WigExponentV-Mmode}
&&\frac{1}{2} (\vt{v}_1^{\top}, \vt{v}_2^{\top})
\begin{pmatrix}
	-({\bf R}_{hh} + {\bf X}_{2M} )^{-1}  &  ({\bf I}_{2M} + {\bf X}_{2M} {\bf R}_{hh})^{-1} \\
	({\bf I}_{2M} + {\bf R}_{hh}{\bf X}_{2M} )^{-1} & {\bf X}_{2M} - {\bf X}_{2M} ({\bf I}_{2M} + {\bf X}_{2M} {\bf R}_{hh})^{-1}
	\end{pmatrix}
	\begin{pmatrix}
	\vt{v}_1 \\
	\vt{v}_2
	\end{pmatrix} \nonumber\\
&=&
- \frac{1}{2} \vt{v}_1^{\top} ({\bf R}_{hh} + {\bf X}_{2M} )^{-1} \vt{v}_1
+ \frac{1}{2} \vt{v}_1^{\top} ({\bf I}_{2M} + {\bf X}_{2M} {\bf R}_{hh})^{-1} \vt{v}_2 
+ \frac{1}{2} \vt{v}_2^{\top} ({\bf I}_{2M} + {\bf R}_{hh}{\bf X}_{2M} )^{-1} \vt{v}_1 \nonumber\\
&&
+  \frac{1}{2} \vt{v}_2^{\top} \big[ {\bf X}_{2M} - {\bf X}_{2M} ({\bf I}_{2M} + {\bf X}_{2M} {\bf R}_{hh})^{-1} \big] \vt{v}_2 \nonumber\\
&=&
- \frac{1}{2} \vt{\gamma}_d^{\top} {\bf R}_{dh} ({\bf R}_{hh} + {\bf X}_{2M} )^{-1} {\bf R}_{hd} \vt{\gamma}_d
- \vt{\gamma}_d^{\top} {\bf R}_{dh} \big[ ({\bf R}_{hh} + {\bf X}_{2M} )^{-1} \vt{y}_h - ({\bf I}_{2M} + {\bf X}_{2M} {\bf R}_{hh})^{-1} \vt{v}_2  \big] \nonumber\\
&&
- \frac{1}{2} \vt{v}_2^{\top} \big[ {\bf X}_{2M} ({\bf I}_{2M} + {\bf X}{\bf R}_{hh})^{-1} - {\bf X}_{2M} \big] \vt{v}_2
+ \vt{v}_2^{\top} ({\bf I}_{2M} + {\bf R}_{hh}{\bf X}_{2M} )^{-1} \vt{y}_h % \nonumber\\
- \frac{1}{2} \vt{y}_h^{\top} ({\bf R}_{hh} + {\bf X}_{2M} )^{-1} \vt{y}_h. \nonumber\\
&=&
- \frac{1}{2} \vt{\gamma}_d^{\top} {\bf R}_{dh} ({\bf I}_{2M} + {\bf X}_{2M} {\bf R}_{hh} )^{-1} {\bf X}_{2M} {\bf R}_{hd} \vt{\gamma}_d
+ \vt{\gamma}_d^{\top} {\bf R}_{dh} ({\bf I}_{2M} + {\bf X}_{2M} {\bf R}_{hh})^{-1} \big( \vt{v}_2 - {\bf X}_{2M} \vt{y}_h \big)  \nonumber\\
&&
- \frac{1}{2} \vt{v}_2^{\top} \big[ {\bf X}_{2M} ({\bf I}_{2M} + {\bf X}_{2M} {\bf R}_{hh})^{-1} - {\bf X}_{2M} \big] \vt{v}_2
+ \vt{v}_2^{\top} ({\bf I}_{2M} + {\bf R}_{hh}{\bf X}_{2M} )^{-1} \vt{y}_h % \nonumber\\
- \frac{1}{2} \vt{y}_h^{\top} ({\bf I}_{2M} + {\bf X}_{2M} {\bf R}_{hh} )^{-1} {\bf X}_{2M} \vt{y}_h. \nonumber\\
\end{eqnarray}
Now the unnormalized Wigner function can be written as 
\begin{eqnarray}\label{eq:Wigner-2-multimode}
W(\vt{\alpha}; \tilde{\rho}_{M}) &=& 
\frac{2^M \mathcal{ P}_0}{\pi^{M} \, \bar{\vt{n}}!}
\bigg[\text{det}
\begin{pmatrix}
	-{\bf R}_{hh} & {\bf I}_{2M} \\
	{\bf I}_{2M} & {\bf X}_{2M}
	\end{pmatrix} \bigg]^{-1/2}
\exp \bigg\{ - \frac{1}{2} \vt{y}_h^{\top} ({\bf I}_{2M} + {\bf X}_{2M} {\bf R}_{hh} )^{-1} {\bf X}_{2M} \vt{y}_h \bigg\} \nonumber\\
&&
\times \exp\bigg\{  
- \frac{1}{2} \vt{v}_2^{\top} \big[ {\bf X}_{2M} ({\bf I}_{2M} + {\bf X}_{2M} {\bf R}_{hh})^{-1} - {\bf X}_{2M}/2 \big] \vt{v}_2
+ \vt{v}_2^{\top} ({\bf I}_{2M} + {\bf R}_{hh}{\bf X}_{2M} )^{-1} \vt{y}_h
\bigg\} \nonumber\\
&&
\times %\int \mathrm d^2 \bar{\vt{\alpha}} \int \mathrm d^2 \bar{\vt{\beta}} ~
\prod_{k=M+1}^{N} \bigg(\frac{\partial^2}{\partial \alpha_k \partial \beta_k^*} \bigg)^{n_k}
\exp\bigg( \frac{1}{2} \vt{\gamma}_d^{\top} {\bf A} \vt{\gamma}_d + \vt{z}^{\top} \vt{\gamma}_d \bigg) \bigg|_{\vt{\gamma}_d = 0},
\end{eqnarray}
where we have defined
\begin{eqnarray}\label{eq:DFAz}
{\bf A} &=& {\bf R}_{dd} - {\bf R}_{dh} ({\bf I}_{2M} + {\bf X}_{2M} {\bf R}_{hh} )^{-1} {\bf X}_{2M} {\bf R}_{hd}, \nonumber\\
\vt{z} &=& \vt{y}_d - {\bf R}_{dh} ({\bf I}_{2M} + {\bf X}_{2M} {\bf R}_{hh})^{-1} {\bf X}_{2M} \vt{y}_h + {\bf R}_{dh} ({\bf I}_{2M} + {\bf X}_{2M} {\bf R}_{hh})^{-1} \vt{v}_2. \nonumber
\end{eqnarray}

The Wigner function can be further simplified. It is observed that 
\begin{eqnarray}
({\bf I}_{2M} + {\bf X}_{2M} {\bf R}_{hh})^{-1} - {\bf I}_{2M}/2 = \frac{1}{2} ({\bf I}_{2M} + {\bf X}_{2M} {\bf R}_{hh})^{-1} ({\bf I}_{2M} - {\bf X}_{2M} {\bf R}_{hh}),
\end{eqnarray}
and therefore
\begin{eqnarray}
&& - \frac{1}{2} \vt{v}_2^{\top} \big[ {\bf X}_{2M} ({\bf I}_{2M} + {\bf X}_{2M} {\bf R}_{hh})^{-1} - {\bf X}_{2M}/2 \big] \vt{v}_2 + \vt{v}_2^{\top} ({\bf I}_{2M} 
+ {\bf R}_{hh}{\bf X}_{2M} )^{-1} \vt{y}_h \nonumber\\
&=&
- \frac{1}{4} \vt{v}_2^{\top} {\bf X}_{2M} ({\bf I}_{2M} + {\bf X}_{2M} {\bf R}_{hh})^{-1} ({\bf I}_{2M} - {\bf X}_{2M} {\bf R}_{hh}) \vt{v}_2 
+ \frac{1}{2} \vt{v}_2^{\top} ({\bf I}_{2M} + {\bf R}_{hh}{\bf X}_{2M} )^{-1} \vt{y}_h + \frac{1}{2} \vt{y}_h^{\top} ({\bf I}_{2M} + {\bf X}_{2M} {\bf R}_{hh})^{-1} \vt{v}_2 \nonumber\\
&=&
- \frac{1}{4} \bigg[ \vt{v}_2 - 2 ({\bf I}_{2M} - {\bf X}_{2M} {\bf R}_{hh})^{-1} {\bf X}_{2M} \vt{y}_h \bigg]^{\top} {\bf X}_{2M} ({\bf I}_{2M} 
+ {\bf X}_{2M} {\bf R}_{hh})^{-1} ({\bf I}_{2M} - {\bf X}_{2M} {\bf R}_{hh}) \nonumber\\
&&
\times \bigg[ \vt{v}_2 - 2  ({\bf I}_{2M} - {\bf X}_{2M} {\bf R}_{hh})^{-1} {\bf X}_{2M} \vt{y}_h \bigg]
\nonumber\\
&&
+ \vt{y}_h^{\top} ({\bf I}_{2M} - {\bf X}_{2M} {\bf R}_{hh})^{-1} ({\bf I}_{2M} + {\bf X}_{2M} {\bf R}_{hh})^{-1} {\bf X}_{2M} \vt{y}_h \nonumber\\
&=&
- \vt{v}^{\top} {\bf X}_{2M} ({\bf I}_{2M} + {\bf X}_{2M} {\bf R}_{hh})^{-1} ({\bf I}_{2M} - {\bf X}_{2M} {\bf R}_{hh}) \vt{v} 
+ \vt{y}_h^{\top} ({\bf I}_{2M} - {\bf X}_{2M} {\bf R}_{hh})^{-1} ({\bf I}_{2M} + {\bf X}_{2M} {\bf R}_{hh})^{-1} {\bf X} \vt{y}_h,
\end{eqnarray}
where we have defined 
\begin{eqnarray}\label{eq:WignerVariable-M}
\vt{v} = \frac{1}{2}\vt{v}_2 - ({\bf I}_{2M} - {\bf X}_{2M} {\bf R}_{hh})^{-1} {\bf X}_{2M} \vt{y}_h =
	\begin{pmatrix}
	\vt{\alpha}^* \\
	\vt{\alpha}
	\end{pmatrix}
- ({\bf I}_{2M} - {\bf X}_{2M} {\bf R}_{hh})^{-1} {\bf X}_{2M} \vt{y}_h. 
\end{eqnarray}
By using the Schur's determinant identity, we find
\begin{eqnarray}
\text{det}
\begin{pmatrix}
	-{\bf R}_{hh} & {\bf I}_{2M} \\
	{\bf I}_{2M} & {\bf X}_{2M}
	\end{pmatrix}
=
\text{det} ({\bf X}_{2M}) \, \text{det} (-{\bf R}_{hh} - {\bf X}_{2M}) = \text{det} ({\bf I}_{2M} + {\bf X}_{2M} {\bf R}_{hh} ). 
\end{eqnarray}
Therefore, the unnormalized Wigner function can now be simplified as
\begin{eqnarray}\label{eq:Wigner-3-Mmode}
W(\vt{\alpha}; \tilde{\rho}_{M}) &=&
\frac{2^M \mathcal{ P}_0}{\pi^{M} \, \bar{\vt{n}}!}
\bigg[\text{det} ({\bf I}_{2M} + {\bf X}_{2M} {\bf R}_{hh} ) \bigg]^{-1/2}
\exp \bigg\{ \frac{1}{2} \vt{y}_h^{\top} ({\bf I}_2 - {\bf X}_{2M} {\bf R}_{hh} )^{-1} {\bf X}_{2M} \vt{y}_h \bigg\} \nonumber\\
&&
\times \exp\bigg\{  
- \vt{v}^{\top} {\bf X}_{2M}({\bf I}_{2M} + {\bf X}_{2M} {\bf R}_{hh})^{-1} ({\bf I}_{2M} - {\bf X}_{2M} {\bf R}_{hh}) \vt{v} 
\bigg\} \nonumber\\
&&
\times %\int \mathrm d^2 \bar{\vt{\alpha}} \int \mathrm d^2 \bar{\vt{\beta}} ~
\prod_{k=M+1}^{N} \bigg(\frac{\partial^2}{\partial \alpha_k \partial \beta_k^*} \bigg)^{n_k}
\exp\bigg( \frac{1}{2} \vt{\gamma}_d^{\top} {\bf A} \vt{\gamma}_d + \vt{z}^{\top} \vt{\gamma}_d \bigg) \bigg|_{\vt{\gamma}_d = 0},
\end{eqnarray}
where
\begin{eqnarray}
{\bf A} &=& {\bf R}_{dd} - {\bf R}_{dh} ({\bf I}_{2M} + {\bf X}_{2M} {\bf R}_{hh} )^{-1} {\bf X}_{2M} {\bf R}_{hd}, \nonumber\\
\vt{z} &=& \vt{y}_d + {\bf R}_{dh} ({\bf I}_{2M} - {\bf X}_{2M} {\bf R}_{hh})^{-1} {\bf X}_{2M} \vt{y}_h + 2\, {\bf R}_{dh} ({\bf I}_{2M} + {\bf X}_{2M} {\bf R}_{hh})^{-1} \vt{v}, \nonumber\\
\vt{v} &=&
	\begin{pmatrix}
	\vt{\alpha}^* \\
	\vt{\alpha}
	\end{pmatrix}
- ({\bf I}_{2M} - {\bf X}_{2M} {\bf R}_{hh})^{-1} {\bf X}_{2M} \vt{y}_h.
\end{eqnarray}

\end{widetext}

\vspace{10 mm}

\bibliography{reference}

\begin{thebibliography}{78}
\bibitem[{nic(Note)}]{niconote}
 (\bibinfo{year}{Note}), \bibinfo{note}{an alternative approach to deriving the
  coefficients can be obtained using operator methods and the notion of loop
  halfnians \cite{nico}.}
\end{thebibliography}

\end{document}